\documentclass{uhthesis}
\usepackage{graphicx,natbib,courier,longtable,array}

\begin{document}

\newcommand\esp{e^{-}\mathrm{s}^{-1}\mathrm{pix}^{-1}}
\newcommand\esr{e^{-}\mathrm{read}^{-1}\mathrm{pix}^{-1}}
\newcommand\um{\mu\mathrm{m}}
\newcommand\us{\mu\mathrm{s}}
\newcommand\na{New Astronomy}
\newcommand\procspie{Proceedings of the SPIE}
\newcommand\aj{Astronomical Journal}
\newcommand\apj{Astrophysical Journal}
\newcommand\pasp{Publications of the ASP}
\newcommand\apjl{Astrophysical Journal, Letters}
\newcommand\apjs{Astrophysical Journal, Supplement}
\newcommand\aap{Astronomy and Astrophysics}
\newcommand\aaps{Astronomy and Astrophysics, Supplement}
\newcommand\mnras{Monthly Notices of the RAS}
\newcommand\farcs{'}
\newcommand\arcsec{''}
\newcommand\earth{♁}
\newcommand\Earth{♁}
\newcommand\plus{\texttt{+}}

\frontmatter
\title{HgCdTe APD Arrays for Astronomy: Natural Guide Star Wavefront Sensing and Space Astronomy}
\author{Dani Atkinson}
\date{9 May 2018}
\chairperson{Donald Hall}
\memberA{Christoph Baranec}
\memberB{Jessica Lu}
\memberC{Jeffrey Kuhn}
\memberD{John Learned}
\maketitle
\makesig

\makecopyright{2018}
{Dani Atkinson}

\dedication{\textit{Dedicated to my four parents Kristyn, Dale, Sophia, and Christopher, and the medical staff of the Queen's Medical Center of Honolulu, HI, and Hilo Medical Center of Hilo, HI, all for their effective intervention in the unexpected events of my life.}}

\makeacknowledgements{This work includes three published papers which were written with co-authors. In the two on the behavior of the SAPHIRA detector this includes adviser Donald Hall, engineer Shane Jacobson, and engineer and Leonardo representative Ian Baker. In the paper following up on \textit{Kepler} Objects of Interest this includes committee member Christoph Baranec, Carl Ziegler, Nicholas Law, Reed Riddle, and Tim Morton. I thank all of these people for their work on the papers which have been included in this dissertation.

I would also like to thank the staff of the Hilo branch of the Institute for Astronomy. This includes Pamela Lau, Sandra Miyata, and Mapuana Hauani'o. Their work has included ensuring paychecks, running to my bank to deposit money for travel, and providing delicious snacks.}

\makeabstract{This dissertation describes work I have conducted over five academic years 2013/14 through 2017/18 as a NASA Space Technology Research Fellow at the University of Hawai'i Institute for Astronomy. The focus has been the characterization and improvement of the Selex Avalanche Photodiode HgCdTe InfraRed Array (SAPHIRA), a $320\times256$@$24\um$ pitch metal organic vapor phase epitaxy mercury cadmium telluride array that provides new capabilities and performance for near infrared (NIR) astronomy. This has involved more than a dozen arrays, working closely with the manufacturer so as to provide feedback for improvement of the next generation.

The investigation has resulted in three lead authored publications in the Astronomical Journal which, as published, constitute the core of this dissertation:

\begin{itemize}
  \item An investigation into the SAPHIRA's dark current, a critical performance characteristic of astronomical detectors that determines the inherent background of observations. This dark current is comparable to other NIR devices. Published as \cite{atkinson2017a}.
	\item The SAPHIRA's ability to detect individual photons, an ability referred to as photon counting. Counting NIR photons gives unique new capabilities in AO and for future time-resolved scientific applications. The SAPHIRA is the only device of this size with these capabilities. Published as \cite{atkinson2018}.
	\item The characterization of \textit{Kepler} Objects of Interest and contaminating companions performed with Near Infrared Camera 2 (NIRC2) on the Keck II telescope, following on their discovery by the Palomar 1.5-m (P1.5-m) telescope's Robotic-Adaptive Optics (Robo-AO) instrument. Published as \cite{atkinson2017b}.
\end{itemize}

These are preceded by chapters providing 1) an introduction, 2) a description of the SAPHIRA array, 3) a more detailed treatment of setup and characterization together with 4) a chapter covering telescope deployments. They are followed by a chapter covering the three NASA Site Experiences performed as part of my Space Technology Research Fellowship. The dissertation concludes with a brief summary of my most significant achievements.}

\tableofcontents
\listoftables
\listoffigures

\mainmatter
\chapter{Introduction}\label{ch:introduction} 

Many astronomical observations focus on the visible and nearby wavelengths out to $1.0\um$, initially due to accessibility with simple Silicon charge-coupled devices \citep{french1975}. However, more difficult observations in the adjacent near infrared (NIR) are vital for many studies, from high redshift galaxies to exoplanetary atmospheres. Seeing NIR requires a more complicated device, and astronomical NIR arrays continue to make great strides (e.g.~\cite{gooding2016,zandian2016,finger2017b,rauscher2018}). The HgCdTe Astronomical Wide Area Infrared Imager (HAWAII) series developed by the University of Hawai'i Institute for Astronomy (UH-IfA) is leading the field of NIR arrays covering wavelengths of $1-5\um$, and is currently in the Hubble Space Telescope's Wide Field Camera 3 \citep{hall2000,hill2003}. Fifteen HAWAII-2RG arrays have been tested and are to be deployed in the James Webb Space Telescope (JWST) when it launches in 2020 \citep{rauscher2011,rauscher2013,rauscher2014}.

Applying NIR to adaptive optics (AO) also contains great advantages over visible detectors. The use of AO compensates for distortion in light caused by the atmosphere, resulting in better images \citep{mccall1978}. AO requires lower noise and faster readout devices than typical NIR arrays. However, NIR wavelengths are more capable in AO than visible light as the same residual wavefront errors produce higher quality images in NIR, which can also run on much dimmer stars \citep{beckers1986,goad1986}. AO typically require information read out from the detector at rates $\geq100\mathrm{Hz}$ as the atmospheric timescale is roughly a few ms, and AO frequently use $\sim1\mathrm{kHz}$ \citep{hardy1998}. Running so quickly produces a diminished signal that leaves a detector photon-starved and very limited by read noise. Meeting photon-starved applications is optimized by moving to avalanche photodiodes (APDs), which use avalanche gain to effectively reduce read noise on the signal. Reducing noise and providing time resolution also makes an APD the detector of choice for observations of faint or quickly changing objects.

APDs suffer from the existence of excess noise. The effect occurs in APDs as the avalanche effect produces noisier measurements than a detection on a conventional device. Excitations produce both electrons and holes, which then travel through the gain medium in opposing directions and cause further excitations, and so on. The process is stochastic and the excess noise increases linearly with gain.

Mercury cadmium telluride (HgCdTe) was pursued for NIR APDs as it was already a superb medium for NIR imagers. HgCdTe APDs were shown to have a high gain, high quantum efficiency, and low excess noise \citep{jack2001}. The superior excess noise performance is attributed to an avalanche that only occurs for electrons, not holes \citep{bryan2012}. The use of linear mode in HgCdTe APDs also allows the detection of multiple photons without requiring a reset, contrary to Geiger mode APDs \citep{jack2011}.

\begin{figure}
   \begin{center}
   \begin{tabular}{c}
   \includegraphics[height=8.0cm]{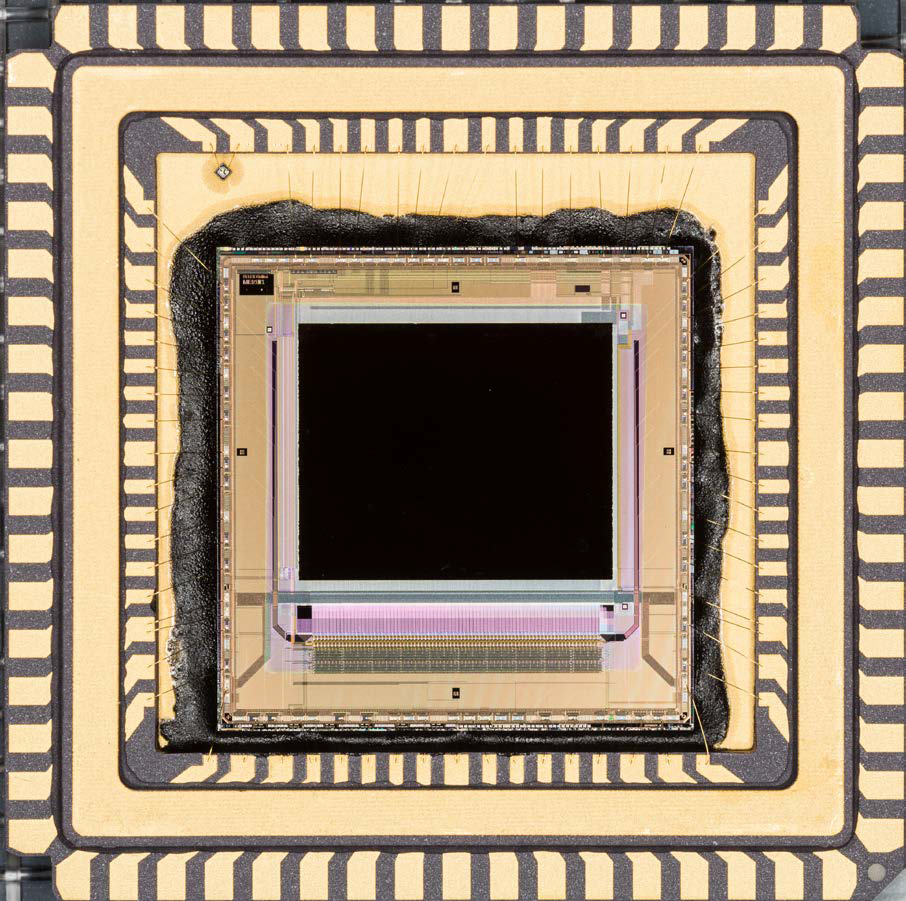}
   \end{tabular}
   \end{center}
	 \vspace{-16pt}
   \caption[The SAPHIRA Detector] 
   { \label{fig:SAPHIRA} 
A standard SAPHIRA detector. The dark rectangle in the center is the APD array, and the temperature sensor is visible in the upper-left corner. In operation it is mounted to a 68-pin leadless chip carrier.}
   \end{figure}

\begin{figure}
   \begin{center}
   \begin{tabular}{c}
   \includegraphics[height=8.0cm]{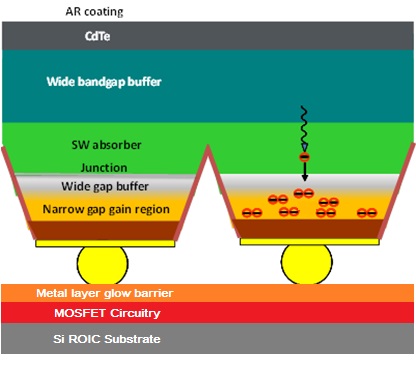}
   \end{tabular}
   \end{center}
	 \vspace{-16pt}
   \caption[Physical Structure of the SAPHIRA] 
   { \label{fig:physical1} 
The physical structure of SAPHIRA pixels. Electrons are pictured at right being produced in the absorber layer by incident photons, then traveling downwards and producing the avalanche in the gain region. The yellow circles are indium bumps bonding the pixels to the ROIC underneath. The depth of individual layers is varied experimentally to improve performance. (\textit{Original figure courtesy Leonardo.})}
   \end{figure}

\begin{figure}
   \begin{center}
   \begin{tabular}{c}
   \includegraphics[height=8.0cm]{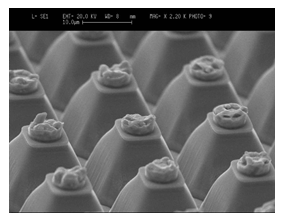}
   \end{tabular}
   \end{center}
	 \vspace{-16pt}
   \caption[Mesa Structure of SAPHIRA Pixel] 
   { \label{fig:mesa} 
The mesa structure of SAPHIRA pixels is clearly visible in this image, with indium bumps at top to connect the pixel to the ROIC.}
   \end{figure}

The Selex Avalanche Photodiode HgCdTe Infrared Array (SAPHIRA) consists of a $320\times256$@$24\um$ pitch HgCdTe APD array hybridized to a complementary metal-oxide-semiconductor (CMOS) readout integrated circuit (ROIC) (see Figure~\ref{fig:SAPHIRA}). The APD consists of an absorption layer sensitive from $0.8\um$ to its cutoff wavelength at $2.5\um$, with a multiplication layer of $3.5\um$ cutoff in a mesa structure (see Figure~\ref{fig:physical1}). Pixels in SAPHIRA have a unique mesa design (see Figure~\ref{fig:mesa}). The ROIC provides 32 outputs reading pixels in adjacent columns in both full frame and sub-array mode.


At UH-IfA the SAPHIRA has been tested over several iterations, pursuing the world's first low-noise NIR detector array capable of both typical charge gain operation and more demanding photon counting. This is where a detector creates a signal  from a single photon large enough to be detected independently, allowing individual photons to be counted. Since 2013, 20 SAPHIRA devices have been investigated at UH-IfA (see Table~\ref{tab:detectors}). The Mark numbers indicated in the table show the version of MOVPE layer design used in its production. The first devices received at UH-IfA were Marks 2 \& 3. Later Marks 13 \& 14 were produced, which extended the wavelength coverage down to $0.8\um$. A high temperature anneal was also introduced to improve dark performance, with a longer anneal used for the Mark 14 relative to the 13. Later productions focus on attempts to improve high bias voltage tunneling current, with some success (see Chapter~\ref{ch:darkcurrent}).

\begin{table}
\caption{Investigated Detectors}\label{tab:detectors}
\centering
A full list of investigated $320\times256$@$24\um$ devices. Note that missing Mark numbers were either limited to a design stage and not produced, or were never sent to UH-IfA because they did not perform as expected. The first 5 numbers of a serial number indicate the source wafer, while the last 2 are the position on the wafer from which the detector came.

\begin{tabular}{ c c c c l }
Serial Number & Mark & ROIC & Received & Description \\
\hline
M02775-10 &  3             & ME-911  & 31 Oct 2013 & One of first received. Was deployed to IRTF,\\
					&								 &				 &						 & Palomar, and Subaru. Accidentally destroyed \\
          &                &         &             & during wirebond removal.\\
M02775-35 &  3             & ME-911  & 11 Mar 2015 & \\
M02815-12 &  2             & ME-911  & 31 Oct 2013 & One of first received.\\
M04055-06 &  5             & ME-911  & 31 Jul 2014 & \\
M04055-39 &  5             & ME-911  & 31 Jul 2014 & \\
M04935-17 &  10            & ME-911  & 11 Mar 2015 & \\
M06495-19 &  12a           & ME-911  & 21 Sep 2015 & \\
M06495-27 &  12a           & ME-911  & 21 Sep 2015 & \\
M06665-03 &  13            & ME-911  & 16 Oct 2015 & Currently deployed to Kitt Peak.\\
M06665-12 &  13            & ME-911  & 16 Oct 2015 & \\
M06665-23 &  13            & ME-1000 & 12 Oct 2016 & Best detector received.\\
M06665-25 &  13            & ME-1000 &  6 Jun 2016 & Currently deployed to Keck.\\
M06715-27 &  14            & ME-911  & 16 Oct 2015 & \\
M06715-29 &  14            & ME-911  & 16 Oct 2015 & \\
M06715-34 &  14            & ME-1000 & 12 Nov 2015 & Currently deployed to Subaru.\\
M09105-27 &  15            & ME-1000 & 14 Mar 2017 & \\
M09215-10 &  18            & ME-1001 & 21 Mar 2017 & \\
M09215-18 &  18            & ME-1001 & 28 Mar 2017 & \\
M09225-11 &  19            & ME-1001 & 14 Mar 2017 & \\
M09225-27 &  19            & ME-1001 & 10 Mar 2017 & \\
\end{tabular}

\end{table}

The ROIC is electronics bonded to the detection medium that operates the device and interacts with an external controller. Different designs of the ROIC were implemented to improve the device's characteristics. Reported are identifiers for the ROICs on the received devices. 

Over the years I have characterized the SAPHIRA detectors I have repeatedly investigated them in the K-band spectrometer (KSPEC) cryostat of IfA Hilo's Detector Lab, previously used for the fifteen H2RG arrays selected for the James Webb Space Telescope. I have used SAPHIRA for astronomical imaging at the Infrared Telescope Facility (IRTF) on Maunakea, used for lucky imaging. It then went to the Robo-AO instrument at the Palomar Observatory's 1.5-meter telescope (Robo-AO/P1.5m). There it provided wavefront sensing and produced data for paper published with me as second author \citep{baranec2015}. I have also used SAPHIRA in AO systems with the Subaru Extreme Adaptive Optics (SCExAO) instrument at the Subaru telescope on Maunakea. The UH-IfA has further sent SAPHIRAs to the new deployment of the Robo-AO instrument at the Kitt Peak National Observatory's 2.1-meter telescope and to the Keck Planet Imager and Characterizer (KPIC) at the Keck Observatory on Maunakea. The SAPHIRA is also being sold for commercial use in the C-RED One, an infrared camera for adaptive optics sold by First Light Imaging, and many other uses are planned \citep{greffe2016}.

This dissertation covers the KSPEC measurements of both dark current and photon counting, and the characterization of contaminating companions in \textit{Kepler} Objects of Interest (KOIs). The SAPHIRA is described in Chapter~\ref{ch:saphira}. For descriptions of the laboratory setup and measurements performed with KSPEC, see Chapter~\ref{ch:setup}. Deployments of the SAPHIRA to several telescopes is described in Chapter~\ref{ch:deployments}.

Chapters~\ref{ch:darkcurrent}, \ref{ch:photoncounting}, and \ref{ch:palomar} are published papers I wrote as lead author. Measurements of dark current in multiple SAPHIRA devices are in Chapter~\ref{ch:darkcurrent}, which is \cite{atkinson2017b}. Photon-counting with a SAPHIRA is presented in Chapter~\ref{ch:photoncounting}, from \cite{atkinson2018}. Chapter~\ref{ch:palomar} covers the use of NIRC/Keck II to characterize KOI contaminating companions initially detected by the Robo-AO instrument at Palomar-1.5m, from \cite{atkinson2017a}. After the publication of this paper, my code for this work was used for further investigations published as \cite{ziegler2016} and \cite{schonhut2017}.

NASA Site Experiences I performed as part of my NASA Space Technology Research Fellowship are described in Chapter~\ref{ch:siteexperiences}. The dissertation as a whole is summarized in Chapter~\ref{ch:conclusions}.


\chapter{SAPHIRA: The HgCdTe APD Array}\label{ch:saphira}
HgCdTe is an amazing material for astronomical detectors as it can detect from optical wavelengths all the way to long-wave infrared radiation. As an alloy its sensitivity is controlled simply by the mixture of HgTe (bandgap 0 eV) and CdTe (bandgap 1.5 eV) that make it up. In addition to the sensitivity range it has proven useful in making low-noise NIR detectors that show a minimum of dark current \citep{loose2003}. These properties have made it extremely useful for astronomical detectors.

The development of APDs using HgCdTe was a major step with the material \citep{delyon1999}. An APD uses a high voltage to cause an avalanche effect and multiply signal. Both electrons and holes (absences of electrons as virtual particles) carry charge, and except for in HgCdTe both move in opposite directions through the medium in response to applied voltage. They accelerate until they collide with and ionize other molecules, producing other charge carrier pairs, increasing the response to an incident photon. Those then accelerate and cause collisions and so on. This amplification of the original particle is called an avalanche. Geiger mode APDs (G-APDs) experience a full breakdown of the medium and require being reset before resuming detection, while linear mode APDs (L-APDs) can detect make multiple detections without a reset. Most APDs in use are G-APDs as L-APDs perform relatively poorly except what is present in this work. Avalanching is inherently noisy, as both an electron and a hole are produced by each avalanche collision in a stochastic process. This excess noise $F$ is defined as the ratio of the square of the mean to the mean square of the APD gain. It measures the signal-to-noise ratio degradation from the avalanche process. It is predicted by the equation





\begin{equation}\label{eq:F}
F = \kappa M + (2 - \frac{1}{M})(1 - \kappa)
\end{equation}

where $M$ is the gain and $\kappa$ is the ratio of hole impact ionization to electron impact ionization \citep{mcintyre1966}. In standard APDs, the excess noise then increases with the gain $M$ as according to Equation~\ref{eq:F}. The ratio $\kappa$ is determined by the gain material in use, which as $\kappa$ approaches zero moves towards a minimum of $F = 2$ by this theory.


Avalanches in HgCdTe are unique as they are single-carrier. Holes produced are virtually immobile and do not cause further avalanches \citep{beck2004,bryan2012}. This is called a ballistic avalanche because the effect only moves in one direction across the detection medium. This is much less noisy than the typical APD and the excess noise has been shown to not rise above $F = 1$, thus having no effect \citep{finger2016}. That makes it a noiseless multiplication. The original theory setting $F = 2$ as the minimum has been previously disproved and a revision to the theory has been attempted \citep{mcintyre1999}. Note that for the higher temperature of $T = 85\mathrm{K}$, a modest increase of $F$ with avalanche gain has been observed for HgCdTe \citep{finger2016}.

\begin{figure}
   \begin{center}
   \begin{tabular}{c}
   \includegraphics[height=8.0cm]{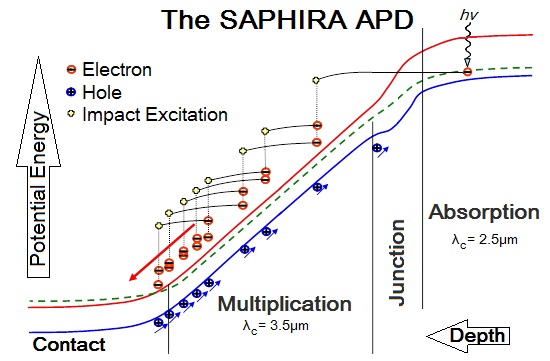}
   \end{tabular}
   \end{center}
	 \vspace{-16pt}
   \caption[Bandgap of the SAPHIRA] 
   { \label{fig:bandgap1} 
The single-carrier electron avalanche is a major noise advantage HgCdTe has over other NIR detection materials. The bias voltage $V_{bias}$ is applied across the multiplication and junction regions. Though photons are depicted here being captured in the absorption region, it is possible for $2.5\um \leq \lambda \leq 3.5\um$ photons to penetrate to and be absorbed in the multiplication layer. (\textit{Original figure courtesy Leonardo.})}
   \end{figure}

The SAPHIRA APD employs a $320\times256$@$24\um$ device structure. MOVPE allows for advanced bandgap design, and different depths can be independently optimized. Photon absorption occurs in the P-type absorber, creating electron-hole pairs. The electrons thermally diffuse to the junction. The high electric field is dropped across a weakly doped N-type region, the multiplication layer. Here electrons are accelerated and generate other electron-hole pairs by impact ionization. The low mobility hole acquires energy from the applied electric field very inefficiently and readily loses it to optical phonons \citep{rogalski2005}. The process is essentially a pure electron cascade with an exponential avalanche gain versus bias voltage profile. The relationship between potential energy and depth is shown in Figure~\ref{fig:bandgap1}, which illustrates that the history-dependent nature of the avalanche gain underpins the low noise figure of HgCdTe.
	
In a SAPHIRA detector where the excess noise $F = 1$, the avalanche gain increases the signal of an observation without increasing the excess noise, increasing the SNR \citep{finger2016}. Read noise is usually dominated by the noise of the input MOSFET in the unit cell. The avalanche gain increases the signal relative to this read noise. This noise is typically Gaussian in output \citep{ohta2007}. Read noise is the dominant noise source in low signal and time resolved observations. The avalanche gain then effectively reduces the read noise relative to the incoming signal.

Electronics bonded to the detector medium are the ROIC, which operates the array and interacts with external controllers. The ROIC for the SAPHIRA was redesigned twice to improve performance. The first received was the ME-911. The ME-1000 was designed to enable a read-reset-read mode requested by ESO. In both the ME-911 and -1000 there was a large glow point source near one corner of the detector's array \citep{atkinson2016}. This was attributed to a circuit mask error in the design resulting in a floating gate. It was resolved in the ME-1001.

It was discovered on receipt of the first arrays that unusual electronic behavior made the detector not function if two readback channels (LSP and SYNC\_OP) were connected. This issue did not occur at ESO or Leonardo and was suspected to result from the length of connecting wires in KSPEC. Disconnecting the channels from the device allowed it to function.

\begin{figure}
   \begin{center}
   \begin{tabular}{c}
   \includegraphics[height=8.0cm]{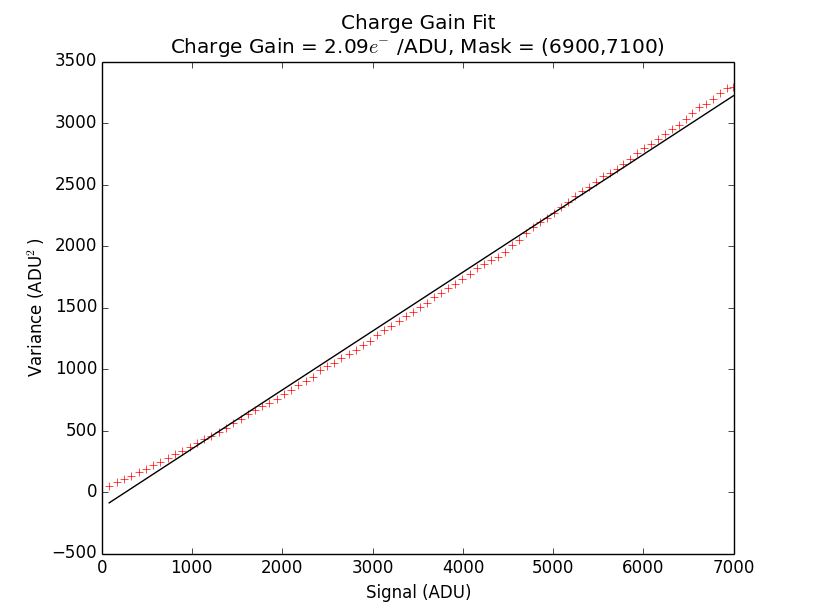}
   \end{tabular}
   \end{center}
	 \vspace{-16pt}
   \caption[Charge Gain Measurement] 
   { \label{fig:chargegain} 
A charge gain measurement for a device on an ME-911 ROIC. The linearity of the data being fit to determines the accuracy of the measurement.}
   \end{figure}

The ROIC is then connected to an external controller which operates the detector, discussed in detail in Chapter~\ref{ch:setup}. The voltage across each individual pixel is converted to analog-to-digital units (ADUs). The charge gain in $e^{-}/\mathrm{ADU}$ is measured by determining the signal versus variance of the photo-charge accumulated when the pixel is illuminated. This has a Poisson noise relationship, which is fit to the plot of signal versus variance behavior from some pixels with an LED on, and is dependent on the pixels selected to perform the measurement. Investigated pixels must be masked for similar behavior, but a strong mask will produce an aberrant result. The mask includes only pixels with a defined range of total signal values, and are selected to produce a linear signal/variance to be fit by Poisson (see Figure~\ref{fig:chargegain}). Starting with a full frame of SAPHIRA data, a linear fit uses approximately 900 or $1\%$ of available pixels. With signal as the x-axis and variance as the y-axis, the charge gain in $e^{-}/\mathrm{ADU}$ is then the inverse of the fit's slope. For the ME-911 on an ARC controller the charge gain was measured as $2.1e^{-}/ADU$, for ME-1000 $1.6e^{-}/ADU$, and for ME-1001 $4.7e^{-}/ADU$. Published papers use an earlier measurement for the ME-911 charge gain of $2.9e^{-}/ADU$, though focus of the papers is on ME-1000 devices. The ME-1001 result is from preliminary measurement on a single Mark 19 device. The ME-1001 has an identical ROIC to a ME-1000, so the difference in charge gain is attributed to a change in capacitance, likely a function of the Mark 19.

Individual pixels have limits to the amount of charge they can hold. This is referred to as well depth or well capacity, as a pixel that cannot take any more charge is a 'full well'. This occurs for SAPHIRA detectors at $>40,000 ADUs$. Note that when at a typical operating voltage of $3.5V$, non-linearity also starts at a depth of $\sim20,000 ADUs$. A reset of the array empties the well.

\section{Summary}
The SAPHIRA is a unique HgCdTe APD $320\times256$@$24\um$ array. HgCdTe provides a single-carrier avalanche to produce a lower excess noise than conventional NIR APDs of other materials. In the SAPHIRA, HgCdTe covers a wavelength rage from $0.8$ to $2.5\um$. The use of MOVPE allows the control of its structure more than is possible with conventional MBE techniques. The SAPHIRA design is an ongoing project to minimize the trap-assisted tunneling that generates dark current. 


\chapter{Setup \& Characterization}\label{ch:setup}
\begin{figure}
   \begin{center}
   \begin{tabular}{c}
   \includegraphics[height=8.0cm]{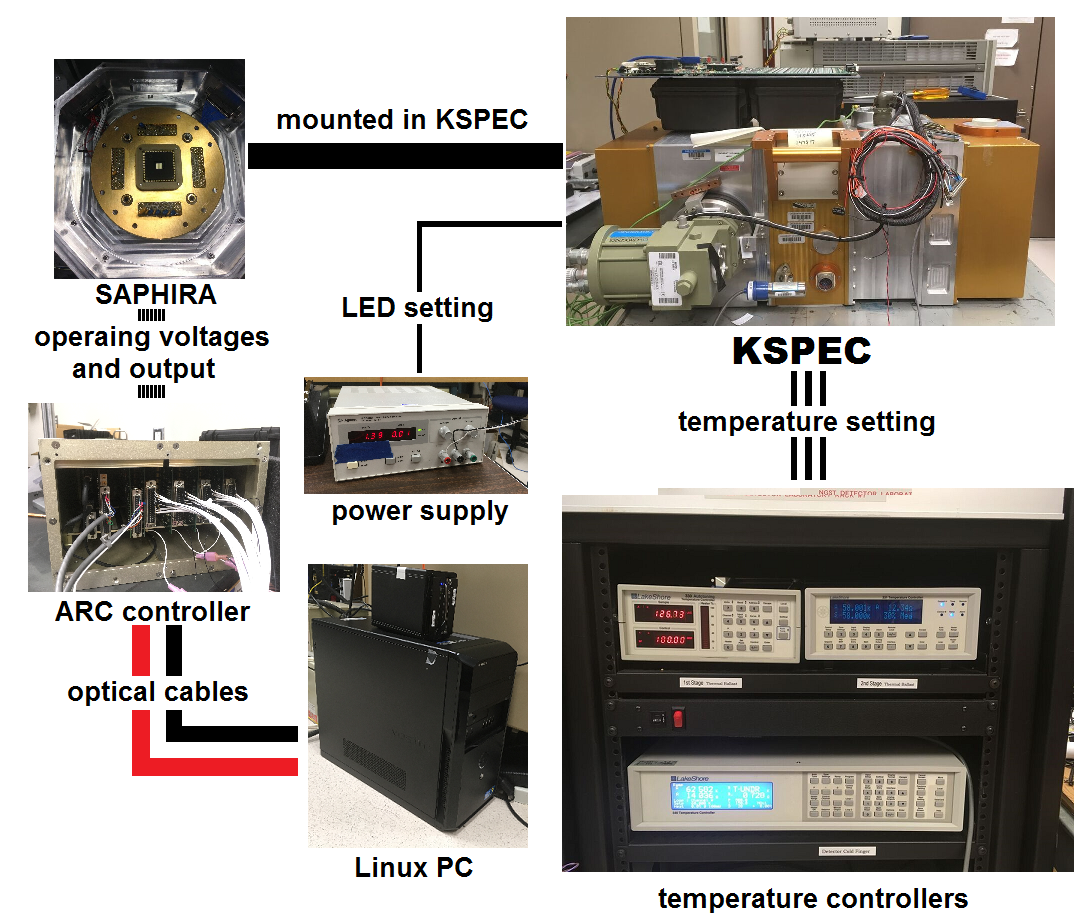}
   \end{tabular}
   \end{center}
	 \vspace{-16pt}
   \caption[Laboratory Setup] 
   { \label{fig:diagram} 
The diagram of the laboratory setup for SAPHIRA characterization. KSPEC is a cryostat in which SAPHIRA is mounted, and temperature is determined by a set of external controllers. Characterization LEDs in KSPEC are run by an external power supply. Voltages in and data out is handled by the ARC controller, which connects to the PC by optical cables.}
   \end{figure}

The laboratory setup for SAPHIRA characterization in the Hilo Detector Lab made use of several connected systems (see Figure~\ref{fig:diagram}). The KSPEC cryostat has a closed-cycle cryocooler, with heaters inside run by external temperature controllers to set the detector temperature. LEDs are powered by an external power supply, operated manually. The Astronomical Research Cameras (ARC) controller provides both voltages to the controller and reads data on the output channels, and is connected to a Linux PC from which the detector is run.

\begin{figure}
   \begin{center}
   \begin{tabular}{c}
   \includegraphics[height=8.0cm]{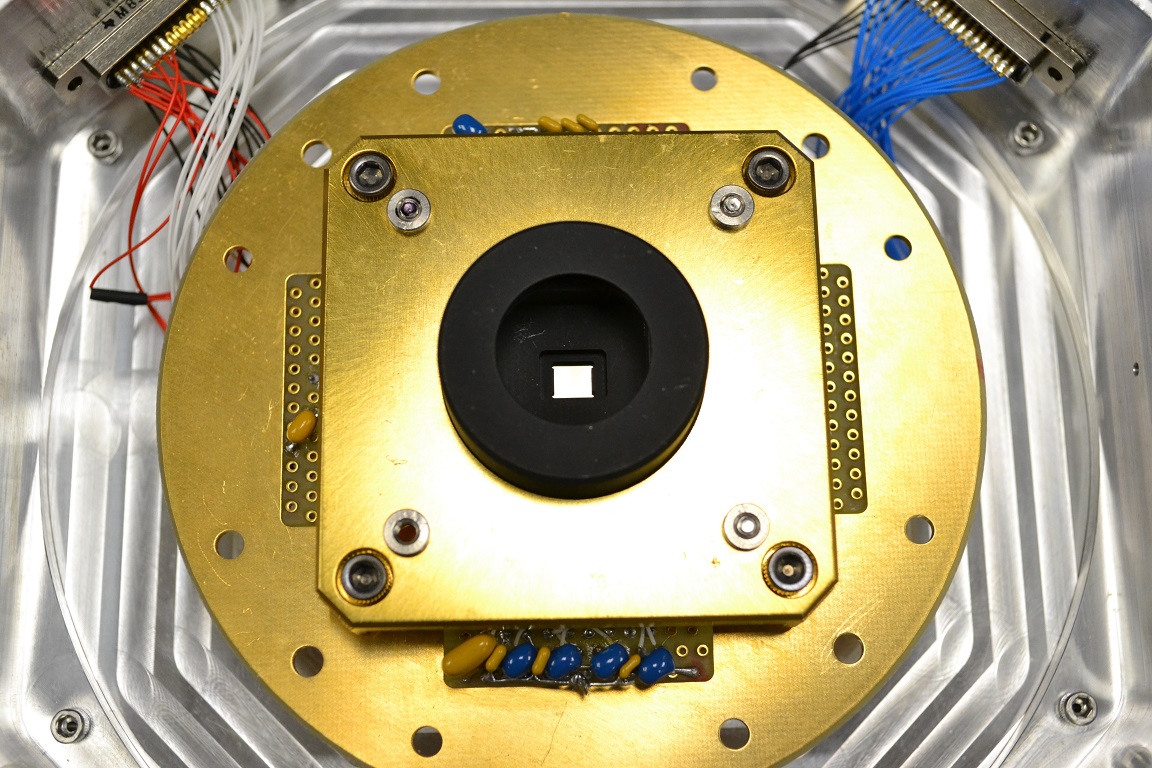}
   \end{tabular}
   \end{center}
	 \vspace{-16pt}
   \caption[The KSPEC Cryostat] 
   { \label{fig:KSPEC1} 
The socket holding SAPHIRA uses springs to maintain constant pressure with the electrical contacts and the clamp assembly in place over the array. The small mask over the surface of the detector is visible at the center.}
   \end{figure}

The SAPHIRA arrays were characterized in KSPEC (see Figure~\ref{fig:KSPEC1}). Introduced in Chapter~\ref{ch:introduction}, KSPEC was originally an $IJHK$-band spectrometer \citep{hodapp1994}. It has since been converted into a detector testbed, originally for the NGST/JWST detector program \citep{hodapp1996}. Making the SAPHIRA work in KSPEC required modifying assembly code from the H2RG to create a working interface. With help from PI Don Hall, Shane Jacobson, and Fred Hee, I found that connections to some of the detector's binary output signals prevented the detector from running correctly. This problem was unique to KSPEC as it did not appear in operation tests at Leonardo or ESO, and was judged to be related to the length of the connecting wires for readbacks LSP and SYNC\_OP. Removing those wires made the SAPHIRA work correctly.

Temperature controllers keep the detector temperature stable by running heating resistance coils with feedback from sensors in KSPEC. Temperature is set manually in 3 controllers mounted in a rack next to the cryostat. The 1st, 2nd, and detector stages are typically set to 100K, around 58K, and 60K, respectively. An additional temperature sensor is part of the device and was used to calibrate the stage controllers. It glows brightly and is not on during typical operation.

The detector itself is operated by a $3^{\mathrm{rd}}$ generation ARC detector controller. The controller uses mostly standard ARC boards (ARC-22, ARC-32) including four 8-channel IR readout boards (ARC-46) \citep{leach2000}. The bias/utility board has been replaced with an extremely low noise design from Australian National University and reproduced at UH-IfA. All voltages applied to the detector are from this board. These boards are installed in a twelve-slot ARC chassis with a single backplane. The ARC-46 boards are inserted with empty board slots between them to minimize interference from adjacent boards. Fiber optic lines connect the controller to a PCIe board (ARC-66) in a Linux desktop computer. The instrument is operated via a set of scripts and code making use of the v3.0 ARC API. The v3.0 ARC API is used as it correctly compiles the scripts whereas the v3.5 ARC API has additional requirements that make it more difficult to use.

I wrote the code and scripts that run on the PC to talk to the detector and perform a wide array of required functions. All is on the computer \texttt{HxRG-vostro1} in the Hilo Detector Lab. In this document, script is simple sets of command line entries referred to as \texttt{scriptname.sh}, while code is in C\plus\plus and is referred to in the format \texttt{./codename} from the SAPHIRA directory. The code uses C\plus\plus to interact with the controller via the ARC API.  All C\plus\plus code is compiled by the script \texttt{build.sh}, which makes calls with needed arguments to G\plus\plus. Code and scripts to perform necessary functions are in \texttt{/users/H4RG/SAPHIRA} and subdirectories, as a new user was not created for SAPHIRA investigations. 


The controller operates on Motorola DSP56000 assembly language code, which I derived from code for H2RG operation provided by both Bob Leach of ARC and Marco Bonati of CTIO. There are three assembly code files that run compiled on the ARC controller, \texttt{tim.asm}, \texttt{timboot.asm}, and \texttt{timIRmisc.asm}. Changes to any of them require the code to be recompiled, which is a run of \texttt{./Wine\_tim}. The code is written to the controller immediately before every detector power-up with \texttt{./loadtim}. Voltages supplied to the detector by the bias board are set in the controller by this code. This includes $VDD$, $VDDA$, $VDDPIX$, and $VDD\_OP$, the 4 necessary channels which set the detector's effective operating voltage (see Table~\ref{tab:voltages}). These are the digital supply, analog supply, pixel array supply, output buffer supply, respectively. All data presented here was taken with $VDDx = 3.5\mathrm{V}$, below which the detector becomes inoperable. Some deployments used higher voltages of $VDDx \sim5\mathrm{V}$. Also supplied are PRV, VCI, and COMMON. PRV simply sets the pixel reset voltage and is one end of the applied bias voltage, while VCI is the voltage clamp input.

\begin{table}
\caption{SAPHIRA Operating Voltages}\label{tab:voltages}
\centering
Voltages applied to the SAPHIRA detector from the ARC controller while running. Other voltages were used on some deployments but those results are not published in this work. Voltages used for adaptive optics applications are also shown. The COMMON voltage is changed to produce a different bias voltage $V_{bias}$ than 2.5V.
\begin{tabular}{ c c c c c }
Name & Typical & AO & Maximum & \\
\hline
VDD    & 3.5V & 5.0V & 7V  & Digital positive supply\\
VDDA   & 3.5V & 4.9V & VDD & Analog positive supply\\
VDDPIX & 3.5V & 4.8V & VDD & Pixel array positive supply\\
VDDOP  & 3.5V & 4.8V & VDD & Output buffer positive supply\\
PRV    & 3.5V & 3.6V & VDD & Pixel Reset Voltage\\
VCI    & 2.5V & 3.1V & PRV & Voltage Clamp Input\\
COMMON & 1.0V & 0.8V & VDDA + 0.3V & Array COMMON Connection\\
VSS    & 0V   & 0V   & 0V  & VSS connection (digital)\\
VSSA   & 0V   & 0V   & 0V  & VSSA connection (analog)\\
\end{tabular}

\end{table}

The bias voltage is set as $V_{bias} = \mathrm{PRV} - \mathrm{COMMON}$. It then determines the avalanche gain, where $V_{bias} = 2.5\mathrm{V}$ is taken as unity. (See Section~\ref{s:avalanche} for a discussion of this assumption.) The gain apparently is insensitive to bias around this value, though at lower voltages capacitance effects bring the effective gain down. Changes to the bias voltage require edits to the assembly code, with a controller reset (\texttt{./reset}), load of the new code (\texttt{./loadtim}), power on sequence (\texttt{./pon}), and setting the detector to idle mode (\texttt{./idle}) bring the detector to a running state. A running detector requires the detector to simply be powered down (\texttt{./poff}) before being ready for a reset. An automated script runs these scripts to power the detector off, reset it, load the new code, and power it back on (\texttt{./cycle}).

The detector can be run with either a full frame $320\times256$ readout or a subarray window. The size and location of a subarray is controlled in the assembly code, where the window is set with calls in the settings where the columns to be read are the first ten bits and the rows read are the next 256 bits. Windows are set with bytes of 8 bits in the code, but the controller will ignore bits written before the full length of the window setting bits. Read and reset windows are set independently.

With the ARC controller the full frame of the detector is read out at 100Hz. Using fewer readouts than the full 32 (16, 8, 4, 2, or 1) is an available setting in the SAPHIRA but was not used in investigations presented in this work (except in studies of NIR glow from the device). Note that setting the number of outputs less than 32 changes the number of bits in the window setting. For the smallest windows with a single read of all outputs ($32\times1$ for the typical 32 outputs), the frame rate is equal to the per pixel rate of $\sim250\mathrm{kHz}$. The subarray can be positioned anywhere in the detector. The typical 32 outputs divide the detector into 10 vertical stripes, each 32 columns across.

Detector temperature is measured on a cold finger in thermal contact with the detector itself, and is accurate to $\sim3{K}$ of the detector's actual temperature. A temperature diode mounted on the leadless chip carrier was used to calibrate the temperature measurements from the cold finger. The operating temperature is 2.3K higher than the measured detector stage temperature at 40K and 2.9K higher at 100K. The diode sensor glows brightly in the NIR when powered and was not operated when other measurements were being taken.

\begin{figure}
   \begin{center}
   \begin{tabular}{c}
   \includegraphics[height=8.0cm]{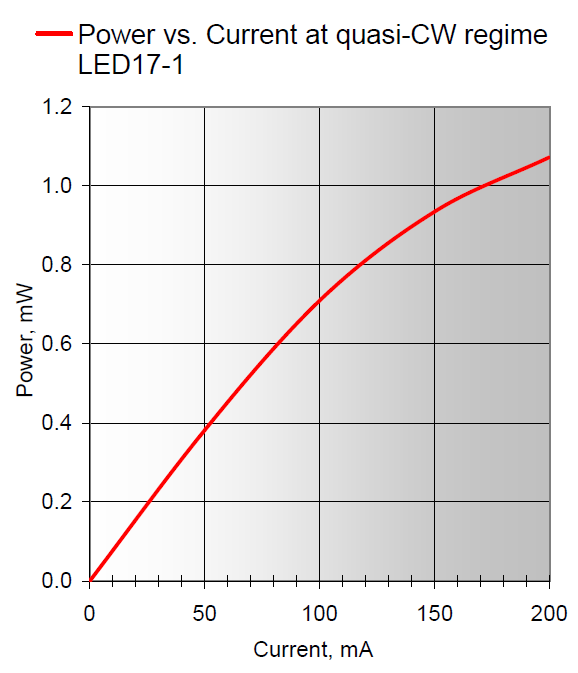}
   \end{tabular}
   \end{center}
	 \vspace{-16pt}
   \caption[LED Power vs. Current] 
   { \label{fig:LED} 
The power for a $\lambda = 1.75\um$ LED vs. current. The LED is a continuous wave device, not pulsed. In characterization the LED was almost entirely set to 1mA, which was the minimum measurable current from the power supply.}
   \end{figure}

Calibration LEDs at $1.05$, $1.30$, $1.75$, and $3.1\um$ are mounted to an integrating sphere in the cryostat, to provide uniform illumination across the detector when active. The $H$-band LED $\lambda = 1.75\um$ was used for presented avalanche gain and photon counting measurements in this work (see Figure~\ref{fig:LED}). The LED power level is determined by a standard power supply next to the cryostat, and is set manually by the operator. Typical current powering the LED is 1mA (the minimum measured by the power supply; lower currents can be explored with alternative setups).

Taking data from the detector requires running the \texttt{./expose} script. The call to this script has two arguments, where the first is the number of frames and the second is the wait time between frames in seconds. A common run is \texttt{./expose 100 0}, which gives 100 frames with no wait from frame to frame. The exposure code skips the wait code if a wait of 0 is requested, otherwise it goes through cycles measuring time elapsed until it passes the requested wait time. Series of consecutive sets of frames can also be run with \texttt{series.sh}, which uses a loop to call \texttt{./expose} a number of runs and wait times that are hard-coded into it. Individual \texttt{./expose} runs called by series are saved into individual files.

The APD can be reset multiple times during a single call to \texttt{./expose}. This creates multiple ramps in a single run, where a ramp is an accumulation of electrons with a reset before and after. Unlike a CCD, the SAPHIRA uses a non-destructive readout (NDRO). The NDRO evaluates the readout from a given set of pixels without a reset, so the pixel may be evaluated many times without resetting it. In use the APD would be reset more often for large bias voltages to prevent not only saturation but also the non-linearity that occurs moving toward saturation, usually at half the range. The number of frames before resetting is controlled by the variable RSTPER in the assembly code, and changes to it require a recompilation and reload of that code as described above.

Data from the SAPHIRA is collected and stored in a 3D \texttt{.fits} file, a datacube. It is then brought into Python code set up specifically to analyze SAPHIRA data. (The Python code is stored at https://github.com/QueenOfLasers/SAPHIRA.) Separate functions perform different types of analysis on the data. Charge gain and possibly avalanche gain corrections are applied at this time. If multiple frames are to be averaged together that is also done here. Then a correlated dual sampling (CDS) can be applied, where adjacent pairs of frames have one subtracted from the other. In most cases this uses only paired frames but can also use a rolling subtraction with every frame subtracted from the one before it. Avalanche gain is measured as the amplitude of signal over a given set of time for every investigated vias voltage. For dark current observations, the current across a very long time frame is the detection. In photon counting, a histogram is generated using \texttt{matplotlib} to show detected photons with the read noise of the detector.

Prominent measurements of the SAPHIRA performed in this work are described below.

\section{Avalanche Gain}\label{s:avalanche}
Measuring the avalanche gain of an APD is critical, and one of the first observations performed in the lab for every detector. The avalanche gain $G$ can be simply defined as the ratio between the number of initial electrons $N_{e^-\mathrm{in}}$ and the number of output electrons $N_{e^-\mathrm{out}}$: 

\begin{equation}
G = \frac{N_{e^-\mathrm{out}}}{N_{e^-\mathrm{in}}}
\end{equation}

A \texttt{.\/expose 200 0} ramp is taken with the LED off and another with the standard $H$-band $\lambda=1.75\um$ LED set to the minimally measured current of 1 mA. Full frames are used to spot anomalies on the device. The avalanche gain frame is produced by subtracting the LED off data from the LED on. The same data is produced for a range of voltages, starting usually at $V_{bias} = 2.5\mathrm{V}$, which is assumed to be unity gain. Note that the existence of unity gain voltage being a threshold below which there is no avalanche effect is assumed. This can be tested with accurate measurements of avalanche gain as a ratio of values to lower voltages than 2.5V. Lower voltages show an effective drop in avalanche gain that is an effect of capacitance in the device. Capacitance has been corrected as best as possible but a drop is still observed. This $2.5\mathrm{V}$ data is taken as unity as avalanche gain is unresponsive to bias voltage there. This means the avalanche gains vs. bias voltage are only relative, and can not be assumed to be absolute. For measuring avalanche gain, bias voltages are typically run with 1V intervals up to 14.5V. The timing code must be recompiled and detector power cycled for each voltage step. 

\begin{figure}
   \begin{center}
   \begin{tabular}{c}
   \includegraphics[height=8.0cm]{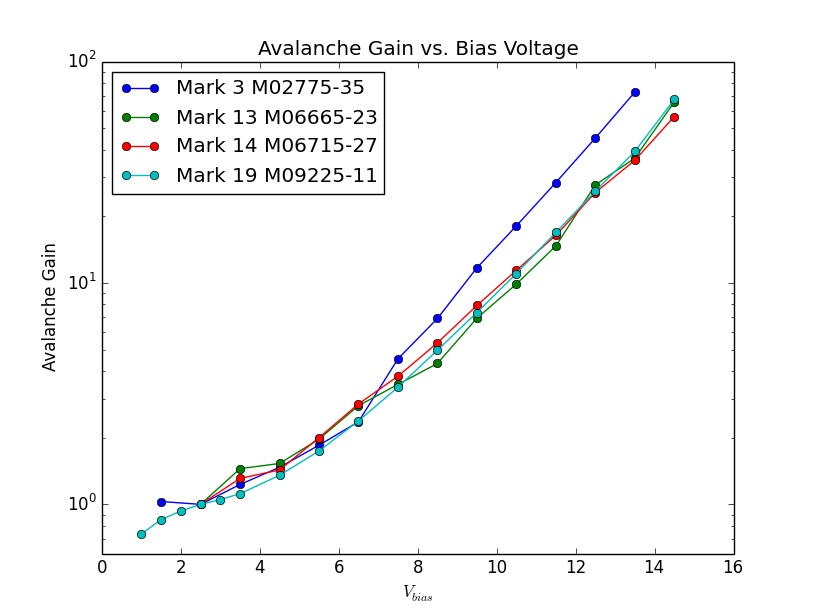}
   \end{tabular}
   \end{center}
	 \vspace{-16pt}
   \caption[Avalanche Gain for Multiple SAPHIRA]{\label{fig:avalanchegain0}
	Measured avalanche gain is higher for Mark 3 and consistent across multiple later SAPHIRA arrays. These values are used to correct dark current measurements to get a clear picture of dark performance independent of gain effects.}
   \end{figure}

The avalanche gain is measured the full amplitude or slope of the LED on - off over time. The measurement is stronger for Mark 3 by a factor of 2, and is otherwise very consistent for later generations of SAPHIRA detectors (see Figure~\ref{fig:avalanchegain0} \& Table~\ref{tab:avalanche}).

\begin{table}
\caption{Avalanche Gains}\label{tab:avalanche}
\centering
The measured avalanche gains plotted in Figure~ref{fig:avalanchegain0} from four different SAPHIRA arrays for a spread of bias voltages.
\begin{tabular}{ c c c c c }
Bias Voltage & Mark 3 M02775-35 & Mark 13 M06665-23 & Mark 14 M06715-27 & Mark 19 M 09225-11\\
\hline
1.0V  &      &      &      & 0.74 \\
1.5V  & 1.03 &      &      & 0.85 \\
2.0V  &      &      &      & 0.93 \\
2.5V  & 1.00 & 1.00 & 1.00 & 1.00 \\
3.0V  &      &      &      & 1.05 \\
3.5V  & 1.23 & 1.45 & 1.31 & 1.12 \\
4.5V  & 1.47 & 1.53 & 1.43 & 1.35 \\
5.5V  & 1.85 & 1.97 & 2.00 & 1.74 \\
6.5V  & 2.36 & 2.79 & 2.84 & 2.38 \\
7.5V  & 4.54 & 3.48 & 3.79 & 3.38 \\
8.5V  & 6.91 & 4.33 & 5.36 & 4.97 \\
9.5V  & 11.7 & 6.93 & 7.91 & 7.35 \\
10.5V & 18.1 & 9.88 & 11.4 & 11.0 \\
11.5V & 28.4 & 14.7 & 16.4 & 16.9 \\
12.5V & 45.1 & 27.7 & 25.6 & 26.0 \\
13.5V & 72.7 & 36.6 & 35.8 & 39.3 \\
14.5V &      & 65.6 & 56.4 & 67.4 \\
\end{tabular}

\end{table}

Using the avalanche gain to correct other measurements is presented in Chapters~\ref{ch:darkcurrent}~and~\ref{ch:photoncounting}.

\section{Dark Current}
In HgCdTe, dark current or junction-related leakage current is dominated by trap-assisted tunneling. It predominantly happens in the multiplication band due to the smaller bandgap. Trap-assisted tunneling is where an electron tunnels to a trap, then uses it to excite to conduction. Electron tunneling is a quantum mechanical phenomenon where an electron penetrates a barrier it could not cross classically, making a probabilistic jump. The probability is strongly dependent on the barrier height, and direct tunneling is thus totally negligible in SWIR HgCdTe, but the presence of a trap makes the tunneling possible \citep{hall2011}.

Trap-assisted tunneling is then the dominant source of dark current in short- and medium-wavelength HgCdTe APDs. The full excitation to conduction occurs as either tunnel-tunnel or tunnel-thermal current, both of which start with quantum tunneling. Tunnel-tunnel dominates dark current at bias voltages above 8V for low temperatures $\mathrm{T} \leq 65K$ and involves the electron tunneling into the trap, then tunneling from the trap into conduction. It is not dependent on temperature in cryogenic cooling. Tunnel-thermal current is sensitive to both voltage and temperature, where the electron is excited from the trap into the conduction band thermally, and dominates at high temperatures $\mathrm{T} > 65K$. Once in the conduction band, the tunneled electrons experience avalanches and gain identically to photoelectrons.

Measuring dark current takes several hours of investigation for two reasons. First, there is a strong settling effect that occurs over multiple hours of observation. It decreases to undetectable at approximately 2 hours and is attributed to glow reducing as the detector is read out at a relatively slow rate. The dark current being measured is also very small and can only be accurately measured over long periods of time. Calls to take dark current data were \texttt{.\/expose 180 5}, so 180 frames were taken with 5 seconds between frames on ramps 15 minutes long. Series.sh was used to take multiple ramps over the course of 4 hours, or 16 calls to .\/expose. This was taken at a range of bias voltages from unity $V_{mathrm{bias}} = 2.5\mathrm{V}$ with either 2V or 1V intervals, selected to accommodate time available and the expected change in dark current.

\begin{figure}
   \begin{center}
   \begin{tabular}{c}
   \includegraphics[height=8.0cm]{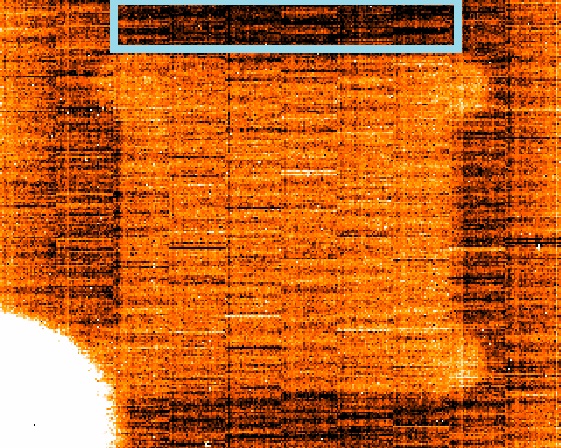}
   \end{tabular}
   \end{center}
	 \vspace{-16pt}
   \caption[Dark Current Across Array]{\label{fig:darkmap0}
	Dark current as measured across a SAPHIRA detector. The glow source in the lower-left is easily visible, with a cylindrical depression in the mask over it at this time. Presented dark measurements are from the outlined area at the top.}
   \end{figure}

\begin{figure}
   \begin{center}
   \begin{tabular}{c}
   \includegraphics[height=8.0cm]{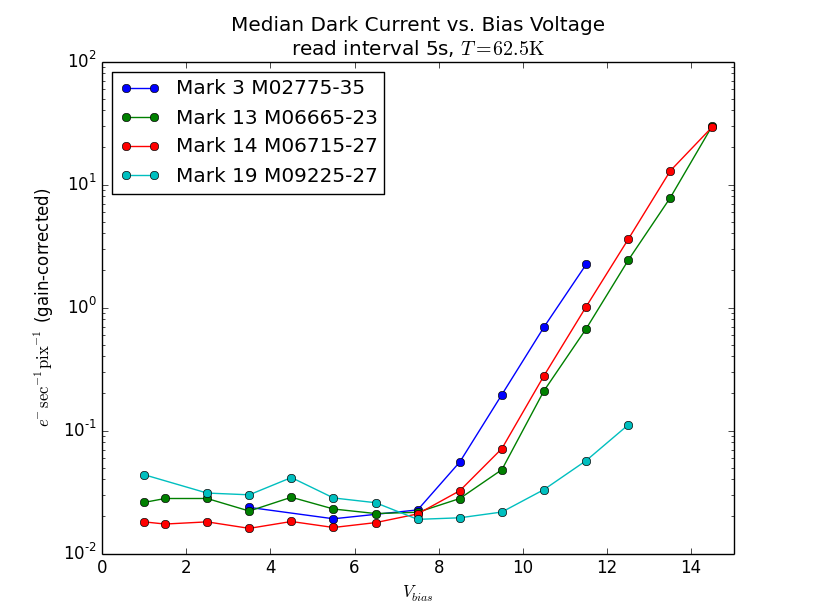}
   \end{tabular}
   \end{center}
	 \vspace{-16pt}
   \caption[Dark Current of 4 Array Iterations]{\label{fig:dark4devices0}
	Dark current measured from 4 different devices in the minimal glow area at the top of the array.}
   \end{figure}

Measured dark current is also very dependent on location on the array, as some areas of the detector are either under the open section of the mask and see the warmer integrating sphere in the KSPEC cryostat, or they receive glow emitted from a corner of the detector's ROIC (see Figure~\ref{fig:darkmap0}). These effects are detailed in Chapter~\ref{ch:darkcurrent}, and an area at the top of the detector was selected to minimize them. Dark measurements are performed several times in sequence on a single device at different bias voltages. Measuring in this region on top, the dark current improved with devices going forward as issues with the early devices were corrected (see Figure~\ref{fig:dark4devices0}). 

\begin{figure}
   \begin{center}
   \begin{tabular}{c}
   \includegraphics[height=8.0cm]{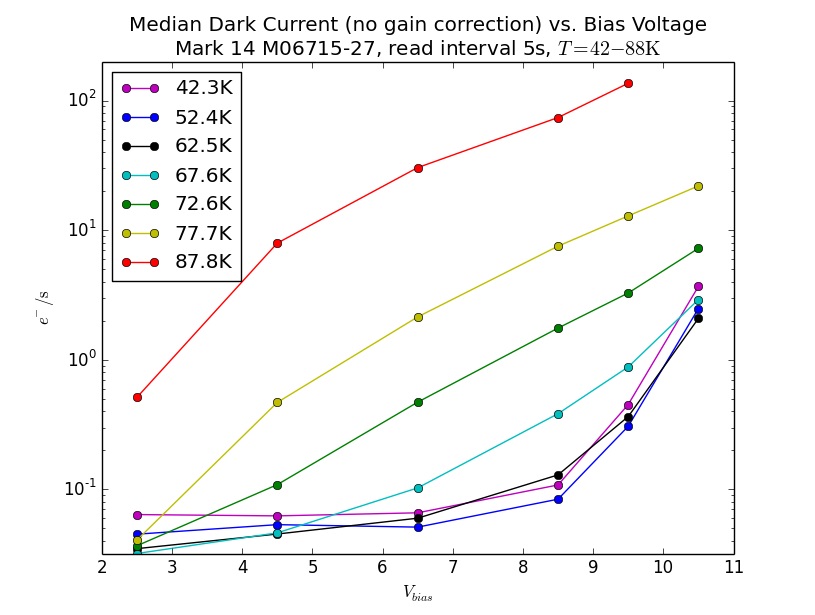}
   \end{tabular}
   \end{center}
	 \vspace{-16pt}
   \caption[Dark Current vs. Temperature] {\label{fig:darkvbiasvtemp0} 
Dark current shows a dependence on both bias voltage and temperature above $\sim60\mathrm{K}$. This is caused by tunnel-thermal behavior. At the highest recorded voltage here $>80\mathrm{K}$, thermal dark current is even visible at the unity bias voltage $2.5\mathrm{V}$. Higher temperatures increase the response of dark current to applied bias voltage.}
   \end{figure}

The steep rise of dark current above $\sim8\mathrm{V}$ is a function of the previously discussed tunnel-tunnel current. The dark current also rises with temperature and voltage at the same time, seen in Figure~\ref{fig:darkvbiasvtemp0}. This is consistent with the tunnel-thermal contribution to trap-assisted tunneling, also previously discussed.

Initial measurements of dark current on the detector found $10\esp$ on the early devices investigated. By changes to the operating voltages the cause of the initial high dark current was attributed to glow, and the state of VDDx = 3.5V was adopted to minimize the contribution of output amplifiers around the perimeter. Further investigation into dark current revealed an unexpected source of glow. A point source was isolated by the addition of a mask that caused emitted glow from the source to make rays across the array's images. The source was seen directly by imaging the detector with an H4RG and relayed to Leonardo, who found it to be a mistaken floating gate \citep{atkinson2016}. The source was masked with tape in existing detectors and was corrected in the ME-1001 ROIC, as mentioned above. Other improvements were made and the measured dark current of the SAPHIRA was found to be $0.025\esp$, an improvement of $\sim3$ orders of magnitude and comparable to HxRG devices. Measurements of this dark current showed that below $V_{bias}\sim8\mathrm{V}$ this value was insensitive to bias voltage. Given similar voltage insensitivities at other temperatures, this dark current is attributed to residual glow. Future reduction of the glow would further reduce the bias voltage above which tunnel-tunnel current dominates, and can be projected by the slope of the observed tunnel-tunnel bias dependency. Current SAPHIRA used very long exposures and limited regions to infer a limiting medium-sourced dark current of $0.0015\esp$. (Literature for the H2RG from Teledyne gives a value of $\leq0.05\esp$ for a $2.5\um$ device at 77K, but arrays frequently perform much better \citep{teledyne2012}. The $5\um$ H2RG arrays for JWST show $\sim0.003\esp$ at 38.5K \citep{rauscher2014}.)

The results of dark current measurements are presented in Chapter~\ref{ch:darkcurrent}.

\section{Pulse Height Distribution}
For measuring the response to single-photon arrival events, a pulse height distribution uses a histogram of the difference between individual reads of single pixels. This can be performed in either individual pairs ($2 - 1, 4 - 3, 6 - 5, \mathrm{etc.}$) or 'rolling' ($2 - 1, 3 - 2, 4 - 3, \mathrm{etc.}$). A window of $32\times1$ is used so the data is taken as fast as possible, near the pixel rate of $\sim250\mathrm{kHz}$. Such a window is set up in assembly code with a setting to columns of \texttt{\$01 \$00} and to rows of \texttt{15 \$00}, \texttt{1 \$01}, and \texttt{16 \$00}. (The \texttt{\$xx} notation is for a hexadecimal number.) This puts the $32\times1$ window in the center of the array. A million frames with LED off and LED on (1 mA) are taken to make a clear pulse height distribution, or \texttt{.\/expose 1000000 0} for 1 million frames with no waiting between frames.

\begin{figure}
   \begin{center}
   \begin{tabular}{c}
   \includegraphics[height=8cm]{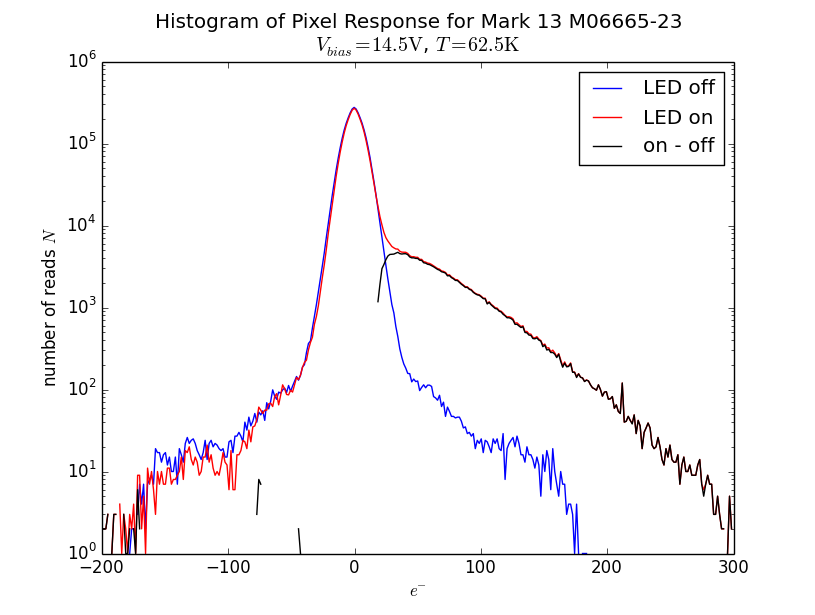}
   \end{tabular}
   \end{center}
	 \vspace{-10pt}
   \caption[Pulse Height Distribution (full)]
   { \label{fig:pchist0} 
A pulse height distribution with logarithmically plotted LED off and LED on measurements. The on - off is plotted also, shown as a simple representation of events from incident photons. The tail towards higher values resulting from photon events is clearly visible.}
	 \vspace{10pt}
\end{figure}

This data is then put on a histogram to see what happens between subsequent frames (see Figure~\ref{fig:pchist0}). To directly see the profile of incident photons, a histogram with the LED off is subtracted from one with the LED on. Pulse height distributions of SAPHIRA devices show a broad tail toward higher values for single photon events that disguise events of multiple simultaneous photons. Expectations of the HgCdTe APD were for entirely Gaussian noise. However, this result is consistent with single-carrier avalanching given the low rate at which events are detected. The rate at which events occur would have to be much higher than was measured to attribute the $on - off$ curve to simultaneous incident photons. Though this work covers using an ARC controller to operate a SAPHIRA, a detector can also be operated at a pixel rate of $>\geq1\mathrm{MHz}$ with the Pizza Box controller being developed at UH-IfA by Charles Lockhart and Eric Warmbier, making it capable of detecting incoming photons as single events in relatively high incident light \citep{lockhart2018}. The Pizza Box is currently in use at the SCExAO deployment of SAPHIRA.


Pulse height distributions are largely for seeing the effect of avalanche gain and thresholding on photon counting, and are used in Chapter~\ref{ch:photoncounting}.

The measured avalanche gain can be confirmed by the pulse height distribution. The observed mean of the LED on - off distribution should reproduce the gain, though part of the curve blending into the observed read noise complicates the measurement. Given the relatively low incident rate of arriving photons, only $2.5\%$ of events consist of multiple photons arriving simultaneously and a coincident event is unlikely. A measurement with a mean of $1e^{-}$ corresponds to a 1.0 gain, however low gains are difficult to measure as the overlap with read noise becomes significant. This measurement is also dependent on charge gain, which in this work requires certain assumptions. This technique was recently suggested, but has not yet been pursued but can be useful going forward.

\section{Summary}
The SAPHIRA arrays at UH-IfA were made to function in the KSPEC cryostat, which required significant modifications to both software and hardware. Control of the applied bias voltages and operation of the detector is achieved by the assembly code running on the ARC controller and C\plus\plus on the connected Linux computer. Temperature is managed by external controllers and LED by a power supply.

Characterization of the SAPHIRA detector over the years has produced measurements of the dark current and avalanche gain over several iterations of the device. Most notably, the initial dark current of $10\esp$ was found to be dominated by glow from the ROIC. It was substantially reduced by dropping operating voltages and masking the device from glow. Isolating this required not just standard dark current measurements of the SAPHIRA but also rays revealed to originate from a single location. The point source was then imaged by a H4RG. This allowed Leonardo to find the error and correct it on future ROICs. After several iterations, the dark current of the SAPHIRA dropped from $10\esp$ to $\sim0.025\esp$, putting it on par with other astronomical HgCdTe arrays.

I proved the SAPHIRA is the first NIR APD array capable of photon counting, and is a leading device in NIR AO applications. I've explored avalanche gain taken with several Marks of the SAPHIRA and found a more stochastic process in HgCdTe then had previously been expected. This was validated by similar measurements performed at ESO by Gert Finger. Despite HgCdTe APDs producing a unique single-carrier avalanche with electrons, the shape of the pulse height distribution is dependent on the simple probability of avalanches. The SAPHIRA is a uniquely capable NIR APD array, but this effect makes it difficult to measure multiple photon events in a single read.

\chapter{Telescope Deployments}\label{ch:deployments} 
The SAPHIRA has been deployed to several telescopes during development, both to test its on-sky functionality and to produce scientific observations. This makes use of the Stirling-Cooler Cryostat built by GL Scientific, a corporation founded by Gerry Lupino that has now become Hawaii Aerospace. This cryostat was originally developed for use with the HxRG detectors but easily accommodates the SAPHIRA. It is light and can be secured to a telescope without requiring much balancing weight. For deployment, funding was secured from the NSF for two traveling systems.

\begin{figure}
   \begin{center}
   \begin{tabular}{c}
   \includegraphics[height=6.5cm]{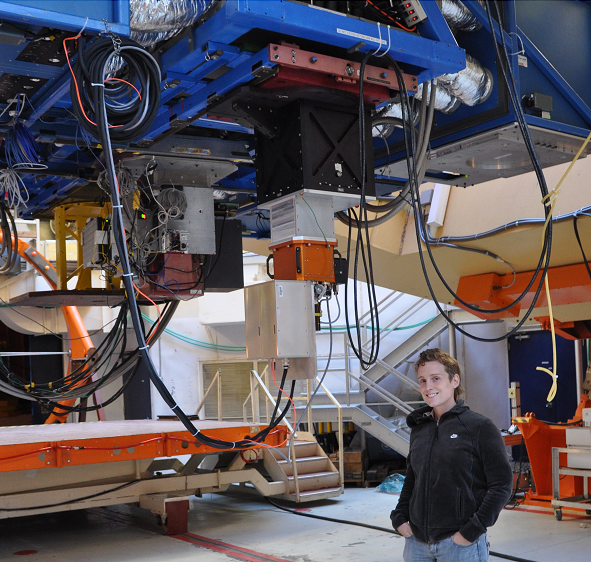}
   \end{tabular}
   \end{center}
	 \vspace{-16pt}
   \caption[First Deployment at the IRTF]{\label{fig:irtf0}
With assistance I brought SAPHIRA in the SCC to IRTF, where it was used for two half nights. This was the worldwide first use of SAPHIRA on-sky.}
   \end{figure}

\begin{figure}
   \begin{center}
   \begin{tabular}{c}
   \includegraphics[height=6.5cm]{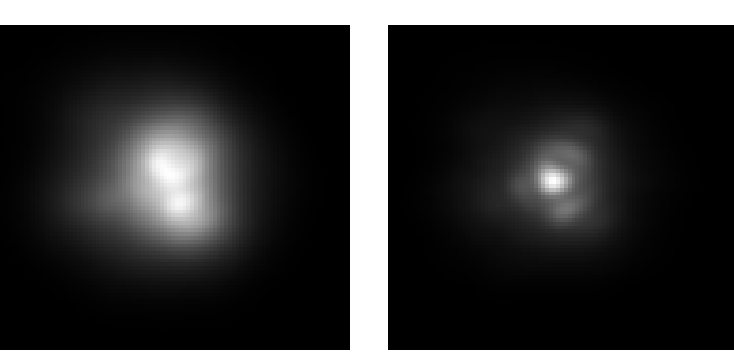}
   \end{tabular}
   \end{center}
	 \vspace{-16pt}
   \caption[Lucky Imaging of 24 Leo Minoris]{\label{fig:24lmi}
Images of the star 24 Leo Minoris taken with SAPHIRA mounted on IRTF. A full composite image with shift-and-add correction applied (\textit{left}) shows lobes from aberrations in the IRTF optics. A composite of $10\%$ best (or lucky) images (\textit{right}) is diffraction-limited with a partial diffraction ring visible. This technique selects for images that don't suffer from either atmospheric distortion or aberrations in the IRTF optics.}
   \end{figure}

The first on-sky use of the SAPHIRA in the world was at the Infrared Telescope Facility (IRTF) on 29 \& 30 Apr 2014, during which I operated the system for both half-nights allotted (see Figure~\ref{fig:irtf0}). The opportunity was made available by the failure of an IRTF instrument. This tested the use of the SAPHIRA as a fast NIR imager. I applied lucky imaging techniques, which uses a bank of images to select for those with minimum atmospheric distortion \citep{tubbs2004}. Diffraction-limited NIR imaging was achieved with IRTF, producing a full-width half-maximum (FWHM) of the point spread function (PSF) of $0.12''$ during a night of $0.4''$ seeing (see Figure~\ref{fig:24lmi}). Images for the composite were selected solely based on being in the top 10\% of PSF FWHM measurements \citep{atkinson2014}. With this application the detector was shown to be capable of a diffraction-limited astronomical observations. I included these pictures during a presentation at the Montréal SPIE conference in mid-2014.

\begin{figure}
   \begin{center}
   \begin{tabular}{c}
   \includegraphics[height=8.0cm]{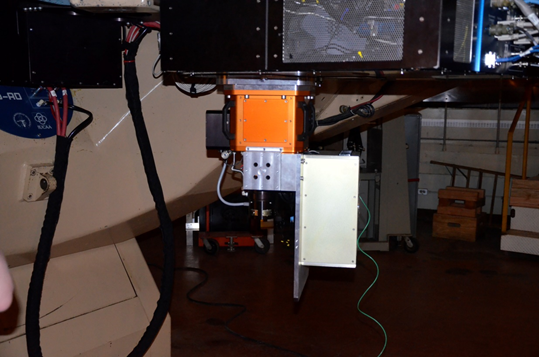}
   \end{tabular}
   \end{center}
	 \vspace{-16pt}
   \caption[Stirling-Cooler Cryostat at Palomar]{\label{fig:scc}
The GLS Stirling-Cooler Cryostat mounted onto the Robo-AO instrument at the Palomar-1.5m telescope. Orange is the cryogenic chamber with the cryopump mounted underneath (\textit{black}), and the controller to the side (\textit{gold}).}
   \end{figure}

Following this, the SAPHIRA in the GLS cryostat was taken from IRTF and installed in the Robo-AO instrument at the Palomar-1.5m observatory in collaboration with Principal Investigator (PI) Christoph Baranec and software lead Reed Riddle (see Figure~\ref{fig:scc}) \citep{baranec2013a}. I commissioned SAPHIRA on Robo-AO for several intermittent weeks. Integrating the SAPHIRA so it worked with the Robo-AO C\plus\plus software revealed a Linux compatibility problem with the ARC software drivers. I took the system to the ARC offices in San Diego where we found the innate problem with the ARC code. Our solution was to add a separate computer to run the SAPHIRA and interface with the Robo-AO system. This was a necessary but workable solution as Robo-AO used the TCP/IP protocol for inter-process communication and lent itself easily to working over a network.

\begin{figure}
   \begin{center}
   \begin{tabular}{c}
   \includegraphics[height=4.5cm]{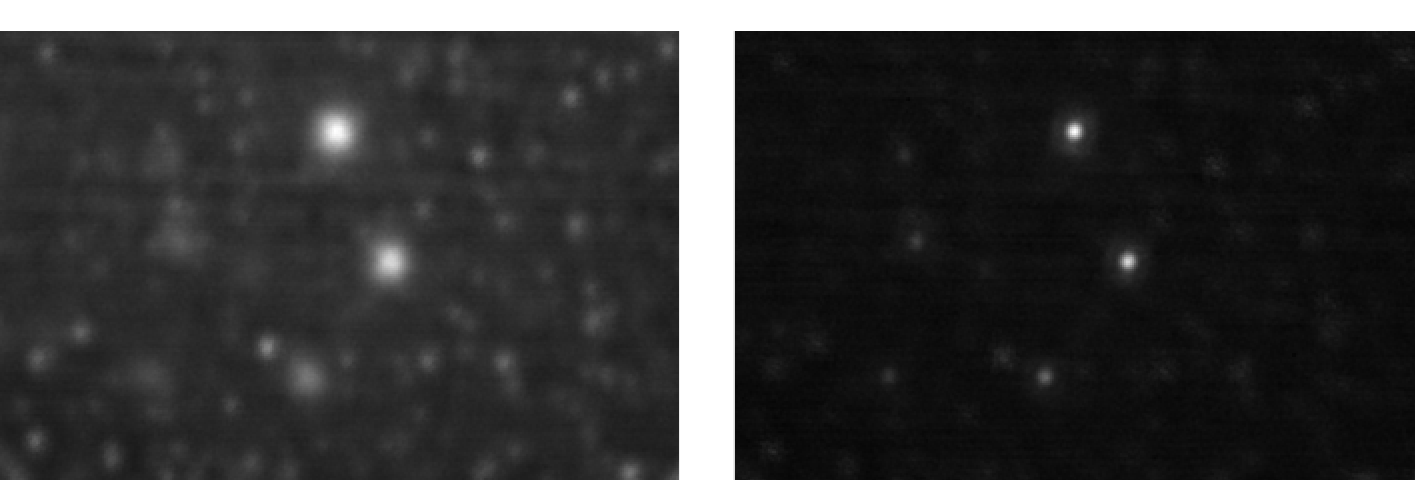}
   \end{tabular}
   \end{center}
	 \vspace{-16pt}
   \caption[Lucky Imaging of M3]{\label{fig:m3}
A subsection of the globular cluster M3, imaged with SAPHIRA on Palomar-1.5m. The image with only shift-and-add correction (\textit{left}) is greatly improved by the lucky technique of selecting only the best $10\%$ of images, which produces a diffraction-limited result.}
   \end{figure}

Working, the SAPHIRA on Robo-AO/P1.5m showed similar capabilities to the IRTF deployment as a science camera. It was again used for lucky imaging (see Figure~\ref{fig:m3}). With the existing working imager and automated adaptive optics, SAPHIRA was integrated to provide tip/tilt guidance driving on the target as a simple natural guide star \citep{baranec2015}. To run quickly enough to provide useful information, SAPHIRA was operated at $\sim\mathrm{kHz}$ with read noise barely at $<1e^{-}$ but otherwise minimal avalanche gain. Robo-AO provided an NIR port that was used for SAHPIRA. It was operated in this manner August 2014-June 2015. 

There the SAPHIRA also produced its first science publication with an imaging of the eclipsing binary T-Cyg1-12664 from the \textit{Kepler} catalog, published in \cite{han2017} where it provided \textit{H}-band differential photometry. In this use, the SAPHIRA operated to freeze tip-tilt motion at $10-30\mathrm{Hz}$. That provided correction to remove tip-tilt motion from images afterward. Avalanche gain was again sufficient to keep read noise $<1e^{-}$, which induced a noise penalty but maximized dynamic range relative to higher gains \citep{baranec2017}.


\begin{figure}
   \begin{center}
   \begin{tabular}{c}
   \includegraphics[height=8cm]{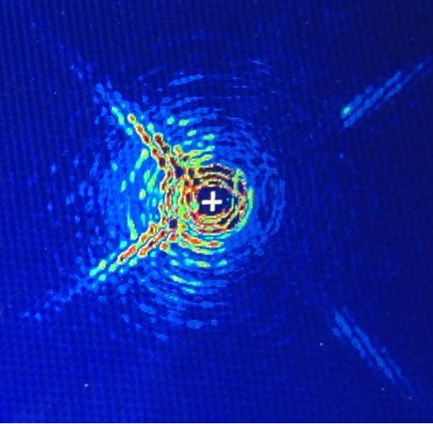}
   \end{tabular}
   \end{center}
	 \vspace{-16pt}
   \caption[Test Image from SCExAO]{\label{fig:scexao}
An image from SAPHIRA of an internal laser, with speckle-nulling applied to the right half only. Comparing with the left side shows the AO suppression of speckles (\textit{image courtesy of Nemanja Jovanovic} and \cite{jovanovic2017a}).}
   \end{figure}

Following Palomar it was brought to the Subaru telescope on Maunakea for use with the Subaru Coronagraphic Extreme Adaptive Optics (SCExAO) instrument in development, a project led by PI Olivier Guyon \citep{goebel2016,jovanovic2017b,goebel2017,goebel2018}. A Pizza Box controller was installed to replace the ARC controller for this. I assisted with commissioning the system on Subaru for several nights. Behind SCExAO the SAPHIRA analyzed the behavior of radial NIR speckles and the diffraction-limited PSF (see Figure~\ref{fig:scexao}). SAPHIRA is still operating with the SCExAO instrument and is supported by UH-IfA graduate student Sean Goebel.

\begin{figure}
   \begin{center}
   \begin{tabular}{c}
   \includegraphics[height=8cm]{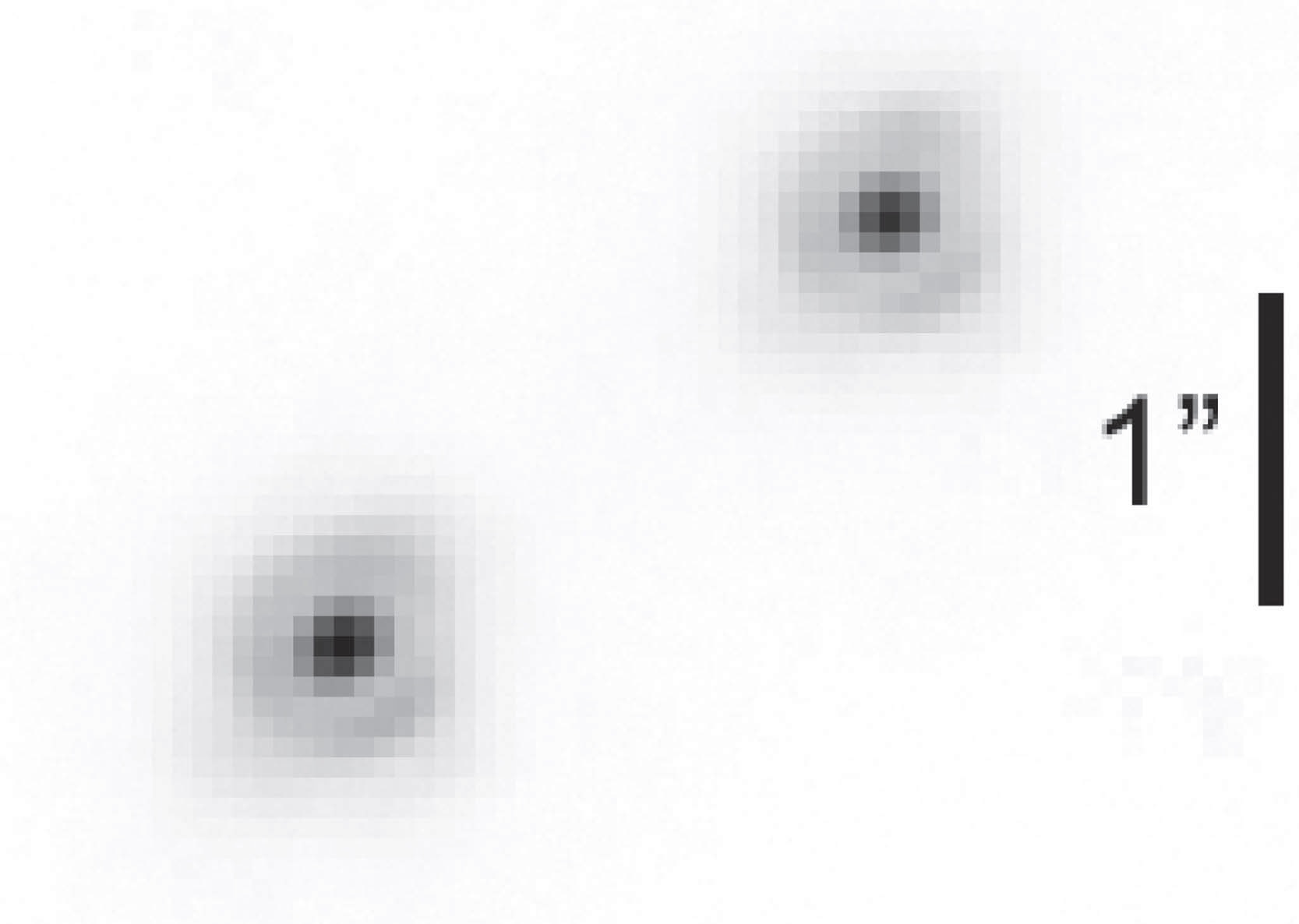}
   \end{tabular}
   \end{center}
	 \vspace{-16pt}
   \caption[Kitt Peak Image of GJ1116 AB]{\label{fig:kp}
A SAPHIRA image at Kitt Peak 2.1m of the binary system GJ 1116 AB in $H$-band. (\textit{From \cite{jensenclem2018}})}
   \end{figure}

In a new SCC acquired by UH-IfA faculty Mark Chun, a SAPHIRA was then brought to the Kitt Peak 2.1-meter telescope for further use with the Robo-AO instrument (see Figure~\ref{fig:kp}) where it is now being used in a survey for substellar companions \citep{salama2016,jensenclem2018}. Another SAPHIRA is currently being tested as a pyramid wavefront sensor (PWFS) detector at the Keck telescope for use with the Keck Planet Imager and Characterizer (KPIC) later this year \citep{mawet2016}. The PWFS has been devised as an upgrade to traditional Shack-Hartmann WFS which use an array of lenslets in front of their detectors to produce multiple measurements of the target wavefront. PWFS uses a single pyramidal prism of the focal plane to produce four pupil images that are used to measure wavefront aberrations \citep{campbell2006}. This has been used previously in the PYRAMIR WFS, deployed at the 3.5-m Calar Alto Observatory in Almería, Spain \citep{peter2010}. A HAWAII I detector was used in PYRAMIR, over which SAPHIRA will improve readout speed necessary for fast wavefront sensing.

The SAPHIRA is planned for use on the new Robo-AO system on the UH-2.2m on Maunakea and Smithsonian-operated telescopes at Paranal. It has been announced for future commercial sale in the C-RED One camera put together by First Light Imaging \citep{greffe2016}.

\section{Future Applications}
All deployments to date have focused on the use of SAPHIRA to drive AO systems. The SAPHIRA's high time-resolution and low noise also make it useful for other NIR astronomical applications.

Observations of high time-resolution in wavelengths above $1\um$ with a full array have not been very accessible to astronomy previously. These capabilities allow new studies of compact objects. Relatively recent ESO/NTT observations of microquasar flares have used a time resolution of 2s ($0.5Hz$), while the SAPHIRA reaches observations $>100\mathrm{Hz}$ in AO use \citep{chaty2015}. Microsecond-scale optical/NIR spectroscopy has been in limited use for ultracompact objects as the development of the microwave kinetic inductance detectors (MKIDs) continues, and a published paper discusses the search for high-frequency variations and quasi-periodic oscillations \citep{szypryt2014}. (My involvement with MKIDs development is described in Chapter~\ref{ch:siteexperiences}.) The ability to perform high time-resolution infrared studies promises to reveal new behavior of compact targets. Other applications in the ms and $\um$ range on Extremely Large Telescopes will allow further observations of stellar occultations and of close binary systems \citep{shearer2010}.

\section{Summary}
I operated the SAPHIRA in its first global on-sky use at the 3-m IRTF telescope on Maunakea for two half-nights. Lucky imaging was demonstrated and the diffraction-limited NIR images from IRTF were produced.

I then helped commission the SAPHIRA camera with the Robo-AO instrument on the 1.5-m telescope at Palomar Observatory. We used the SAPHIRA camera as a NIR tip-tilt sensor to drive an image stabilization loop within Robo-AO. This let us observe targets on the visible camera that were otherwise too faint for the standard post facto image registration techniques. This was the first ever use of SAPHIRA to provide AO correction. The SAPHIRA also acted as a science camera behind the AO system, observations from which were published.

Palomar was followed by a deployment to the 8-m Subaru Telescope on Maunakea. It was installed and tested with the SCExAO instrument, an extreme adaptive optics system. I supported this work over the course of several nights which involved the integration of the camera with the instrument. It provided fast imaging in the NIR of residual speckles post AO, improving the novel SCExAO system.

\chapter{Dark current in the SAPHIRA series of mercury cadmium telluride APD arrays}\label{ch:darkcurrent} 
This chapter was published as \cite{atkinson2017a}.

We present the dark current performance of the SAPHIRA series of HgCdTe APD arrays, characterized as a function of bias voltage and temperature. We measure a gain-normalized dark current in multiple SAPHIRA arrays of $0.025\esp$ from unity gain ($V_{bias} = 2.5\mathrm{V}$) up to an avalanche gain of $\sim5$ ($V_{bias} = 8V$). Under a restricted subarray and long exposures we set an implied upper limit on intrinsic dark current in the SAPHIRA of $0.0015\esp$. These values are still dominated by glow, NIR illumination generated by the ROIC.

\section{Introduction}
The Selex Avalanche Photodiode for HgCdTe Infrared Arrays, or SAPHIRA, has quickly become the detector of choice for NIR wavefront sensing and fringe tracking. Manufactured by Leonardo (formerly Selex), the metal organic vapor phase epitaxy (MOVPE) process grows HgCdTe on GaAs and enables sophisticated solid state engineering of the avalanche photodiode (APD) architecture. The current generation of MOVPE SAPHIRA arrays have demonstrated quantum efficiencies $>80\%$ and avalanche gains $>500$ \citep{atkinson2016,finger2016}.

Our pursuit of wavefront sensing and adaptive optics applications has seen deployment of SAPHIRA to 4 different telescopes to date \citep{atkinson2014,baranec2015,goebel2016,salama2016,jensenclem2018}. While rapid operation and photon counting are critical for use in high-bandwidth adaptive optics applications, dark current is the most significant limiting factor for broader, low-background astronomical observations.

The University of Hawai'i (UH)-Leonardo collaboration on SAPHIRA development aims to achieve the dark current necessary for low-background astronomy at a bias voltage high enough to provide useful avalanche gain. A detector with both dark current comparable to the best conventional HgCdTe arrays ($\sim0.001\esp$) and sub-electron read noise would be a boon to low-background science observations. Such a device would be useful in both burgeoning fields like spectroscopic exoplanet studies and new observational parameter spaces like NIR high time-resolution astrophysics.

This work reports measurements of dark current for four SAPHIRA arrays as a function of temperature and avalanche gain. We discuss glow from readout electronics and demonstrate that glow from the readout integrated circuit (ROIC) dominates dark current at avalanche bias voltages up to 8V (gain $\sim5$). Uniformity over the SAPHIRA array and the appearance of hot pixels with bias voltage are also examined.

Section~\ref{darktechniques} gives background information, describes the laboratory setup for SAPHIRA characterization, and details our measurement techniques. Section~\ref{darkmeasurements} of this paper presents dark current as a function of temperature and bias voltage $V_{bias}$ at both device and per-pixel levels. We analyze these results using other information about SAPHIRA in Section~\ref{darkanalysis}. Finally, the cause of the observed dark currents, limitations thereof, and ongoing improvements to SAPHIRA are discussed in Section~\ref{darkdiscussion}; we then state our conclusions in Section~\ref{darkconclusions}.

\section{Setup \& Operation}\label{darktechniques}
\begin{figure}
   \begin{center}
   \begin{tabular}{c}
   \includegraphics[height=5.0cm]{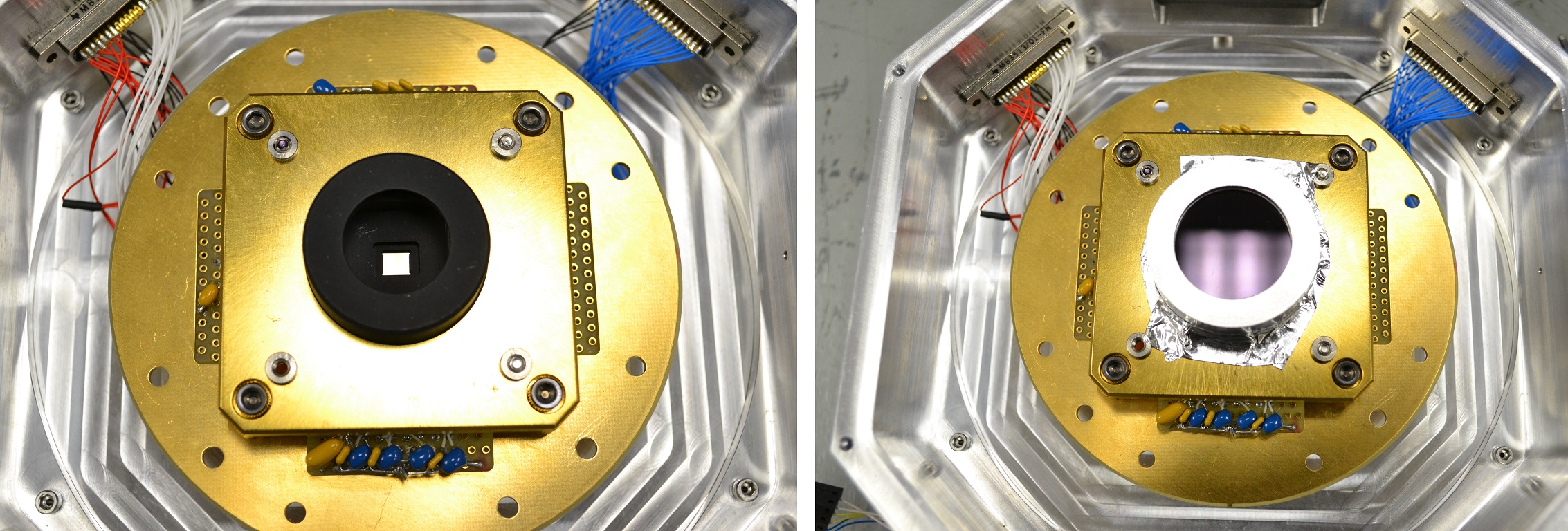}
   \end{tabular}
   \end{center}
	 \vspace{-16pt}
   \caption[The KSPEC Cryostat] 
   { \label{fig:KSPEC2} 
Contacts in the socket accommodating SAPHIRA use springs to maintain constant pressure with the clamp assembly in place over the array. A mask rests on top of the detector to block background and glow sources (see Section~\ref{darkglow}). The shortpass filter installed over this reduces the incident background yet further (\textit{right}).}
   \end{figure}

Measurements of dark current for SAPHIRA arrays use KSPEC (pictured in Figure~\ref{fig:KSPEC2}). It is a former $IJHK$-band spectrometer that has been converted to an ultra-low background NIR detector testbed, originally for service in the NGST/JWST detector development program \citep{hodapp1994, hodapp1996, hall2000}. KSPEC contains a two-stage Gifford-McMahon cryocooler that provides adequate cooling capacity to three temperature-controlled stages: first stage ballast for the thermal shielding and cryogenic integrating sphere, followed by the second stage ballast for a core stabilizing mass, which in turn cools the subsequent detector stage mounted to a cold finger in thermal contact with SAPHIRA's ceramic leadless chip carrier. Stainless steel bellows between the cryogenic integrating sphere and the detector enclosure allow dark conditions while maintaining thermal isolation between the two temperature ballasts. With this arrangement, the detector enclosure maintains millikelvin stability through week-long observations. NIR LEDs of $1.05\um$, $1.30\um$, $1.70\um$, and $3.1\um$ (each with $\sim10\%$ spectral bandwidth) are mounted in a cryogenic integrating sphere positioned. These provide uniform illumination across the surface of an installed detector. 

An additional temperature diode on SAPHIRA's leadless chip carrier was used to calibrate temperature measurements. We observe that when the detector is powered and operating the temperature measured by this diode is 2.3K higher than the detector-stage temperature of 40K and 2.9K higher at $100\mathrm{K}$. Unfortunately, the sensor glows brightly in the NIR when powered and is not operated during dark observations. Subsequent reporting of the detector temperature uses the detector stage measurement but takes into account this temperature calibration.

KSPEC has a low inherent background, measured with SAPHIRA to be $\sim0.1\esp$ in the 0.8 to $3.5\um$ bandpass. The addition of a cooled shortpass blocking filter $\lambda_c = 1.85\um$ further suppresses thermal NIR background from the integrating sphere.

The detector controller is a generation 3 Astronomical Research Cameras (ARC) system, making use of an ARC-22 fiber optic timing board, ARC-32 clock board, and four ARC-46 eight-channel IR video boards \citep{leach2000}. Instead of the standard ARC bias/utility board, a custom low-noise bias supply board designed at Australia National University and populated at UH provides supply and bias voltages. These cards are installed in a twelve-slot ARC chassis operating on a common backplane. Video cards are mounted with one empty slot in between each to reduce noise and mutual interference across the backplane. Two fiber optic lines connect the controller to an ARC-66 PCIe board in a Linux desktop computer from which the instrument is operated via a set of command-line scripts making use of the v3.5 ARC API.

The ARC controller uses timing code written in Motorola DSP56000 assembly language \citep{kloker1986}. Code for operating the SAPHIRA was developed at UH, building on samples for H2RG operation supplied by Bob Leach of ARC and Marco Bonati of CTIO. The timing code is loaded to the controller just prior to detector power-up. 

The SAPHIRA, like any NIR detector array, must be hybridized to a matching ROIC that provides the necessary electrical interface \citep{finger2012}. The SAPHIRA ROIC provides four modes of read operation, output channels selectable to 4, 8, 16, or 32, and a serial programmable interface with other user-configurable options. Typical operation makes use of all 32 outputs, which read adjacent pixels in a row and can be brought to bear on any subarray. To date, three versions of this ROIC have been used:

\begin{itemize}
  \item The ME911 was the first ROIC design to host a MOVPE SAPHIRA array. An unexpected glow source was discovered and later attributed to a mask error in production, leaving a floating gate. 
	\item The ME1000 design was commissioned by ESO to enable a new mode optimized for high-speed performance, read-reset-read in a row-by-row process \citep{finger2016}. Metal layers in the ROIC were reconfigured to shield glow from the output amplifiers. A source follower was also eliminated, increasing the ROIC gain.
	\item The ME1001 modified metal layers in the ME1000 to correct the floating gate, eliminating the prominent glow source from previous ROICs.
\end{itemize}

\begin{figure}
   \begin{center}
   \begin{tabular}{c}
   \includegraphics[height=8.0cm]{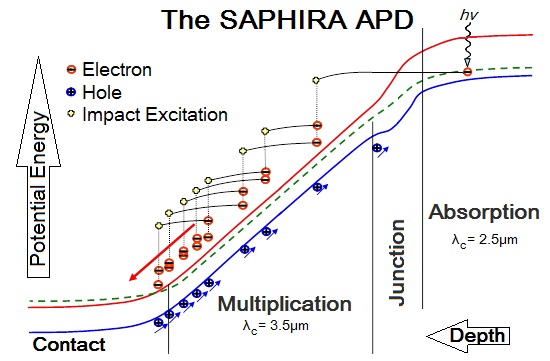}
   \end{tabular}
   \end{center}
	 \vspace{-16pt}
   \caption[SAPHIRA Bandgap] 
   { \label{fig:bandgap2} 
The single-carrier electron avalanche is a major noise advantage HgCdTe has over other NIR detection materials. The bias voltage $V_{bias}$ is applied across the multiplication and junction regions. Though photons are depicted here being captured in the absorption region, it is possible for photons $2.5\um \leq \lambda \leq 3.5\um$ photons to penetrate to and be absorbed in the multiplication layer. (\textit{Original figure courtesy Leonardo.})}
   \end{figure}

\begin{figure}
   \begin{center}
   \begin{tabular}{c}
   \includegraphics[height=8.0cm]{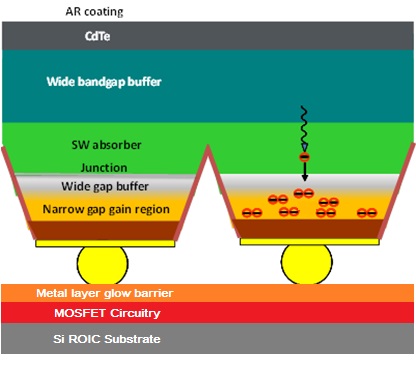}
   \end{tabular}
   \end{center}
	 \vspace{-16pt}
   \caption[SAPHIRA Physical Structure] 
   { \label{fig:layout} 
Physical structure of the SAPHIRA APD. Note that photons with $\lambda < 1.4\um$ (up to Mk12) and $\lambda < 0.8\um$ (for Mk13 and later) are absorbed in the wide bandgap buffer. Glow originating in the ROIC unit cell is heavily attenuated by staggered metal layers in the ROIC. The metal contact pad on the APD leaves only the mesa trench as a path for any glow photons and these will be preferentially absorbed in the multiplication region with significantly reduced gain.}
   \end{figure}

The SAPHIRA APD comprises several layers of HgCdTe with differing bandgaps. Though the exact bandgap structure is proprietary to Leonardo, a simple diagram is shown in Figure~\ref{fig:bandgap2} \textbf{a physical diagram in Figure~\ref{fig:layout}}. There are three regions of note. Near the top of the device, the absorption layer captures incident photons ($\lambda \leq 2.5\um$) and converts them to electrons. Photons with longer wavelengths ($2.5\um \leq \lambda \leq 3.5\um$) will penetrate the absorption layer and the junction to be absorbed directly into the multiplication layer, and thus must be rejected by a suitable cooled blocking filter. By diffusion, electrons from the absorption layer reach the junction, a relatively shallow area with a wider bandgap to reduce dark current due to generation-recombination and tunneling effects. A bias voltage $V_{bias}$ is applied across the junction and multiplication layer, accelerating electrons to excite other electrons by impact. This titular electron avalanche multiplies the signal produced by the initial photoelectron. (Photons absorbed directly into the multiplication layer experience diminished gain.) Leonardo tailors the bandgap profile to improve performance and target specific applications. These different versions are denoted by their Mark number.

To date we have evaluated 18 SAPHIRA arrays, with 5 more in hand awaiting characterization. These arrays span many iterations of design of the APD's bandgap structure.  Measurements from four arrays are discussed in this paper:

\begin{itemize}
	\item M02775-35, a Mark 3 array mounted on an ME911 ROIC. Five Mark 3 arrays were initially installed in the GRAVITY instrument at VLT \citep{finger2012,kendrew2012,finger2014,finger2016}.
	\item M06665-23, a Mark 13 array with an ME1000 ROIC. Mark 13/14 arrays extended wavelength sensitivity down to $\sim0.8\um$. A short anneal process at elevated temperature was introduced with Mark 13/14, greatly improving uniformity at high $V_{bias}$.
	\item M06715-27, a Mark 14 array on an ME911 ROIC. Mark 14 received a longer anneal than Mark 13 but is otherwise identical.
	\item M09225-27, a Mark 19 array on an ME1001 ROIC. Mark 19 was designed to move the onset of tunneling current to higher bias voltages.
\end{itemize}

\subsection{Glow Suppression}\label{darkglow}
Glow is the emission of NIR photons from MOSFETS in a detector's ROIC, attributed to direct hot electron transitions in the conduction band \citep{deluna2005}. In Si CMOS transistors in saturation, the rate of emission is amplified by 4 orders of magnitude at 80K relative to 300K \citep{lanzoni1991, sullivan2015}. Glow can be a dominant background/dark source in NIR arrays at cryogenic temperatures.

\begin{figure}
   \begin{center}
   \begin{tabular}{c}
   \includegraphics[height=8.0cm]{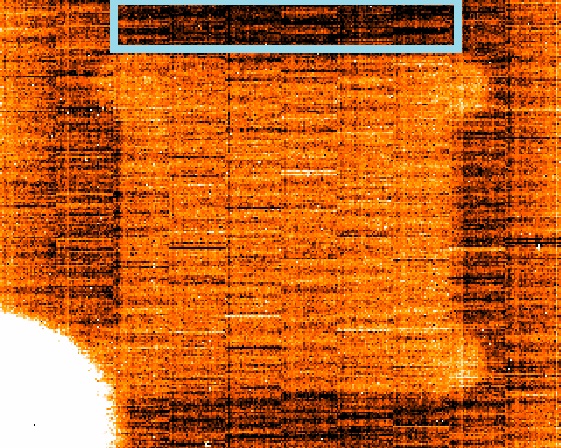}
   \end{tabular}
   \end{center}
	 \vspace{-16pt}
   \caption[Dark Current Across Detector]{\label{fig:darkmap}
As demonstrated in this dark current map, the 196x32 region along the top edge of the array (\textit{shown in cyan}) was selected for dark current measurements. This region was chosen minimize the effect of glow, as is visible from the lower-left point source, the output amplifiers on either side, and background, seen in the mask aperture. The distinctive shape of the mask aperture is visible in the center of the image, which is from the Mark 14 device M06715-27, $T = 62.5\mathrm{K}$, and $V_{bias} = 2.5\mathrm{V}$, and was made prior to installation of the shortpass filter.}
   \end{figure}

\begin{figure}
   \begin{center}
   \begin{tabular}{c}
   \includegraphics[height=6.0cm]{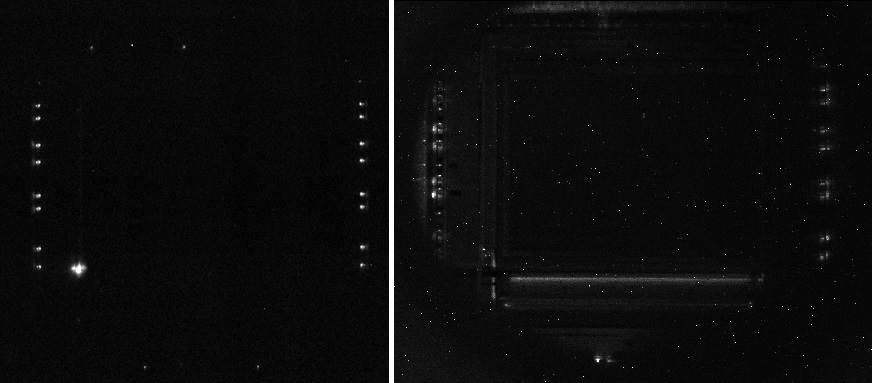}
   \end{tabular}
   \end{center}
	 \vspace{-16pt}
   \caption[Glow Along the SAPHIRA] 
   { \label{fig:glowcorner} 
As imaged in the J-band by a HAWAII 4RG array, the output amplifiers are visible in an ME911 based array (\textit{left}) but the glow is dominated by a point source to the lower-left of the APD array. A much longer exposure of a bare ME1001 ROIC (\textit{right}) shows no sign of the lower-left corner glow source but vestigial glow from the output amplifiers, even though they are heavily shielded in the ME1000 and ME1001. At these faint glow levels, glow from the fast (column) shift register is evident below the active area. There is no sign of glow from the unit cells in the active area.}
   \end{figure}

Early SAPHIRA dark current measurements at UH found a linear relationship with both operating voltage and number of outputs. This implied that the dark current measured was dominated by glow from the output amplifiers which are located on both sides and around the corners of the ROIC. We designed and fabricated a fiberglass mask with a distinctively shaped aperture to isolate the array from glow originating in output amplifiers around the perimeter of the ROIC. Upon installation the mask greatly reduced background light on masked pixels, revealing an unexpected glow source in the lower left corner (see Figure~\ref{fig:darkmap}). With no circuitry in that area of the ROIC that would explain the glow source, we imaged a SAPHIRA array onto another infrared camera in the $J$- and $H$-bands ($1.2\um$ and $1.6\um$, respectively). This confirmed both glow from the output amplifiers and the bright lower-left point source (see Figure~\ref{fig:glowcorner}). The position of the point source was determined to a precision of $10\um$. With this information, Leonardo then identified the source as a MOSFET gate left floating due to a mask verification error.

	
As an interim measure, the lower-left point source was masked with a dot of black tape in subsequent Mark 13/14 SAPHIRA arrays. A modified version of the mask with a cavity to accommodate the increased height of the tape was fabricated. With the tape applied, we observe a factor of $\sim100$ reduction in glow from the lower-left corner. The floating MOSFET gate was corrected in the ME1001 ROIC. Imaging of a bare ME1001 ROIC shows that this entirely eliminated the lower-left point source (see Figure~\ref{fig:glowcorner}) leaving only the residual glow from the output amplifiers and glow from the fast (column) shift registers evident. There is no sign of glow from the unit cells in the APD array.

\section{Measurements of Dark Current}\label{darkmeasurements}
In SAPHIRA, settling appears to different degrees both immediately post-reset and over long periods. This phenomena is either thermal or voltage, and it resets almost immediately between long series of readouts. We are unable to identify the source as thermal or voltage as the effect is fast enough to confuse the two. None of the SAPHIRA ROICs include reference pixels, which are commonly used to correct for settling and other short- and long-term effects. Reference pixels in NIR arrays require replacing the HgCdTe photodiode with a matching capacitor to the same bias voltage. In the absence of reference pixels, we use a series of reads for 2-4 hours to accommodate the settling effect. Measuring dark current also requires the array to be at the given temperature for a day or more, to allow the interior of the cryostat to reach thermal equilibrium.

\begin{figure}
   \begin{center}
   \begin{tabular}{c}
   \includegraphics[height=8.0cm]{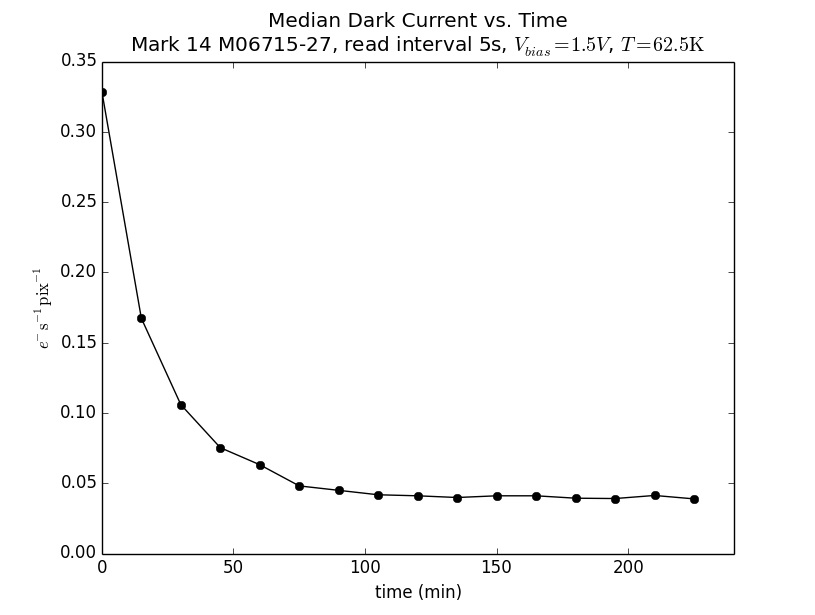}
   \end{tabular}
   \end{center}
	 \vspace{-16pt}
   \caption[Measurement of Dark vs. Time]{\label{fig:darkramp}
Measurements of dark current in a SAPHIRA stabilize over the course of 2+ hours due to voltage settling effects. After settling, measurements are stable to $\pm0.001\esp$.} 
   \end{figure}

The typical dark current measurement ramp consists of an initial reset of the array, followed by 180 reads of the full $320\times256$ array conducted at intervals of 5 seconds. The standard ramp then requires 15 minutes. Longer ramps are necessary for measuring exceptionally low dark currents and are noted appropriately. Readout of the full array takes 10ms, for a readout duty cycle of $0.02\%$. Dark current is determined by averaging together two sets of typically 20 frames each at the beginning (post-settling) and end of a ramp (CDS 20/20), measuring the change between them and adjusting for elapsed time and the conversion or charge gain ($e^{-}/\mathrm{ADU}$). The median in a subarray underneath the mask is relatively isolated from local glow soruces and is taken as the measured dark current (see Figure~\ref{fig:darkmap}). Electrical settling after a change to bias voltages and power cycle distorts measurements, and requires $\sim2$ hours to decay. After settling, dark current measurements are stable to $\pm0.001\esp$ (see Figure~\ref{fig:darkramp}).
	
\begin{figure}
   \begin{center}
   \begin{tabular}{c}
   \includegraphics[height=8.0cm]{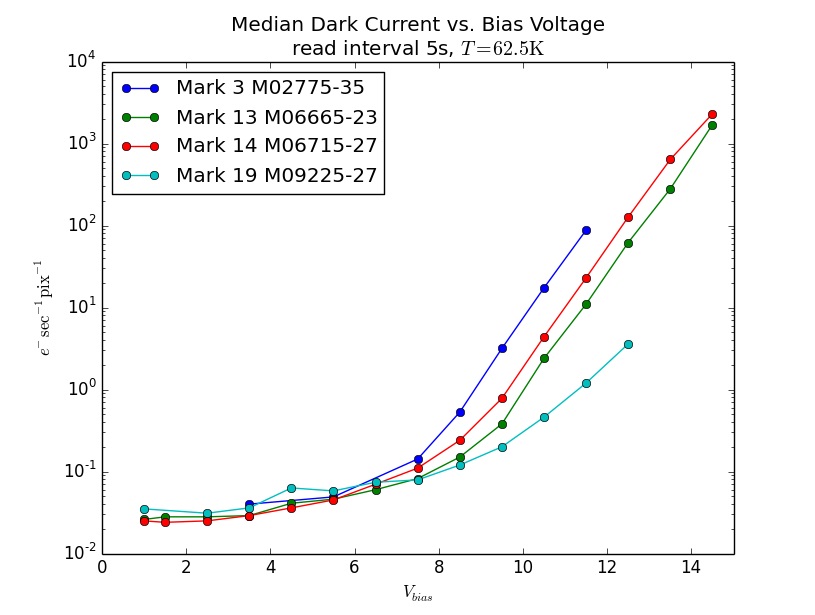}
   \end{tabular}
   \end{center}
	 \vspace{-16pt}
   \caption[Dark Current for Multiple Detectors]{\label{fig:darkcomparisonacrosstypes}
The median dark current vs. $V_{bias}$ function is consistent across multiple production iterations of the SAPHIRA series. Mark 19 arrays show a reduced increase in dark current at high $V_{bias}$.}
   \end{figure}

\begin{figure}
   \begin{center}
   \begin{tabular}{c}
   \includegraphics[height=8.0cm]{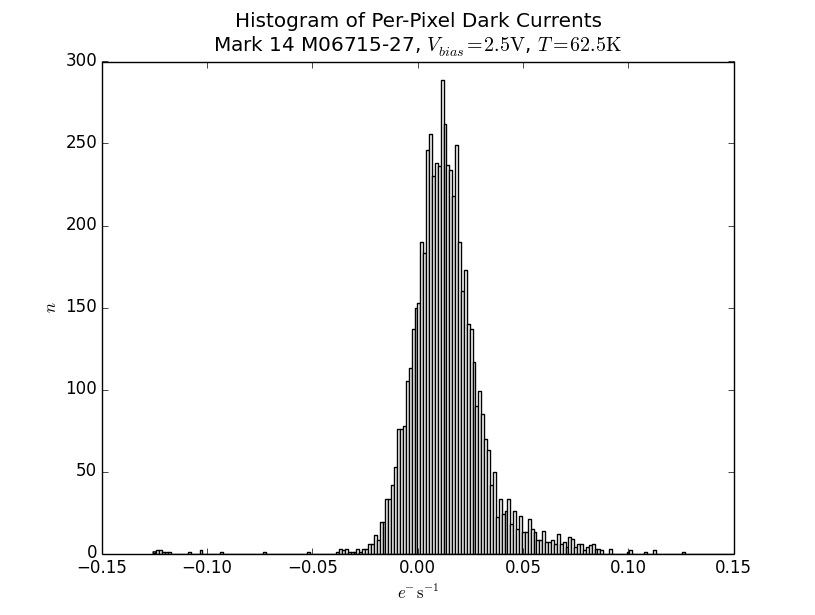}
   \end{tabular}
   \end{center}
	 \vspace{-16pt}
   \caption[Dark Current for $V_{bias} = 2.5\mathrm{V}$]{\label{fig:histogram25V}
For low $V_{bias}$, measured dark currents fall into a normal distribution, dominated by read noise. A small population of 'hot' pixels is evident.}
   \end{figure}

The unity gain baseline is at $V_{bias} = 2.5\mathrm{V}$. Lower bias voltages experience increased capacitance due to only partial depletion and experience correspondingly reduced charge gain. With the suppression of glow (see Section~\ref{darkglow}), and a read interval of 5s, we measure a baseline dark current for SAPHIRA arrays across multiple iterations of $0.025\esp$ (see Figure~\ref{fig:darkcomparisonacrosstypes}). Uncertainty in per-pixel measurements in the specified subarray is $\pm 0.028 e^{-}\mathrm{s}^{-1}$, and is dominated by read noise (see Figure~\ref{fig:histogram25V}).

\begin{figure}
   \begin{center}
   \begin{tabular}{c}
   \includegraphics[height=8.0cm]{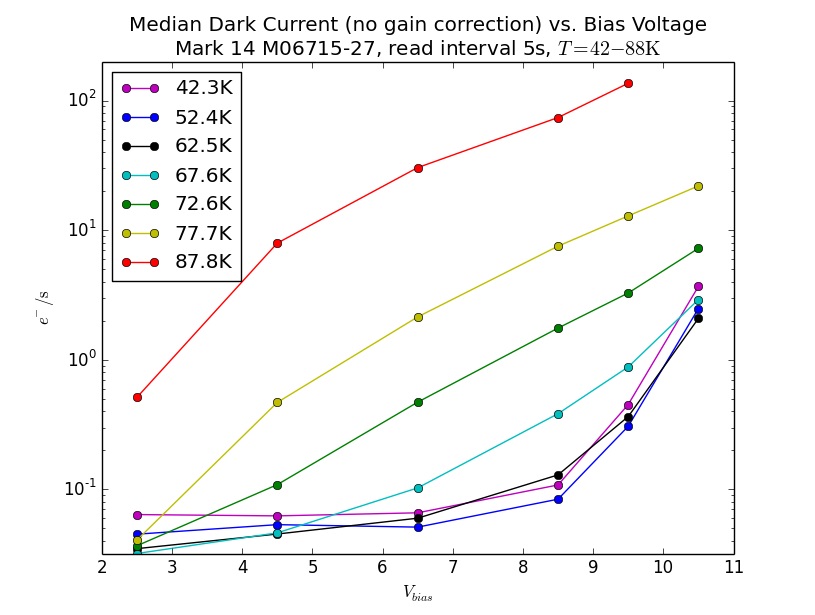}
   \end{tabular}
   \end{center}
	 \vspace{-16pt}
   \caption[Dark Current vs. Temperature] {\label{fig:darkvbiasvtemp} 
At above 60K, dark current both increases with temperature and shows an increased response to $V_{bias}$. There appears to be no change in dark current vs. $V_{bias}$ for temperatures between 40 and 60K.}
   \end{figure}

\begin{figure}
   \begin{center}
   \begin{tabular}{c}
   \includegraphics[height=8.0cm]{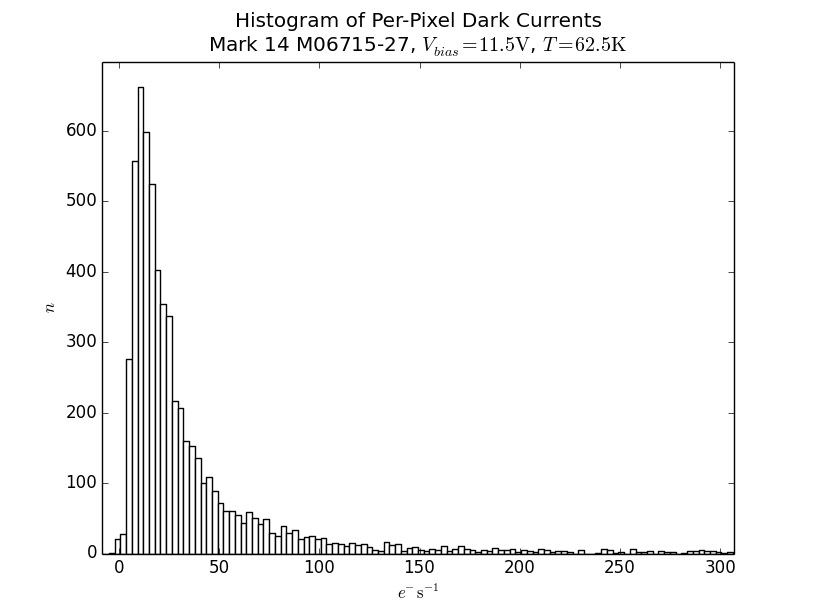}
   \end{tabular}
   \end{center}
	 \vspace{-16pt}
   \caption[Dark Current for $V_{bias} = 11.5\mathrm{V}$]{\label{fig:histogram115V}
In contrast to Figure~\ref{fig:histogram25V}, hot pixels stand out at high $V_{bias}$, and what was a normal distribution has a clear rightward tail.}
   \end{figure}

\begin{figure}
   \begin{center}
   \begin{tabular}{c}
   \includegraphics[height=8.0cm]{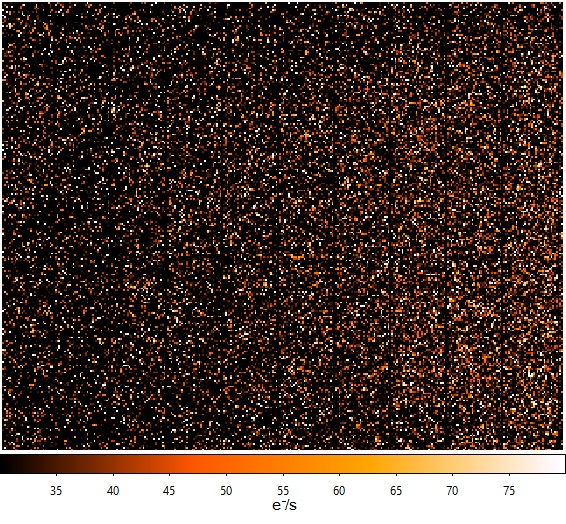}
   \end{tabular}
   \end{center}
	 \vspace{-10pt}
   \caption[Dark Current Across Detector Showing Hot Pixels]
   { \label{fig:hotpixels} 
Hot pixels are scattered randomly across the array except for a slight left/right gradient. Dark vertical striping is due to a single dead readout channel. Image is from Mark 13 M06665-23, $V_{bias} = 14.5\mathrm{V}$, $T = 62.5\mathrm{K}$.}
\end{figure}

We also measure dark current as a function of temperature (see Figure~\ref{fig:darkvbiasvtemp}). Dark current increases exponentially with temperature above $\sim62.5\mathrm{K}$, but is unchanged from 62.5K down to 42.3K. Note also that temperatures above 60K show an increase in dark current as a function of bias for even low bias voltages, as expected for thermal-tunnel current (see Section~\ref{darkanalysis}), but this effect is not evident at lower temperatures. As $V_{bias}$ is increased to produce even modest avalanche gains, at $T \geq 62.5K$ the arrays exhibit sharply increasing dark current (see Figures~\ref{fig:histogram115V}~and~\ref{fig:hotpixels}).

\begin{figure}
   \begin{center}
   \begin{tabular}{c}
   \includegraphics[height=8.0cm]{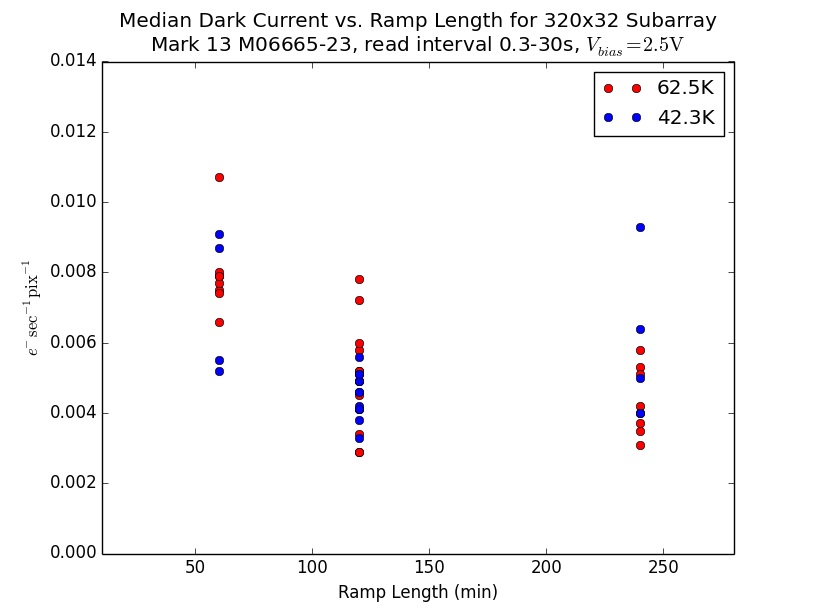}
   \end{tabular}
   \end{center}
	 \vspace{-16pt}
   \caption[Dark Current Upper Limit] {\label{fig:darkvramplength} 
By restricting the subarray to $320\times32$ we put an upper limit on the SAPHIRA's intrinsic dark current of $0.0048 \pm 0.0013\esp$ at $T = 62.5\mathrm{K}$, measured across all 2- and 4-hour ramps. For $T = 42.3\mathrm{K}$ we measure $0.0050 \pm 0.0016\esp$, which is consistent with the $T = 62.5\mathrm{K}$ result.}
   \end{figure}

By restricting the readout to a $320\times32$ subarray over extended sequences of twelve long ($>1\mathrm{hr}$) ramps, we can place even more stringent upper limits on baseline dark current. We measure a dark current of $0.0048 \pm 0.0013\esp$ at $T = 62.5\mathrm{K}$ over two sequences each with six consecutive 2+ hour ramps. A third sequence at $T = 42.3\mathrm{K}$ yields $0.0050 \pm 0.0016\esp$ (see Figure~\ref{fig:darkvramplength}). These are the strictest upper limits yet measured on dark currents in Mark 13 MOVPE arrays. The difference in respective population mean from 62.5K to 42.3K is $0.0002 \pm 0.0006$, showing no significant change with temperature.

\section{Analysis}\label{darkanalysis}
\begin{figure}
   \begin{center}
   \begin{tabular}{c}
   \includegraphics[height=8.0cm]{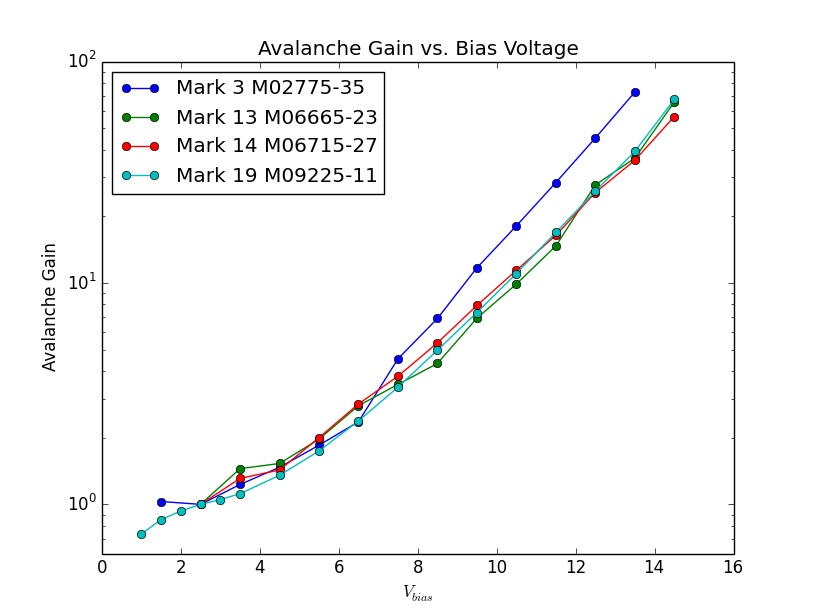}
   \end{tabular}
   \end{center}
	 \vspace{-16pt}
   \caption[Avalanche Gain for Multiple SAPHIRA]{\label{fig:avalanchegain}
	Measured avalanche gain is remarkably consistent across multiple SAPHIRA arrays. These values are used to correct dark current measurements to get a clear picture of dark performance independent of gain effects.}
   \end{figure}

\begin{figure}
   \begin{center}
   \begin{tabular}{c}
   \includegraphics[height=8.0cm]{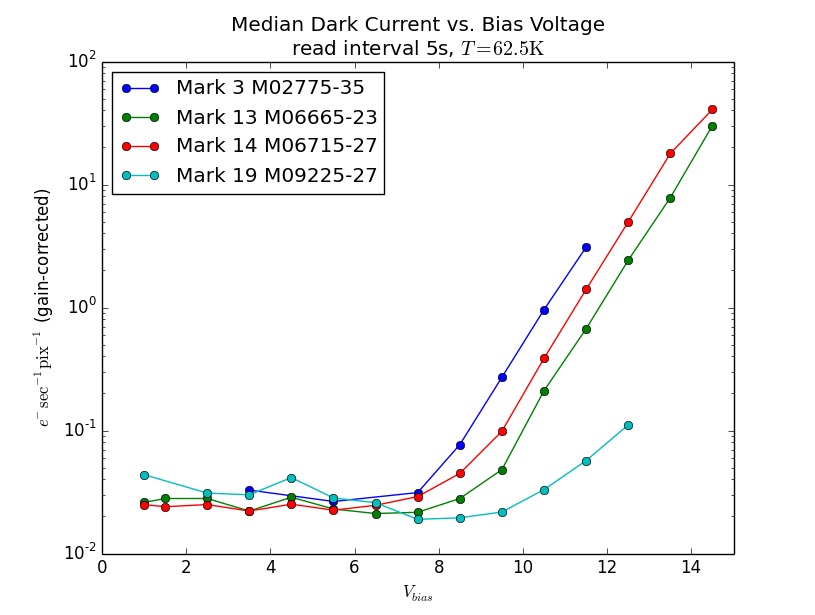}
   \end{tabular}
   \end{center}
	 \vspace{-16pt}
   \caption[Gain-Corrected Dark Current]{\label{fig:darkgc}
	With gain correction, dark current is insensitive to bias voltage for $V_{bias} < 8V$. The shape of the dark vs. $V_{bias}$ curve for high $V_{bias}$ is largely unchanged.}
   \end{figure}

APD observations are normalized using the avalanche gain to produce an input-referred measurement. Similarly, it is necessary to normalize dark current by avalanche gain for a direct comparison. Avalanche gains for the four SAPHIRAs are shown in Figure~\ref{fig:avalanchegain}, and the gain-corrected dark currents in Figure~\ref{fig:darkgc}. This clearly separates gain-corrected dark current vs. $V_{bias}$ into two distinct regimes above and below $V_{bias} = 8\mathrm{V}$ (corresponding to a  gain of $\sim5$). For $V_{bias} < 8\mathrm{V}$, gain corrected dark current is unchanged with increasing bias voltage. Three different iterations of SAPHIRA all show this consistent low-bias dark current of $0.025\esp$ up to $V_{bias} \sim 7.5\mathrm{V}$, a result confirmed by measurements of a Mark 14 SAPHIRA at ESO \citep{finger2017a}. This demonstrates that any increase in dark current at low bias voltage is entirely due to avalanche gain, and does not reflect a change in intrinsic dark current. Note also that the Mark 19 device deviates slightly with a mild decline with $V_{bias}$ at moderate values.

As the measured dark current experiences full avalanche gain, the electons must travel the full depth of the multiplication layer, and thus originate in the absorption layer. If the source were true thermal dark current, we would expect it to originate uniformly throughout the narrower-bandgap multiplication layer, and thus receive lower avalanche multiplication. The constant gain-normalized dark current with bias voltage, taken with the insensitivity to temperature from 62.5K to 42.3K strongly indicates that these dark current measurements are still dominated by glow. 

The reduction in dark current by a factor of 5 to $0.0048\esp$ with a combination of long (2+ hour) ramps and a small subarray readout ($320\times32$), as well as the insensitivity to temperature below 60K, also support these measurements as glow-dominated instead of intrinsic dark current in the MOVPE pixel. This intrinsic dark current should continue the exponential decrease with temperature evident in Figure~\ref{fig:darkvbiasvtemp} from 87.8-67.6K to lower temperatures, a factor of 100 from 62.5K to 42.3K. Our measured two sigma upper limit of $0.0014\esp$ on the drop from 62.5K to 42.3K constrains the intrinsic dark current at 62.5K to be no more than this. This limit of $0.0014\esp$ is by far the lowest such limit yet measured.
	
Above $V_{bias}\sim8\mathrm{V}$, the median dark current climbs rapidly (see Figures~\ref{fig:darkcomparisonacrosstypes}~and~\ref{fig:darkgc}), though the onset voltage for individual pixels ranges from 6 to 11V. We attribute this to the onset of tunnel-tunnel current in individual pixels. The goal of the Mark 19 design was to push the onset of tunneling to higher $V_{bias}$. We do not observe a significant change in onset voltage, though we do see a much slower increase in dark with bias voltage.

\section{Discussion}\label{darkdiscussion}
Previous generations of APDs, particularly those manufactured by liquid phase epitaxy, experience a dramatic increase in dark current with bias voltage even at moderate gains. Bandgap engineering in the critical junction area of the MOVPE SAPHIRA allows bias voltages up to 7.5V to maintain a measured dark current of $0.025\esp$, providing a gain of 5 without the drawback of elevated dark current.

We assert that the measured dark current in the 0.025 to $0.005\esp$ range is dominated by ROIC glow. This is based primarily on the absence of any detectable change in dark current from 62.5K down to 42.3K and, for dark currents $\sim0.025\esp$, the absence of any change in gain-corrected dark current with bias voltage up to 7.5V. It is unlikely this glow originates within the unit cell of the ROIC as the metal layers at the top of the ROIC are staggered to form an opaque barrier over the circuitry below (see Figure~\ref{fig:layout}). This is supported by the absence of any observable glow from the active pixel area of the bare ME1001 ROIC (see Figure~\ref{fig:glowcorner}) while glow from the fast shift register just below the active area and the output amplifiers is clearly evident. We also note that as the lower surface of the HgCdTe is obscured by the indium bump and its contact pad (see Figure~\ref{fig:layout}) any glow from the unit cell of the ROIC would be absorbed in the mesa trench, preferentially in the gain region due to the geometry. It seems more plausible that the glow is due to photons from the observed glow sources scattering between the lower surface of the mask and the reflective surface of the ROIC; entering the edge of the wide band gap buffer and propagating within it (Figure~\ref{fig:layout}); or possibly escaping from around the mask’s edge contact with the 68-pin LCC and being reflected back onto the unmasked area.

All indications are that the measured dark current at low bias voltage ($V_{bias} < 8\mathrm{V}$) is still glow-limited, and not a function of material quality. There are a few reasons for this. The insensitivity to temperature below 60K is one, and indicates that the source of dominant dark current is likely not HgCdTe. The second is the sensitivity to overall running voltage. Although dark current decreases as the voltage goes down to 3.5V, we are not able to explore lower voltages as the detector stops running at $<3.5V$. Last, we note that the lowest dark current is confined to the top center of the detector, farthest from the output amplifiers on the sides and the fast shift register below the APD array. This implies that it originates from the detector's electronics. Revisions to the SAPHIRA ROIC have suppressed glow; changes to the metal layers of the ME1000 ROIC reduced visible glow from the output amplifiers, and the ME1001 ROIC eliminated the glow point source in the lower left corner of the ROIC (see Section~\ref{darkglow}).

Measurements of dark current $<0.01\esp$ are demanding as they require sub-millivolt stability over hours-long integrations. Though future reductions in glow may in fact reduce the measured dark current further, such an effect will be difficult to characterize without the use of reference pixels. Future iterations of SAPHIRA that include reference pixels on the array would allow the operator to calibrate for voltage instability effects.

Outside the glow-limited regime of $T\leq60\mathrm{K}$ and $V_{bias} < 8\mathrm{V}$, we observe two distinct dark current sources. We attribute the rapid increase in dark current above $V_{bias} = 8\mathrm{V}$ to tunnel-tunnel current. The electron tunnels into a trap as an intermediate step before tunneling to the conduction band. This effect is bias voltage dependent but has no dependence on temperature. The second effect is evident for $T>60\mathrm{K}$, and is attributed to thermal-tunnel current. In this process the electron is thermally excited into a trap, then tunnels into the conduction band. Thus thermal-tunnel has both thermal and voltage components, as seen in Figure~\ref{fig:darkvbiasvtemp}.

The onset of tunnel-tunnel current above $V_{bias} = 8\mathrm{V}$ is consistent for Marks 3, 13, and 14. The goal of the Mark 18/19 generation of SAPHIRA detectors was to push the onset of this tunneling current to yet higher bias voltages. As observed in Figure~\ref{fig:darkcomparisonacrosstypes}, Mark 19 did not substantially push back the onset of tunneling, but did serve to suppress the increase with $V_{bias}$ due to the introduction of a bandgap gradient in the multiplication region. This design is also responsible for the mild decrease in gain-corrected dark current up to $V_{bias}$ above 9.5V, as the majority of dark current is generated in the narrow-bandgap material near the end of the multiplication region. Further development will be necessary to unify the reduced tunneling with the broader operational characteristics of the Mark 13/14 arrays.

The current state-of-the-art for low background NIR astronomy is Teledyne's H2RG and H4RG arrays. Science-grade H2RG arrays regularly achieve dark currents of $\leq0.01\esp$ with a cutoff wavelength of $2.5\um$ and pitch $18\um$, and with a typical bias voltage of 250mV reach $0.002\esp$ \citep{blank2012}. In a $320\times32\times24\um$ SAPHIRA subarray at $V_{bias} = 2.5\mathrm{V}$ we measure $0.0048 \pm 0.0013\esp$ in ramps of 2 or more hours and set a two sigma upper limit of $0.0015\esp$ for the intrinsic dark current in the MOVPE pixel, a value fully consistent with the best reported dark currents in $2.5\um$ cutoff H2RG arrays but at 10 to 35 times the avalanche bias voltage. 

Future SAPHIRA development has bifurcated into separate designs optimized for two different goals. First, an adaptive optics array will depend on strong avalanche gains at high $V_{bias}$ and fast readouts at the expense of dark current. This follows on the current APD designs at Leonardo. Second, a future array optimized for low-background astronomical observations will use only moderate avalanche gains to achieve sub-electron read noise and realize the ultimate dark current limits of Leonardo's MOVPE APD arrays. The UH-Leonardo collaboration is pursuing dark current optimization of $1k\times1k$ class arrays on the new ROIC design, to include reference pixels and a reduced readout rate as a tradeoff for further suppression of glow.

\section{Conclusions}\label{darkconclusions}
We have demonstrated that SAPHIRA arrays achieve an avalanche gain of 5 with a glow-limited dark current of $0.025\esp$, advancing one of the primary goals of SAPHIRA development. With further improvement of the ROIC glow useful avalanche gains at yet lower dark currents are readily achievable. Using a $320\times32$ array read at 5-second intervals over a $>2$-hour ramp we measure a dark current of $0.0048 \pm 0.0013\esp$ at $T = 62.5\mathrm{K}$ and $V_{bias} = 2.5\mathrm{V}$ (unity gain). The implied upper limit on intrinsic detector dark current is $0.0015\esp$. This validates the MOVPE manufacture process as a means of making astronomical HgCdTe arrays on par with conventional molecular beam epitaxy. In the process we have shown not only that prior SAPHIRA dark current measurements were entirely glow-dominated, but that our present measurements likely are as well. We have pushed glow mitigation and dark current measurements as far as possible with the current SAPHIRAs. New ROICs and arrays are needed. The UH-Leonardo collaboration is now focused on developing large-format arrays and further suppressing glow to measure the true dark current limits of Leonardo's MOVPE APD detector arrays.
	
\chapter{Photon Counting Properties of SAPHIRA APD arrays}\label{ch:photoncounting} 
This chapter was published as \cite{atkinson2018}.

We measure the ability of the Leonardo SAPHIRA avalanche photodiode array to perform photon counting. The current SAPHIRA arrays achieve $>90\%$ single photon efficiency (independent of quantum efficiency expected to be also $>90\%$) and a time resolution of $125\us$ with a dark current of $21\esp$. Our characterization of several iterations of the SAPHIRA detector over the past 3 years of its development have also revealed a broader pulse height distribution than was originally expected.

\section{Introduction}
Originally developed for fringe-tracking at the European Southern Observatory (ESO), the Selex Advanced Photodiode HgCdTe Infrared Array (SAPHIRA) is now the premier detector for near-infrared (NIR) wavefront sensing in adaptive optics \citep{finger2010,atkinson2014,atkinson2016,hall2016a}. The ongoing collaboration between the SAPHIRA's manufacturer Leonardo (previously Selex) and the University of Hawai'i Institute for Astronomy (UH-IfA) has evaluated the detector's ability to count NIR photons. Prior photon counting in the NIR was limited to frame and dark count rates of $>10$MHz and $>100,000\esp$ respectively, prohibitively high for astronomy \citep{beck2015}. The new capabilities of the SAPHIRA will also enable observations in the largely unexplored area of NIR high time resolution astronomy.

We have previously shown the SAPHIRA to have a baseline dark current of $<=0.001\esp$ up to a bias voltage $V_{bias}$ of 8.5V \citep{atkinson2016,atkinson2017a,finger2016,hall2017}. This work presents the current results from the ongoing photon counting development effort, and thoroughly details the SAPHIRA's ability to discriminate individual photon events with sub-millisecond time resolution. We will present that this ability has a more complex limitation; individual photons can be detected, but small numbers of photons simultaneously arriving are difficult.

Section~\ref{pcsetup} lays out our laboratory setup for investigations of SAPHIRA photon counting. Section~\ref{pcgainprofile} describes characterization of the detector's electron avalanche and presents measurements of the pulse height distribution and other properties relevant to photon counting. Section~\ref{pcanalysis} discusses the nature of the ballistic avalanche and the relationship between the avalanche pulse height distribution and the excess noise. In Section~\ref{pcdiscussion} we discuss the pulse height distribution further and the aims of future SAPHIRA development. Section~\ref{pcconclusions} provides a summary of our results.

\section{Setup}\label{pcsetup}
\begin{figure}
   \begin{center}
   \begin{tabular}{c}
   \includegraphics[height=8cm]{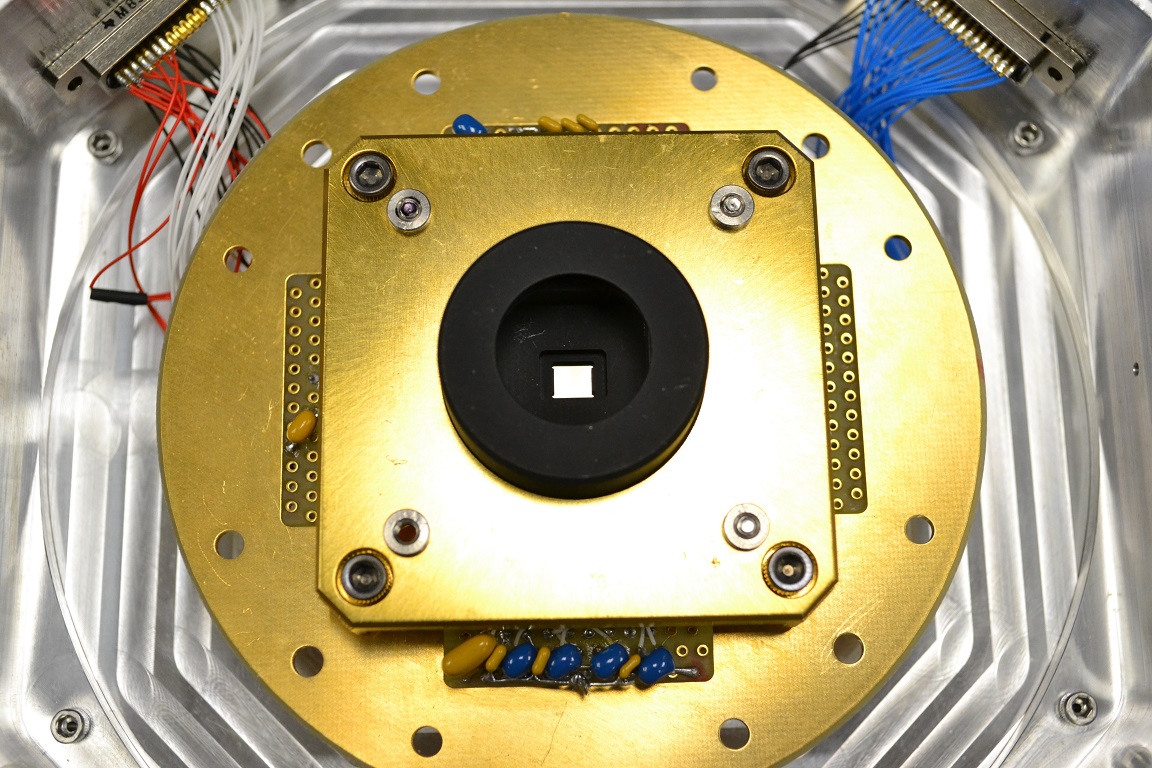}
   \end{tabular}
   \end{center}
	 \vspace{-16pt}
   \caption[The KSPEC Cryostat] 
   { \label{fig:KSPEC3} 
The socket holding SAPHIRA use springs to maintain constant pressure with the electrical contacts and the clamp assembly in place over the array.}
   \end{figure}

We characterized the SAPHIRA arrays in the KSPEC cryostat (see Figure~\ref{fig:KSPEC3}). KSPEC was originally an $IJHK$-band spectrometer, since converted into a detector testbed for the NGST/JWST detector program \citep{hodapp1994,hodapp1996}.

The detector itself is operated by a third-generation Astronomical Research Cameras (ARC) detector controller. The controller uses mostly standard ARC boards (ARC-22, ARC-32) including four 8-channel IR readout boards (ARC-46) \citep{leach2000}. The bias/utility board has been replaced with an extremely low noise version designed at Australian National University and reproduced at UH-IfA. These are installed in a twelve-slot ARC chassis with a single backplane. Fiber optic lines connect to the ARC-66 PCIe board in a Linux desktop computer. The instrument is operated via a set of command-line scripts making use of the v3.5 ARC API.

The controller uses Motorola DSP56000 assembly language code to operate, which we derived from code for HAWAII-2RG operation provided by both Bob Leach of ARC and Marco Bonati of CTIO. To propagate changes in the code, it is written to the controller immediately before every detector power-up.

Detector temperature is measured on a cold finger in thermal contact with the detector itself, and is accurate to $\sim3{K}$ of the detector's actual temperature. A temperature diode mounted on the leadless chip carrier was used to calibrate the temperature measurements from the cold finger. The operating temperature is 2.3K higher than the measured detector-stage temperature at 40K and 2.9K higher at 100K. The diode sensor glows brightly in the NIR when powered and was typically not operated when measurements were being taken.

Calibration LEDs at $1.05\um$, $1.30\um$, $1.75\um$, and $3.1\um$ are mounted to an integrating sphere in the cryostat, to provide uniform illumination across the detector when active. Typically the $H$-band LED $\lambda = 1.75\um$ was used for photon counting measurements. When in use the LED power level is manually set by the operator.

Our ability to count photons is most directly characterized by observing the detector's high-gain response to individual events. At the minimum readout subarray size, 32x1 pixels read out through 32 output channels, the frame rate is almost identical to the pixel rate of 250kHz save for some minor overhead. Absorbed light from the integrating sphere LED ($\lambda = 1.75\um$) at the lowest setting is estimated at $\sim6000 \mathrm{photons} s^{-1} pix^{-1}$, estimated from operating the detector at the lowest gain.

\begin{figure}
   \begin{center}
   \begin{tabular}{c}
   \includegraphics[height=8.0cm]{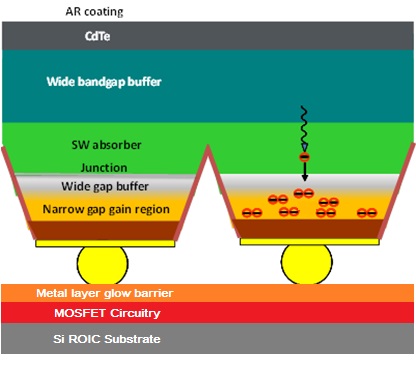}
   \end{tabular}
   \end{center}
	 \vspace{-16pt}
   \caption[Physical Structure of the SAPHIRA] 
   { \label{fig:physical2} 
The broad physical structure of SAPHIRA pixels, manufactured by MOVPE. The depth of individual layers is varied experimentally to improve performance. (\textit{Original figure courtesy Leonardo.})}
   \end{figure}

\begin{figure}
   \begin{center}
   \begin{tabular}{c}
   \includegraphics[height=8cm]{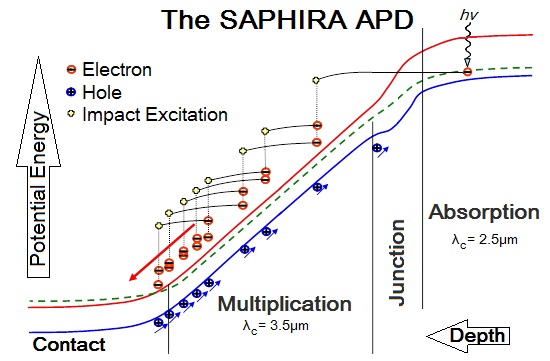}
   \end{tabular}
   \end{center}
	 \vspace{-16pt}
   \caption[Bandgap of the SAPHIRA] 
   { \label{fig:bandgap3} 
The single-carrier electron avalanche is a major noise advantage HgCdTe has over other NIR detection materials. The bias voltage $V_{bias}$ is applied across the multiplication and junction regions. Though photons are depicted here being captured in the absorption region, it is possible for $2.5\um \leq \lambda \leq 3.5\um$ photons to penetrate to and be absorbed in the multiplication layer. (\textit{Original figure courtesy Leonardo.})}
   \end{figure}

A SAPHIRA detector is a 320 x 256 array, with larger arrays planned for the future. The metal organic vapor phase epitaxy device structure employed in the APD is shown in Figure~\ref{fig:physical2}. Photon absorption occurs in the P-type absorber, creating electron-hole pairs. The electrons thermally diffuse to the junction. The high field is dropped across a weakly doped N-type region, the multiplication layer. Here electrons are accelerated and generate other electron-hole pairs by impact ionization. The low mobility hole acquires energy from the applied electric field very inefficiently and readily loses it to optical phonons \citep{rogalski2005}. The process is essentially a pure electron cascade with an exponential gain versus bias voltage profile. The potential energy with depth is shown in Figure~\ref{fig:bandgap3}, which illustrates that the history-dependent nature of the avalanche gain underpins the low noise figure of HgCdTe.

Multiple iterations of the SAPHIRA detector were investigated. This work presents marks 13, 14, and 19. Mark 13 and 14 are the same APD design but were subjected to different high temperature anneals. Mark 19 was an experimental design aimed at reducing tunnel current at low temperature.

\section{Measurements}\label{pcgainprofile}
The photon-counting performance of the detector is analyzed by pulse height distributions. These are direct measures of the difference between successive reads of the same pixel, i.e. frames $2-1$, $3-2$, $4-3$, and so on. The data set for an individual pixel is then a list of all subtracted read pairs, and the full data set cube is those individual sets for all pixels. Data was generally taken with 100,000 frames consecutively, with a reset separating ramps applied every 10,000 or 2,000 frames depending on the incident light level, so there are either 10 or 50 ramps and resets in a data set. To reach the maximum readout rate the window observed is the minimum $32\times1$, a single read per frame.

\begin{figure}
   \begin{center}
   \begin{tabular}{c}
   \includegraphics[height=8cm]{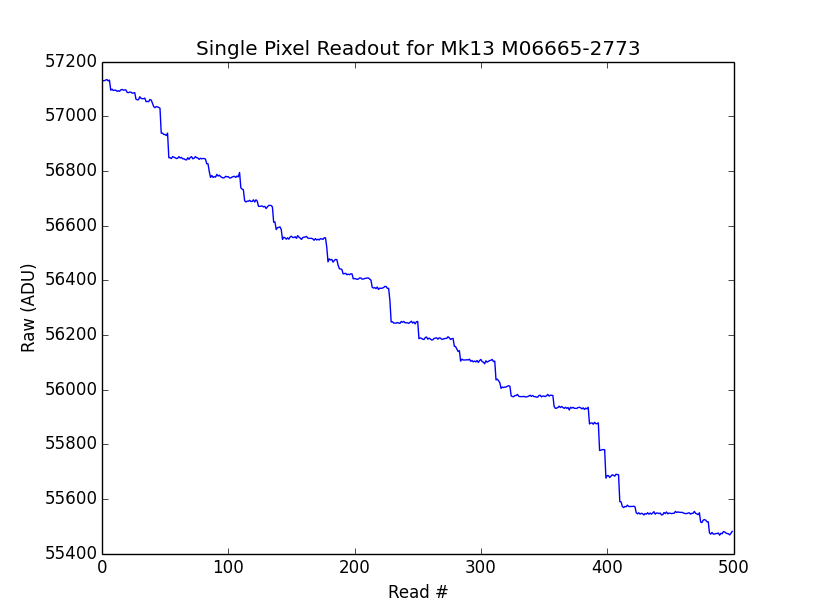}
   \end{tabular}
   \end{center}
	 \vspace{-16pt}
   \caption[Demonstration of Photon Events] 
   { \label{fig:rawpix1} 
Single pixel read out 500 times with an LED on. The incident photons can be seen as large jumps in the ADU value, but note that the size of the jump varies.}
   \end{figure}

\begin{figure}
   \begin{center}
   \begin{tabular}{c}
   \includegraphics[height=8cm]{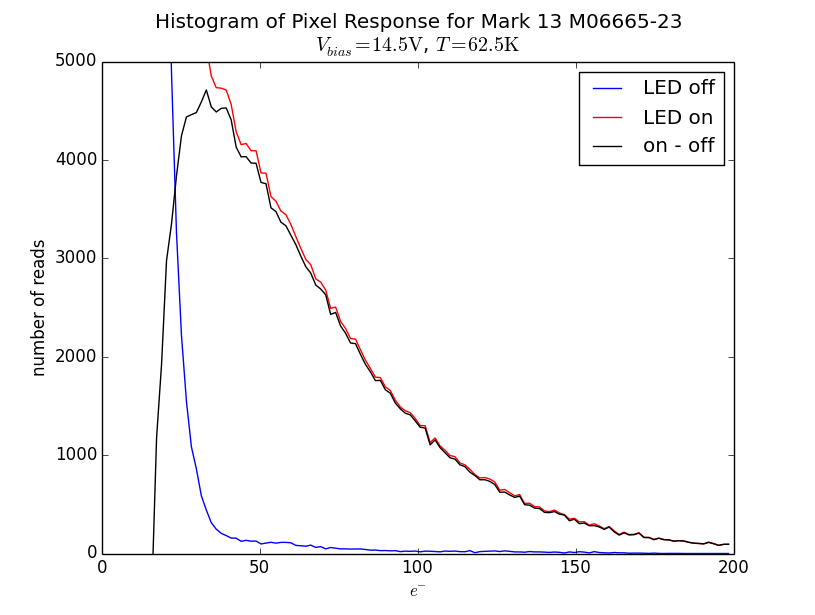}
   \end{tabular}
   \end{center}
	 \vspace{-16pt}
   \caption[Pulse Height Distribution (zoomed)] 
   { \label{fig:pchistlinear1} 
A linear pulse height distribution for a mark 13 SAPHIRA array at $V_{bias} = 14.5\mathrm{V}$, mean avalanche gain = 66, with LED source off (blue), on (red), and on - off (black). No averaging or filtering has been applied. The subtraction gives us the pulse height distribution, which has a peak around 0.5 and a sharp lower bound at 0.25.}
   \end{figure}

\begin{figure}
   \begin{center}
   \begin{tabular}{c}
   \includegraphics[height=8cm]{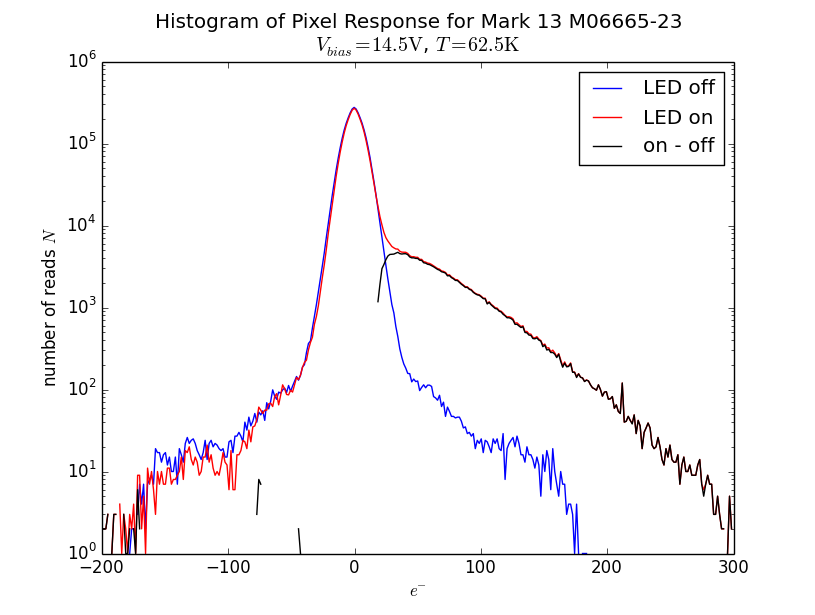}
   \end{tabular}
   \end{center}
	 \vspace{-10pt}
   \caption[Pulse Height Distribution (full)]
   { \label{fig:pchist1} 
The full logarithmic distribution of the off and on curves from Figure~\ref{fig:pchistlinear1}. The non-normally distributed section of the LED off distribution at $>50e^{-}$ is symmetric about zero and is a result of rare electrical glitches in the readout.}
	 \vspace{10pt}
\end{figure}

\begin{figure}
   \begin{center}
   \begin{tabular}{c}
   \includegraphics[height=8cm]{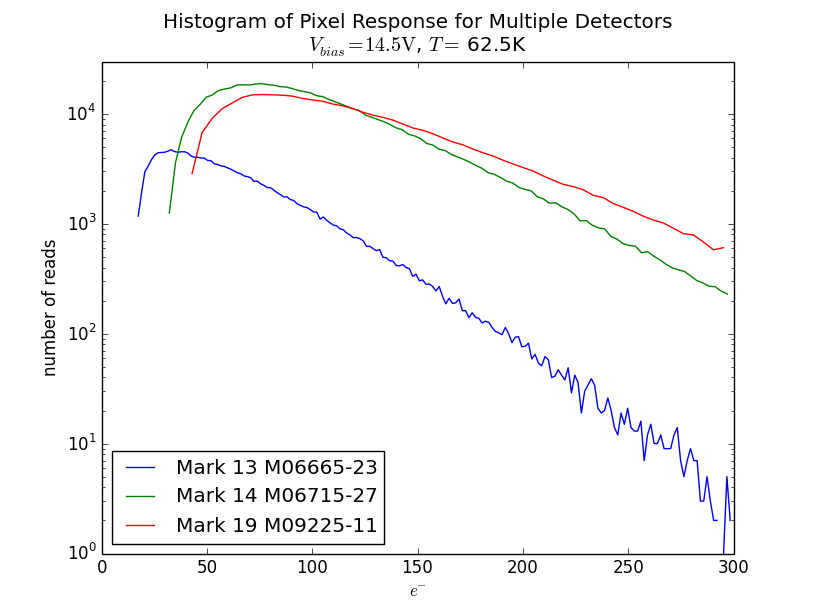}
   \end{tabular}
   \end{center}
	 \vspace{-10pt}
   \caption[Pulse Height Distributions of Multiple Detectors]
   { \label{fig:pchistmultiple1} 
Logarithmic distribution of LED on - off curves from multiple SAPHIRA detectors. The different amplitudes (heights) of Mark 13 vs. Marks 14/19 is due to a different incident light level during the observation. The upward tails of the distributions are universal. Note also the increased curve at the low end of the Mark 14/19 devices, this is due to increased read noise in those detectors.}
\end{figure}

With the LED operated at minimum voltage, a number of frames show no incident photons from the pixels, while some see a single photon (see Figure~\ref{fig:rawpix1}). A pulse height distribution is shown in Figures~\ref{fig:pchistlinear1}~and~\ref{fig:pchist1} as overlapping histograms with and without incident light. The on - off curve shows the avalanches that individual photons produce as a pulse height distribution. The shape is verified in other work \citet{finger2017a}. Similar behavior is observed in multiple SAPHIRA detectors (see Figure~\ref{fig:pchistmultiple1}). (Note that the presented work makes use of different indicent light levels for different detectors.)

We estimate dark current as full avalanche events, as initiates anywhere within the detector's depth and thus does not experience a full avalanche. With this technique, mean dark current at $V_{bias} = 14.5\mathrm{V}$ and $T = 62.5\mathrm{K}$ is measured as $\sim20\esp$ and thus insubstantial to photon counting measurements, which take approximately 0.4 seconds and thus have only 8 equivalent dark current events. Measurements of the gain are normalized to the $V_{bias} = 2.5\mathrm{V}$, which is assumed to be a gain of 1. 

\begin{figure}
   \begin{center}
   \begin{tabular}{c}
   \includegraphics[height=8cm]{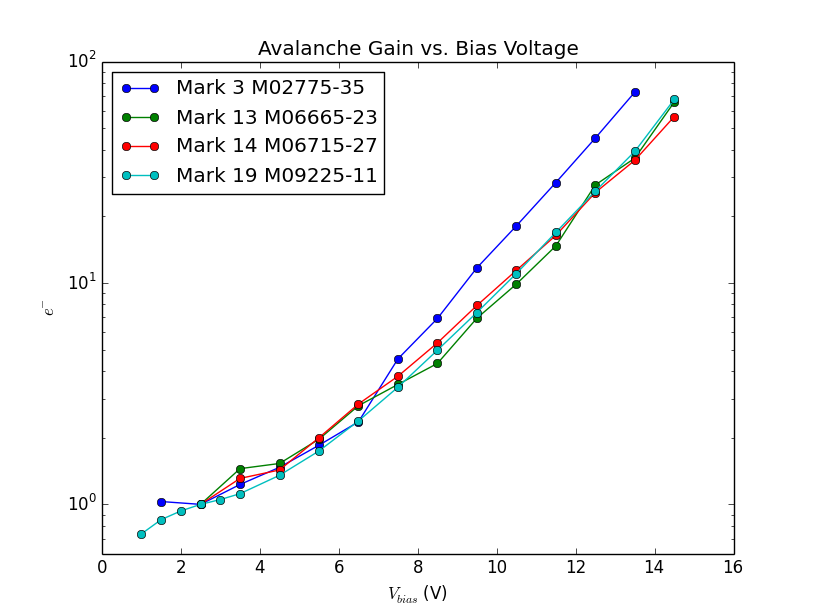}
   \end{tabular}
   \end{center}
	 \vspace{-6pt}
   \caption[Avalanche Gain for Multiple SAPHIRA]
   { \label{fig:avgain1} 
Measured avalanche gain of multiple SAPHIRA devices plotted against bias voltage $V_{bias}$. At very low bias voltages there is a capacitance effect that makes avalanche gain appear lower.}
	 \vspace{10pt}
\end{figure}

Given a measured avalanche gain of 66 at $V_{bias} = 14.5V$ for a mark 13 SAPHIRA, the mean single photon avalanche is $66e^{-}$. We can determine from the incident light that approximately 2.5\% of these events are coincident multi-photon avalanches. Given that the ratio between 1 and $2e^{-}$ results is a factor of $<10$, the single-photon avalanche must still be the dominant part of the distribution at $2e^{-}$. Therefore we attribute the shape of the curve entirely to single photon avalanches. Other SAPHIRA devices show similar avalanche gain vs. $V_{bias}$ curves (see Figure~\ref{fig:avgain1}).

\begin{figure}
   \begin{center}
   \begin{tabular}{c}
   \includegraphics[height=8cm]{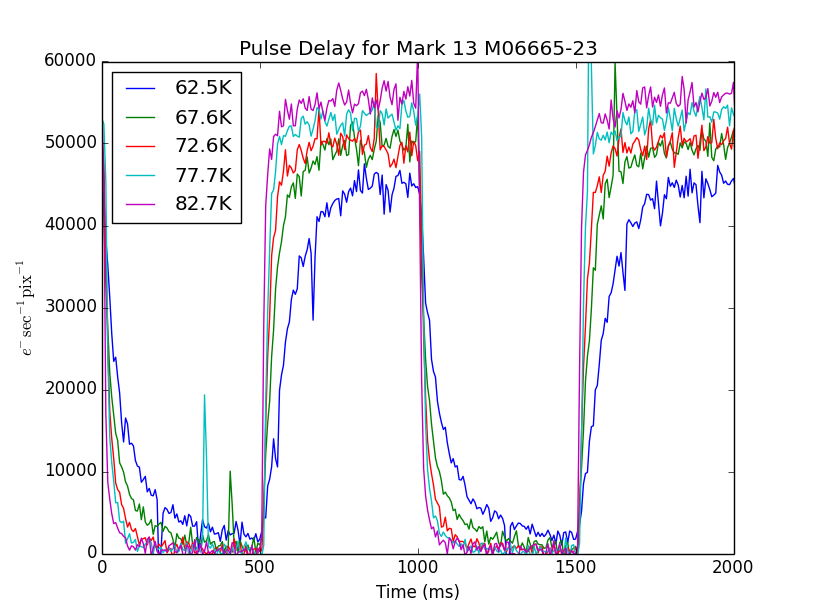}
   \end{tabular}
   \end{center}
	 \vspace{-10pt}
   \caption[Time Delay]
   { \label{fig:timedelay1} 
Response time presented as a function of temperature. Incident light is an LED source driven by a square wave at 1Hz with $50\%$ duty cycle.}
	 \vspace{10pt}
\end{figure}

We also measure a delay in response time, the detector's lag in detecting a change in the incident light. Our observations use a calibration LED in the integrating sphere driven by a square wave to generate light with a sharp cut-on and -off. We see this effect appearing at $T_{det} \sim80\mathrm{K}$ and growing with decreasing temperature (see Figure~\ref{fig:timedelay1}). Measured delays are specific to individual devices, with some devices having a time constant $\tau > 100 \mathrm{ms}$ though $50-100\mathrm{ms}$ is typical at 60K. For photon counting applications, this delay easily dominates the detector's bandwidth when operated at low temperatures. Only a fixed number of electrons in a given time interval experience the delay. This does not ameliorate the effect for astronomical applications given typically low light levels, but does point toward an explanation as avalanche electrons are absorbed and released on non-trivial timescales by traps in the HgCdTe.

\section{Analysis}\label{pcanalysis}
We present the pulse height distribution having a distinct shape, with a sharp lower bound and a broad tail upwards. Although this allows the easy detection of individual photons, multiple photons being absorbed into the same read cannot be differentiated due to the tail, as would be capable with a fully deterministic gain.

Photon counting with any detector requires setting a detection threshold, a value above which a result is assumed to represent a photon event (or events). We statistically evaluate the efficacy of threshold via two categories of error: false negatives, avalanches lower than the threshold and undetected; and false positives, null results higher than the threshold value and erroneously interpreted as photons.  We characterize the probability of false negatives as a threshold efficiency (TE) analogous to a device's quantum efficiency (QE). The false positive (FP) rate is given in units of $\esr$ as it occurs as a function of read noise. The FP rate becomes $\esp$ when the readout rate is applied, which is functionally equivalent to dark current.

\begin{figure}
   \begin{center}
   \begin{tabular}{c}
   \includegraphics[height=8cm]{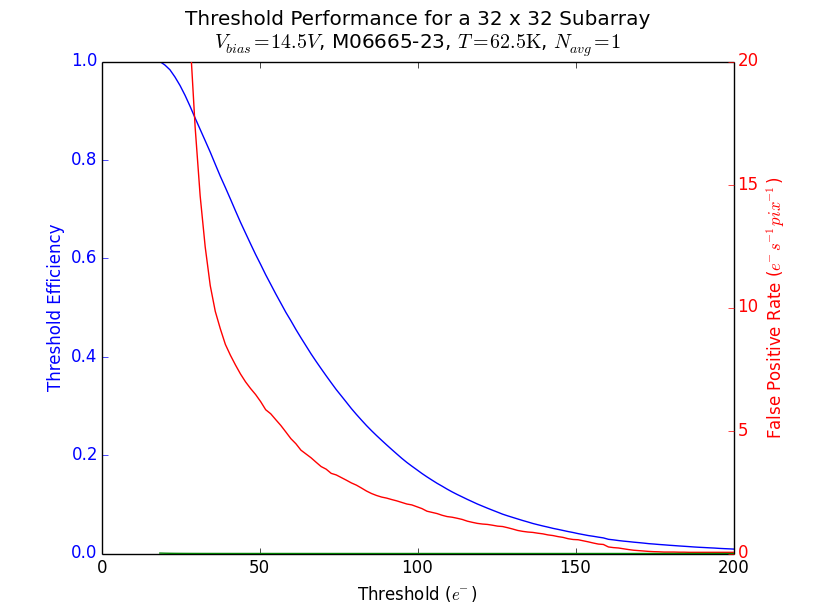}
   \end{tabular}
   \end{center}
	 \vspace{-10pt}
   \caption[Threshold Efficiency and False Positives]
   { \label{fig:threshold} 
Evaluation of threshold values for threshold efficiency and false positive rate. Selected values are shown in Table~\ref{tab:TEvFPTable1}.}
	 \vspace{10pt}
\end{figure}

\begin{figure}
   \begin{center}
   \begin{tabular}{c}
   \includegraphics[height=8cm]{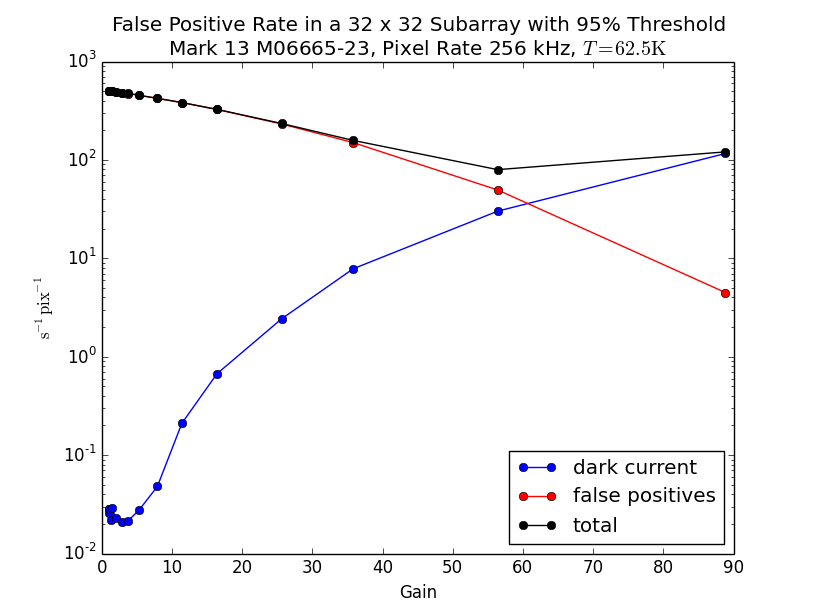}
   \end{tabular}
   \end{center}
	 \vspace{-10pt}
   \caption[False Positives and Dark Current]
   { \label{fig:FP} 
False positive rate and dark current as a function of detector gain for a $32\times32$ subarray of M06665-23. This is an extrapolation from $32\times32$ data, for which both measurements should be identical when adjusted for time.}
\end{figure}

Results for the Mark 13 M06665-23 are presented in Figures~\ref{fig:threshold}~and~\ref{fig:FP}. A selected sample of thresholds with their TE and FP rate are also shown in Table~\ref{tab:TEvFPTable1}. With threshold such that $\mathrm{TE} > 90\%$, the false positive rate is $21\esp$, very near to the $20\esp$ median dark current of a SAPHIRA device at $V_{bias} = 14.5V$. The other two detectors discussed in this paper show worse read noise behavior. At $\mathrm{TE} > 90\%$ the Mark 19 has an FP rate of $178\esp$, and the Mark 14 shows $492\esp$.

\begin{table}
\caption{Values of Threshold Efficiency and False Positive Rate}\label{tab:TEvFPTable1}
\centering
\begin{tabular}{ c c c c }
Threshold & TE & 
\multicolumn{2}{c}{False Positive Rate} \\
&& $\esr$ & $\esp$ \\
\hline
58 & $51\%$ & $6.3*10^{-4}$ & 5.21 \\
36 & $82\%$ & $1.3*10^{-3}$ & 10.9 \\
28 & $91\%$ & $2.5*10^{-3}$ & 20.9 \\
26 & $95\%$ & $4.2*10^{-3}$ & 34.8 \\
20 & $99\%$ & $0.011$       & 93.8 \\

\vspace {10pt}
\end{tabular}

Selected thresholds and matching values for the threshold efficiency (TE) and false positive (FP) rates. The $\esr$ and $\esp$ rates reported depend on the maximum read rate of $250\mathrm{kHz}$ applied to a 32 x 32 window as well as averaging.
\end{table}

\section{Discussion}\label{pcdiscussion}
Given the measured pulse height distribution of the SAPHIRA APD and similarity with other investigations of devices we expect the shape to represent a fundamental property of HgCdTe APDs \citep{finger2014,finger2016}. The linear operation of SAPHIRA APDs is an advantage over developing Geiger-mode NIR APD arrays which require millisecond resets after each detection. Our high TEs (combined with the high QEs of HgCdTe detectors), the linear operation, and the low dark current also compare favorably with Geiger-mode APDs. The SAPHIRA is also not subject to crosstalk, a limiting issue on Geigers.

Additionally, an ongoing redesign to the bandgap structure of the device seeks to push back the onset of dark (tunneling) current to higher bias voltages. This would greatly reduce the dark current penalty for operating an array at the high bias voltages necessary for photon counting, and would improve SAPHIRA performance in applications of NIR high time resolution astronomy.

\section{Conclusions}\label{pcconclusions}
We measure the overall performance of the SAPHIRA APD and demonstrate its capability to count photons and limitations thereof. The pulse height distribution shows that individual IR photons can be easily detected but two or more photons being absorbed in a single read are indistinguishable. Although the ballistic avalanche does greatly reduce amplification noise relative to other APDs, its stochastic nature makes the avalanche pulse height distribution still relatively broad. We have shown that despite this, the SAPHIRA is capable of counting $>90\%$ of photon events with a time resolution of $125\us$, incurring a mild increase in dark current relative to the dominant tunneling current. Also shown is that the tail of the pulse height distribution prevents the accurate detection of multiple photons in a single read. Maintaining a high time resolution helps identify incident photons for relatively rapid arrivals.

Though still in development, we find present iterations of SAPHIRA to be capable of efficient NIR photon counting, a unique capability among astronomical detector arrays. Future versions of the SAPHIRA are expected to resolve issues such as temperature-dependent time delay. The SAPHIRA is then the first NIR APD capable of both high efficiency and low dark current in an array format.

\chapter{Probability of Physical Association of 104 Blended Companions to \textit{Kepler} Objects of Interest Using Visible and Near-Infrared Adaptive Optics Photometry}\label{ch:palomar} 
This chapter was published as \cite{atkinson2017b}.

We determine probabilities of physical association for stars in blended Kepler Objects of Interest, and find that $14.5\%^{+3.8\%}_{-3.4\%}$ of companions within $\sim4\arcsec$ are consistent with being physically unassociated with their primary. This produces a better understanding of potential false positives in the Kepler catalog and will guide models of planet formation in binary systems. Physical association is determined through two methods of calculating multi-band photometric parallax using visible and near-infrared adaptive optics observation of 84 KOI systems with 104 contaminating companions within $\sim4\arcsec$. We find no evidence that KOI companions with separation of less than $1\arcsec$ are more likely to be physically associated than KOI companions generally. We also reinterpret transit depths for 94 planet candidates, and calculate that $2.6\% \pm 0.4\%$ of transits have $R > 15R_{\earth}$, which is consistent with prior modeling work.

\section{Introduction}
The \textit{Kepler} mission had a simple observing strategy: it observed a 105 deg$^2$ field in Cygnus near-continuously with an unfiltered wideband camera. Its main data output were the light curves of target stars, in which it found transits and measured their depth and timing. The conversion of this transit information to planetary characteristics requires the stellar parameters of the host, which the \textit{Kepler} telescope could not provide itself. Stellar characterization is then dependent on data from other sources, typically photometric observations performed for the Kepler Input Catalog in the visible and by 2MASS in the near-infrared \citep{brown2011,liebert1995,huber2014}.

A complication arises from the vulnerability of \textit{Kepler}'s relatively large 4\arcsec pixels to the misinterpretation of unresolved binaries as single stars \citep{borucki2010}. These unseen companions dilute the transit by making it appear shallower relative to its host star, and thus the transiting object's size is underestimated. Photometric characterization of the host star is also distorted by the blended light.

Many of these blended and contaminating companions can be identified in the \textit{Kepler} data by careful examination of the light curve data for irregularities, including secondary transits (indicative of an eclipsing binary) and shifts in the star's centroid coincident with observed transits \citep{batalha2010,tenenbaum2013}. These techniques have proven largely successful in screening out many false positives, and though it has been shown that of the remaining contaminated KOIs the great majority ($>90\%$) are not false positives, many transiting planets are still larger than interpreted \citep{morton2011,fressin2013,santerne2013,ciardi2015,desert2015}. Further validation then requires finding contaminating companions either indirectly, e.g. with transit photometry \citep{colon2012,colon2015} or directly, e.g. high angular resolution imaging \citep{morton2012}. The necessary sub-arcsecond resolution to find these contaminating companions can be achieved on the ground by several techniques, most notably lucky/speckle imaging \citep{horch2012,lillo-box2012,lillo-box2014} and adaptive optics \citep{adams2012, adams2013, dressing2014}.

With 6395 Kepler Objects of Interest (KOI) to vet \citep{coughlin2015}, we adopted a strategy to conduct a comprehensive survey with Palomar 1.5-m/Robo-AO \citep{baranec2013b,baranec2013a} and use Keck II/NIRC2 to follow up on secure and likely detections of companions. To date we have reported the optical detection of 53 contaminating companions within $2.5\arcsec$ in a sample of 715 KOIs \citep{law2014} and 426 companions within $\sim4\arcsec$ to 2598 KOIs \citep{baranec2016, ziegler2016}. The addition of near-infrared observations to the existing visible data permits us to perform characterization of the detected stars, estimate their photometric parallax and the likelihood of physical association between primary/companion pairs, and calculate reinterpreted sizes for individual planet candidates. In several cases we have also found additional unseen companions.

Section~\ref{kepobservations} of this paper describes the observations made and the image reduction process for Keck II/NIRC2 data. Section~\ref{kepanalysis} describes the derivation of photometric and stellar characteristics from the objects studied, including techniques used for fitting to stellar type and results thereof. Section~\ref{kepdiscussion} discusses the spectral fit results in context of the entire KOI catalog. The paper concludes in Section~\ref{kepconclusions} with an overview of our findings and an outline of future avenues of investigation.

\section{Observations and Data Reduction}\label{kepobservations}

   \begin{figure}
   \begin{center}
   \includegraphics[height = 6.0cm]{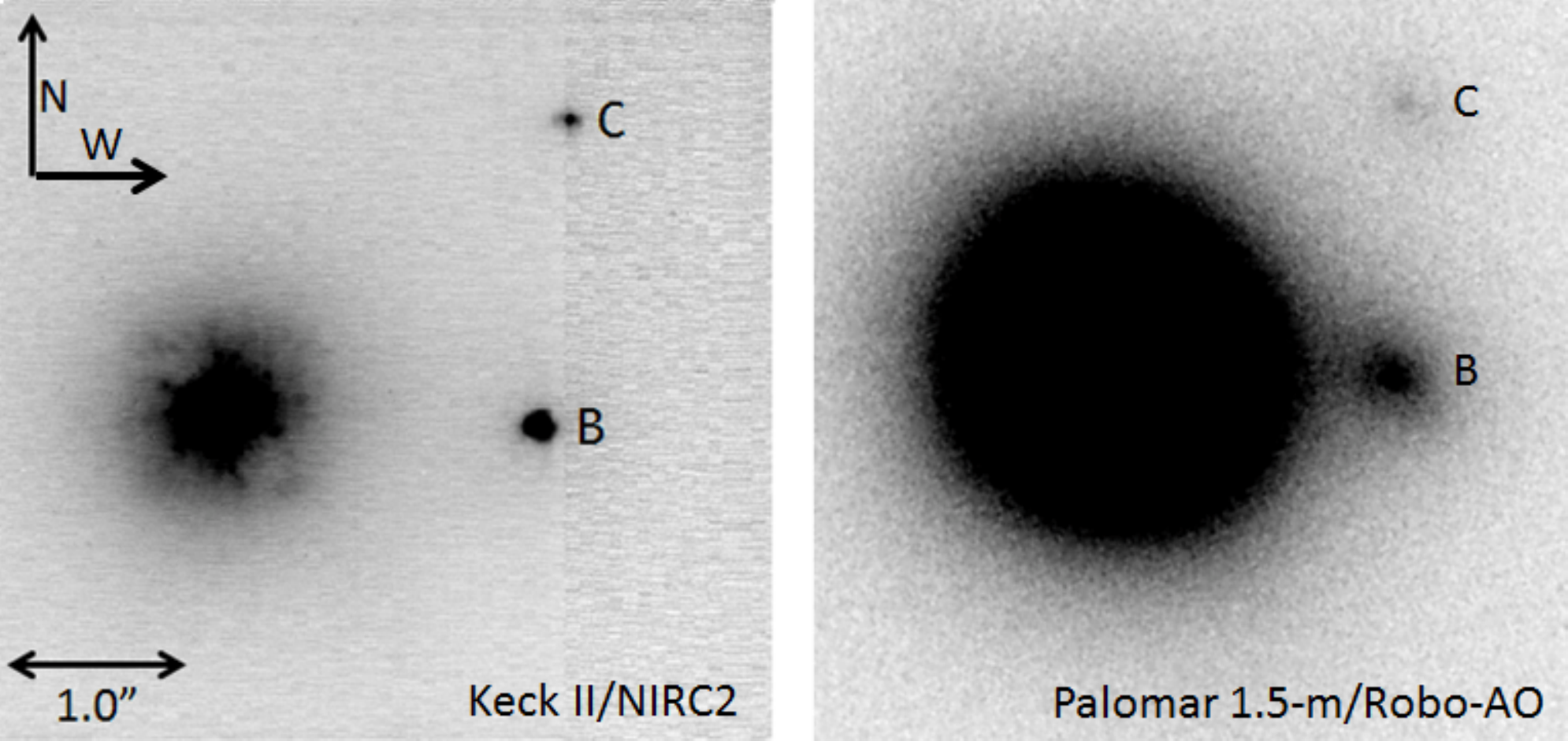}
   \end{center}
   \caption[KOI-268 with Contaminating Companion]{ \label{fig:example} 
Images of KOI-268 from both Keck II/NIRC2 (\textit{left}) and Palomar 1.5-m/Robo-AO, presented as an example. Visible in both images are companions B and C at separations from A of 1\farcs75 and 2\farcs53, respectively.}
   \end{figure} 

The initial observations identifying companion candidates are from multiple Robo-AO observing runs on the Palomar Observatory 1.5-m telescope, spanning July to September 2012, April to October 2013, June to September 2014, and June 2015. Observations were in either Sloan-\textit{i} or a long-pass 600nm (LP600) filter, the latter being similar to the \textit{Kepler}-bandpass when combined with the EMCCD's quantum efficiency curve for red/cool stars. Images were automatically reduced by the Robo-AO observing pipeline \citep{law2014}.

The near-infrared observations are from the NIRC2 instrument on the 10-m Keck II telescope, conducted 2013 June 24, August 24 and 25, 2014 August 17, 2015 July 25, and August 4 in the \textit{J}, \textit{H}, \textit{K}, and/or $K_p$ filters in the narrow mode of NIRC2 (9.952 mas pixel$^{-1}$; \citealt{Yelda2010}). For KOIs brighter than $m_V \sim 13$ we typically used the KOI as the guide star in natural guide star mode, and for fainter KOIs we used the laser guide star, with the KOI as the tip-tilt-focus guide star \citep{wizinowich2006, vandam2006}. An initial 30s exposure was taken for each target, and we waited for the low-bandwidth wavefront sensor to settle if the laser was used. The integration time and number of coadds per detector readout were adjusted to keep the peak of the stellar PSF counts less than 8,000 ADU per single integration (roughly half the dynamic range where sensitivity of the detector is linear), while maintaining a total exposure time of 30s. Dithered images were then acquired with the primary centered in the 3 lowest noise quadrants using the `bxy3 2.5' command, for a total exposure time of 90s. Occasional dither failures, particularly on 4 Aug 2015, resulted in exposures where the target is centered on the detector.

Images are first sky-subtracted and then flat-fielded. A pipeline developed for this investigation is then used to automatically pick out companion stars by spatially binning pixels and selecting the locations of the brightest bins as candidates. These candidates then have their radius measured in the eight cardinal and intermediate directions from their local centroid. This measurement steps in the given direction until it finds a value consistent with the measured background value within a specified confidence interval (initially $3\sigma$). If the median radius is both larger than a specified cutoff value (typically 3 pixels) and the standard deviation of the measured radii smaller than the same cutoff, the candidate is accepted as a star. The brightest star in the field is assumed to be the primary unless manually specified otherwise (the narrow $10\arcsec$ NIRC2 field makes this a rare occurrence). If a companion is not found, the background confidence interval and radius cutoff are adjusted to optimize for close ($<0\farcs5$) companions and the procedure repeated. If the procedure fails to find a companion, or if it finds multiple companions, a warning message notifies the operator to review the source image. 

For the majority of targets the pipeline correctly locates the primary and any present companions, but manual validation is necessary for many targets largely because the speckles in the point spread function (PSF) are mistaken for stars.  To avoid this misinterpretation the pipeline cross-references stars found in different filters for the same object, and discards any objects that do not appear in multiple filters. In some cases, images were only taken with one filter (typically $K_{p}$), and thus cross-referencing is not possible. These targets are then manually vetted and removed if visually confirmed to be associated with the primary PSF (i.e. coincident with rays projecting from the primary and presenting a PSF inconsistent with other imaged stars).

The separation and phase angle of each companion are measured from these individual reduced images, with uncertainties measured from the variability in measurements across all available images, and corrected for distortion using the most recent solution \citep{service2016}. Last, images are co-added into a single composite image for each target and filter for use in photometry.

\subsection{Aperture Photometry}
For the majority of our Keck data, the diffraction-limited resolution makes simple aperture photometry sufficient for measuring the contrast between the two stars. To account for the overlap of the stars' PSF envelopes, a matching aperture from the location opposite each star relative to its companion is used to estimate background subtraction, if available. In cases where the corresponding Robo-AO results were unavailable the method was also applied to those images, using the known position of the companion taken from the Keck analysis.

Systematic error from aperture size is our primary source of uncertainty, and is measured as the standard deviation of contrast across a range of aperture sizes from 1 to 3 FWHM in 0.5 FWHM increments. Injected companions are used to estimate further uncertainties typically yielding an error of $5\%$.

\section{Analysis}\label{kepanalysis}
\subsection{Photometric Classification of Stars}\label{fitting}
By combining our contrast measurements with extant \textit{JHK} photometry for the blended system (from the Exoplanet Archive\footnote{http://exoplanetarchive.ipac.caltech.edu Most \textit{JHK} magnitudes are from the 2MASS catalog \citep{liebert1995}.}) we derive the multi-color photometry for all components of each system. For blended magnitudes lacking a reported uncertainty (\textit{i},\textit{Kep}), one was estimated based on the measurement's reported source as recommended by the guide supplied by MAST \citep{mikulski}.\footnote{The documentation on Kepler magnitude sources at archive.stsci.edu/kepler/help/columns.html under heading `Kepmag Source' describes the respective uncertainties for the Kepler magnitude sources.}

Multi-color photometry allows characterization of the stars, necessary as the existing data on these objects is drawn from a blended target. Effective temperature is relatively strongly correlated with color-color photometry, but stellar radius (upon which our measurement of a transiting planet's size is dependent) exhibits a much weaker correlation for late-type stars. To demonstrate the systematic biases inherent in photometric type-fitting, we present fitting results from 2 different photometric datasets.

The first dataset is a set of metallicity- and age-agnostic stellar SEDs, originally assembled from a heterogenous set of models and data for an investigation of the Praesepe and Coma Berenices open clusters by \citet{kraus2007}, henceforth KH07, and which has previously been used for photometric fitting of exoplanet host stars (e.g. \citealt{bechter2014,wang2014,wang2015}). Photometric values for the \textit{Kepler}-band were computed by the method described in \citet{brown2011} using an arithmetic combination of \textit{gri} colors. The list of types and magnitudes from KH07 is expanded with missing types linearly interpolated from existing data, and an additional 9 intermediate types also interpolated between each two adjacent integer stellar types (e.g. type G2.5 is linearly halfway between G2 and G3). This makes a table of 521 entries from B8 to L0 to be fitted to as standards, and with matching absolute magnitudes and radii. The interpolated decimal types are not reported directly but those entries are used to refine radius/distance estimates. Radii for spectral types are drawn from \citet{habets1981}.

To fit types we use a Monte Carlo technique, generating a Gaussian distribution for each of the photometric combinations $J-K$, $H-K$, $i-K$, and $Kep-K$, if information in the respective filters is present. $K$ was chosen as the baseline as it produces the most precise contrast measurements and occupies the longest wavelength. Extinction is corrected for during this fitting process, relying on the canonical $A_V$ for each target in the Kepler catalog and adjusted for the various filters/bandpasses via the standard relations from \citet{cardelli1989}: $A_{Kep} = 0.896A_V$, $A_i = 0.321A_V$, $A_J = 0.158A_V$, $A_H = 0.100A_V$, $A_K = 0.060A_V$.

Each time beginning in the center of the list of standards, the Gaussian-generated photometry is compared to each canonical type's set of magnitudes to measure the error for all available data, as
\begin{equation}
R_{f1-f2} = (m_{f1,*} - m_{f2,*}) - (M_{f1,std} - M_{f2,std})
\end{equation}
where $m_{f\#,*}$ and $M_{f\#,typ}$ are the star's measured apparent magnitude and the standard's absolute magnitude in filters $f1$ and $f2$. The quality of the fit against the given standard is then
\begin{equation}
R^2 = \sum\limits_{J,H,i,Kep} \frac{R^2_{filter-K}}{w^2_{filter-K}}
\label{rsquared}
\end{equation}
where $w_{filter-K}$ is the weight of the respective filter combination as
\begin{equation}
w_{filter-K} = \sqrt{\sigma^2_{filter} + \sigma^2_K}
\end{equation}
Note that Equation~\ref{rsquared} does not require normalization as the number of filters used is consistent for a given star. $R^2$ is also measured for comparisons to standards in both directions (earlier- and later-type), and moved if a lower $R^2$ is found in either. The process repeats until it finds a minimum $R^2$. The type, radius, and absolute magnitudes are recorded and the next member of the Gaussian-generated list is fit to the standards in the same way. After fitting every entry on the list, the means of type, radius, and absolute magnitudes are taken as the fitted values, and the standard deviation of the latter two are their uncertainties.

The second set is the \textit{Kepler} Input Catalog's (KIC) primary standard stars as reported in \citet{brown2011}, henceforth B11. The advantage of this catalog is that these stars are in the \textit{Kepler} field and therefore reasonably representative of stars in our sample. The subset of standard stars with which each studied KOI component's photometry is consistent to $1\sigma$ was used to compute a mean and standard deviation for the star's stellar parameters. Subsets are typically $10\%$ of the full list of primary standard stars (or $\sim30$ stars) for each fitted object. Given the relatively poor correlation in the KIC standards of any of the measured stellar characteristics with NIR-only color comparisons, only $Kep - K$ and $i - K$ measurements were used. For stars without available $Kep$- or $i$-band measurements (with LP600 approximating $Kep$), a fit is not produced. If the photometry for a target fits two or fewer primary standard stars, its results are omitted. The effective temperature is then fit to a modeled stellar catalog to produce absolute magnitudes \citep{pecaut2013}, and compared to apparent magnitudes in turn to estimate distance. The KIC standard magnitudes were corrected for extinction/reddening as calculated from \citet{schlafly2011}.

The fitted values for both methods are displayed in Table~\ref{stellarparams}. The two methods are in broad agreement, although disparities are apparent for M-type companions in particular as the KIC primary standards contain few objects in that temperature range and exhibit a systemic overestimation of late-K and M-type radii that is not corrected for here \citep{muirhead2012}. While the range of potential fit types for KH07 covered the full main sequence from B8 to M9, almost all stars fit to late types F-M. 

Notably, two stars are presented as type B8, and five are too poorly restrained to produce KH07 fits. For the former, B8 is the end of the fitting range, and indicates they are too blue (adjusted for reddening) to fit to our list of main sequence stars. These stars are then very distant O/B types, and at 2 of 159 total stars analyzed by KH07 make up $\sim1\%$ of the sample. As we are unable to fit above B8 their properties are not well constrained, and we present radius and distance as lower limits.

KH07 fitting fails to fit five stars in the sample due to relatively poor photometric constraints in one or more observation bands.

\subsection{Uncertainties of Fitted Characteristics}
As described above, the photometric uncertainties arise largely from sampling the contrast for a range of photometric apertures and the inherent 5\% error measured by injection tests. For the stellar type fitting to KH07, the full Monte Carlo fit measures the uncertainty of derived characteristics. Gaussian distributions matching each apparent magnitude measurement and uncertainty thereof are grouped into sets, and each set has its type and other fitted characteristics calculated by the method described above. The measured uncertainty in the derived characteristics is then the standard deviation of the full set of measurements. Uncertainties from the B11 fits are simply the measured standard deviations of the respective parameters of $1\sigma$ consistent KIC standard stars. Uncertainties may be underestimated due to the granularity of the data being fit.

B11 also reports that use of photometric fitting on \textit{Kepler} primary standard stars results in uncertainties of approximately 200K for effective temperature and 0.2 dex for stellar radius without prior constraints on stellar age or metallicity. We take this to be generally applicable to all our photometric type-fitting, but it is not factored into the uncertainties reported in Table~\ref{stellarparams}.

\begin{longtable}{c c c c c c c c c}
\setlength{\extrarowheight}{3pt}\\
\caption{Fitted Stellar Parameters}\label{stellarparams}\\
\caption*{Stellar parameters as a result of two different fitting techniques. $\sigma_{unassoc}$ is the certainty (in standard deviations) that each companion is physically unassociated with its host due to their respective distances. The Kraus \& Hillenbrand fit yields a stellar type and corresponding radius \citep{kraus2007}. Values are interpolated between the table items in the source for improved precision. The fit to the KIC primary standards from Brown yields effective temperature and stellar radius for all stars with sufficient photometry, as produced by comparison to stars with similar color-color measurements among the 279 entries in the KIC Primary Standard catalog \citep{brown2011}. As noted, photometric type-fitting of \textit{Kepler} targets has been found to have a limiting accuracy of $\pm$200K and $\sim$0.2dex respectively, which is largely a function of age/composition and is not taken into account here. For each primary/companion pair a distance measurement was produced from the measured apparent and fitted absolute magnitudes, and used to generate a confidence of non-association between the two objects.}\\
& \multicolumn{4}{|c}{via \citet{kraus2007}} & \multicolumn{4}{|c}{via \citet{brown2011}}\\
\hline
object & $SpT$ & $R/R_{Sun}$ & $dist (pc)$ & $\sigma_{unassoc}$ & $T_{eff}$ & $R/R_{Sun}$ & $dist (pc)$ & $\sigma_{unassoc}$\\
\endfirsthead
 & \multicolumn{4}{|c}{via \citet{kraus2007}}  & \multicolumn{4}{|c}{via \citet{brown2011}}\\
\hline
object & $SpT$ & $R (R_{Sun})$ & $dist (pc)$ & $\sigma_{unassoc}$ & $T_{eff}$ & $R (R_{Sun})$ & $dist (pc)$ & $\sigma_{unassoc}$\\
\endhead
\endfoot

\endlastfoot
0190A & G0 & $1.18^{+0.04}_{-0.07}$ &  $921^{+127}_{-109}$ &      &  $6041^{+116}_{-126}$ & $1.04^{+0.03}_{-0.03}$  & $1019^{+52}_{-93}$ &     \\
0190B & K3 & $0.93^{+0.03}_{-0.02}$ &  $716^{+89}_{-65}$   & 1.46 &  $4904^{+345}_{-265}$ & $0.83^{+0.13}_{-0.07}$  & $779^{+122}_{-77}$ & 1.56\\
0191A & G0 & $1.07^{+0.05}_{-0.03}$ &  $934^{+82}_{-69}$   &      &  $5855^{+109}_{-184}$ & $1.00^{+0.02}_{-0.03}$  & $1115^{+93}_{-115}$ &     \\
0191B & F4 & $1.01^{+0.14}_{-0.07}$ & $2804^{+1007}_{-492}$& 3.75 &  $5110^{+378}_{-472}$ & $0.94^{+0.09}_{-0.12}$  & $2573^{+651}_{-386}$ & 3.67\\
0268A & F3 & $1.29^{+0.06}_{-0.06}$ &  $258^{+33}_{-34}$   &      &  $6136^{+143}_{-111}$ & $1.04^{+0.04}_{-0.03}$  & $230^{+18}_{-12}$ &     \\
0268B & K4 & $0.85^{+0.04}_{-0.06}$ &  $315^{+33}_{-33}$   & 1.22 &  $4807^{+423}_{-283}$ & $0.79^{+0.16}_{-0.06}$  & $392^{+81}_{-41}$ & 3.62\\
0268C & K3 & $0.95^{+0.31}_{-0.08}$ &  $904^{+3016}_{-290}$& 2.21 &  $4889^{+660}_{-352}$ & $0.82^{+0.28}_{-0.08}$  & $829^{+354}_{-133}$ & 4.46\\
0401A & G4 & $1.05^{+0.02}_{-0.02}$ &  $589^{+31}_{-44}$   &      &  $5890^{+152}_{-206}$ & $1.01^{+0.02}_{-0.03}$  & $744^{+70}_{-96}$ &     \\
0401B & K7 & $0.79^{+0.05}_{-0.08}$ &  $690^{+74}_{-78}$   & 1.20 &  $5000^{+970}_{-554}$ & $0.88^{+0.33}_{-0.17}$  & $1185^{+716}_{-315}$ & 1.37\\
0425A & F4 & $1.22^{+0.05}_{-0.04}$ & $1374^{+212}_{-138}$ &      &  $6088^{+113}_{-165}$ & $1.02^{+0.04}_{-0.03}$  & $1425^{+95}_{-115}$ &     \\
0425B & F2 & $1.20^{+0.06}_{-0.07}$ & $1947^{+343}_{-253}$ & 1.74 &  $6013^{+116}_{-137}$ & $1.03^{+0.03}_{-0.03}$  & $2024^{+122}_{-177}$ & 2.98\\
0511A & F3 & $1.25^{+0.02}_{-0.04}$ & $1067^{+77}_{-118}$  &      &  $6128^{+73}_{-112}$  & $1.04^{+0.03}_{-0.03}$  & $1061^{+38}_{-59}$ &     \\
0511B & K4 & $0.84^{+0.04}_{-0.07}$ &  $995^{+109}_{-115}$ & 0.45 &  $4796^{+456}_{-303}$ & $0.79^{+0.18}_{-0.06}$  & $1280^{+290}_{-149}$ & 1.42\\
0511C & K6 & $0.62^{+0.13}_{-0.16}$ & $2345^{+491}_{-609}$ & 2.08 &                       &                         &        \\
0628A & F5 & $1.24^{+0.02}_{-0.02}$ &  $876^{+40}_{-46}$   &      &  $6085^{+113}_{-192}$ & $1.04^{+0.04}_{-0.03}$  & $856^{+51}_{-104}$ &     \\
0628B & K9 & $0.64^{+0.08}_{-0.11}$ & $1070^{+140}_{-180}$ & 1.05 &                       &                         &        \\
0628C & K1 & $0.85^{+0.05}_{-0.07}$ & $2247^{+254}_{-286}$ & 4.75 &  $4744^{+533}_{-329}$ & $0.74^{+0.21}_{-0.07}$  & $2809^{+820}_{-385}$ & 5.03\\
0687A & F6 & $1.24^{+0.02}_{-0.04}$ &  $837^{+56}_{-72}$   &      &  $6012^{+119}_{-158}$ & $1.02^{+0.04}_{-0.03}$  & $806^{+55}_{-76}$ &     \\
0687B & K4 & $0.89^{+0.05}_{-0.07}$ &  $699^{+116}_{-83}$  & 1.01 &  $5486^{+417}_{-300}$ & $1.02^{+0.15}_{-0.07}$  & $1114^{+311}_{-176}$ & 1.67\\
0688A & F4 & $1.30^{+0.01}_{-0.01}$ & $1204^{+49}_{-54}$   &      &  $6158^{+86}_{-158}$  & $1.03^{+0.02}_{-0.02}$  & $1057^{+50}_{-87}$ &     \\
0688B & K0 & $0.99^{+0.07}_{-0.04}$ & $1166^{+239}_{-125}$ & 0.16 &  $5122^{+342}_{-342}$ & $0.87^{+0.09}_{-0.09}$  & $1174^{+249}_{-142}$ & 0.78\\
0712A & F1 & $1.26^{+0.04}_{-0.17}$ & $1100^{+124}_{-221}$ &      &  $6228^{+78}_{-169}$  & $1.07^{+0.03}_{-0.03}$  & $1045^{+52}_{-76}$ &     \\
0712B & K1 & $0.95^{+0.05}_{-0.03}$ &  $637^{+82}_{-54}$   & 1.96 &  $5204^{+472}_{-318}$ & $0.89^{+0.16}_{-0.09}$  & $762^{+220}_{-118}$ & 1.22\\
0931A & F3 & $1.27^{+0.04}_{-0.04}$ & $1720^{+195}_{-186}$ &      &  $6140^{+74}_{-106}$  & $1.05^{+0.03}_{-0.04}$  & $1654^{+57}_{-94}$ &     \\
0931B &    &                        &                      &      &  $5943^{+139}_{-225}$ & $1.01^{+0.04}_{-0.03}$  & $6491^{+602}_{-817}$ & 5.91\\
0984A & G5 & $1.03^{+0.03}_{-0.02}$ &  $267^{+21}_{-19}$   &      &  $5806^{+191}_{-304}$ & $1.00^{+0.05}_{-0.06}$  & $326^{+48}_{-44}$ &     \\
0984B & G5 & $1.03^{+0.03}_{-0.02}$ &  $273^{+22}_{-19}$   & 0.21 &  $5717^{+185}_{-302}$ & $0.99^{+0.04}_{-0.07}$  & $317^{+44}_{-46}$ & 0.14\\
0987A & G7 & $1.01^{+0.01}_{-0.01}$ &  $270^{+13}_{-6}$    &      &  $5656^{+241}_{-333}$ & $0.99^{+0.08}_{-0.08}$  & $325^{+55}_{-49}$ &     \\
0987B & M0 & $0.74^{+0.05}_{-0.05}$ &  $458^{+47}_{-37}$   & 4.79 &                       &                         &        \\
1066A & F6 & $1.13^{+0.05}_{-0.05}$ & $1342^{+152}_{-108}$ &      &  $5932^{+102}_{-157}$ & $1.02^{+0.02}_{-0.03}$  & $1535^{+110}_{-145}$ &     \\
1066B & G9 & $0.85^{+0.03}_{-0.04}$ & $2894^{+245}_{-259}$ & 5.17 &  $4608^{+256}_{-285}$ & $0.74^{+0.06}_{-0.05}$  & $3333^{+289}_{-419}$ & 4.15\\
1067A & F3 & $1.29^{+0.01}_{-0.01}$ & $1431^{+60}_{-65}$   &      &  $6146^{+80}_{-205}$  & $1.04^{+0.03}_{-0.03}$  & $1280^{+72}_{-125}$ &     \\
1067B & K2 & $0.89^{+0.03}_{-0.02}$ & $2103^{+214}_{-175}$ & 3.63 &  $4833^{+409}_{-282}$ & $0.80^{+0.15}_{-0.06}$  & $2520^{+493}_{-269}$ & 4.45\\
1112A & F5 & $1.22^{+0.04}_{-0.06}$ & $1140^{+141}_{-160}$ &      &  $6160^{+88}_{-87}$   & $1.07^{+0.04}_{-0.03}$  & $1232^{+63}_{-43}$ &     \\
1112B & K8 & $0.73^{+0.04}_{-0.04}$ & $1528^{+120}_{-104}$ & 2.21 &                       &                         &        \\
1151A & F9 & $1.10^{+0.08}_{-0.05}$ &  $501^{+76}_{-44}$   &      &  $5776^{+148}_{-270}$ & $0.99^{+0.02}_{-0.06}$  & $534^{+61}_{-71}$ &     \\
1151B & K8 & $0.71^{+0.19}_{-0.30}$ &  $738^{+373}_{-289}$ & 0.79 &  $4620^{+488}_{-418}$ & $0.72^{+0.20}_{-0.07}$  & $991^{+240}_{-177}$ & 2.44\\
1214A & G8 & $1.00^{+0.02}_{-0.02}$ &  $656^{+47}_{-26}$   &      &  $5600^{+256}_{-230}$ & $0.99^{+0.08}_{-0.07}$  & $836^{+136}_{-102}$ &     \\
1214B & B8 & $\geq1.5$              & $\geq5591$           & 7.29 &                       &                         &        \\
1274A & G7 & $1.00^{+0.01}_{-0.01}$ &  $363^{+14}_{-9}$    &      &  $5602^{+250}_{-319}$ & $0.97^{+0.09}_{-0.08}$  & $443^{+76}_{-65}$ &     \\
1274B & K9 & $0.63^{+0.17}_{-0.26}$ &  $585^{+191}_{-254}$ & 0.87 &                       &                         &        \\
1375A & F5 & $1.25^{+0.01}_{-0.02}$ &  $809^{+28}_{-41}$   &      &  $6136^{+89}_{-103}$  & $1.05^{+0.04}_{-0.03}$  & $811^{+31}_{-47}$ &     \\
1375B & K3 & $0.88^{+0.21}_{-0.19}$ & $1862^{+1419}_{-480}$& 2.19 &  $5097^{+433}_{-375}$ & $0.86^{+0.16}_{-0.09}$  & $2305^{+609}_{-325}$ & 4.58\\
1442A & G4 & $1.08^{+0.05}_{-0.03}$ &  $322^{+26}_{-14}$   &      &  $5791^{+137}_{-294}$ & $0.99^{+0.02}_{-0.06}$  & $352^{+40}_{-50}$ &     \\
1442B & M7 & $0.30^{+0.16}_{-0.01}$ &  $350^{+65}_{-29}$   & 0.72 &                       &                         &        \\
1447A & A6 & $1.45^{+0.04}_{-0.03}$ & $1206^{+62}_{-54}$   &      &                       &                         &     \\
1447B & K3 & $0.91^{+0.03}_{-0.02}$ &  $517^{+43}_{-34}$   & 9.98 &  $4852^{+430}_{-216}$ & $0.77^{+0.17}_{-0.06}$  & $595^{+122}_{-59}$ &        \\
1536A & F5 & $1.27^{+0.01}_{-0.01}$ &  $550^{+20}_{-18}$   &      &  $6229^{+71}_{-89}$   & $1.08^{+0.03}_{-0.03}$  & $551^{+19}_{-25}$ &     \\
1536B & K3 & $0.80^{+0.24}_{-0.34}$ & $1614^{+1388}_{-600}$& 1.77 &  $5117^{+449}_{-439}$ & $0.87^{+0.15}_{-0.10}$  & $2169^{+607}_{-349}$ & 4.63\\
1546A & F2 & $1.22^{+0.05}_{-0.09}$ & $1259^{+177}_{-256}$ &      &  $6116^{+75}_{-124}$  & $1.05^{+0.03}_{-0.03}$  & $1319^{+55}_{-80}$ &     \\
1546B & K0 & $0.93^{+0.06}_{-0.04}$ &  $939^{+155}_{-98}$  & 1.07 &  $5111^{+384}_{-377}$ & $0.87^{+0.12}_{-0.10}$  & $1108^{+264}_{-148}$ & 0.76\\
1546C &    &                        &                      &      &  $5873^{+300}_{-429}$ & $1.00^{+0.06}_{-0.10}$  & $5685^{+1077}_{-1181}$ & 3.69\\
1546D & F7 & $0.91^{+0.10}_{-0.08}$ & $2527^{+798}_{-415}$ & 2.81 &  $5208^{+386}_{-492}$ & $0.92^{+0.09}_{-0.12}$  & $3175^{+834}_{-522}$ & 3.54\\
1613A & F5 & $1.27^{+0.05}_{-0.05}$ &  $404^{+60}_{-46}$   &      &                       &                         &     \\
1613B & G4 & $1.07^{+0.11}_{-0.09}$ &  $419^{+101}_{-89}$  & 0.14 &                       &                         &        \\
1700A & G8 & $0.97^{+0.02}_{-0.02}$ &  $602^{+36}_{-33}$   &      &  $5228^{+216}_{-261}$ & $0.89^{+0.06}_{-0.07}$  & $648^{+80}_{-68}$ &     \\
1700B & K3 & $0.89^{+0.03}_{-0.03}$ &  $632^{+52}_{-40}$   & 0.56 &  $4614^{+409}_{-272}$ & $0.74^{+0.15}_{-0.06}$  & $685^{+119}_{-75}$ & 0.34\\
1784A & F7 & $1.16^{+0.05}_{-0.06}$ &  $751^{+76}_{-73}$   &      &  $5939^{+100}_{-135}$ & $1.01^{+0.03}_{-0.03}$  & $817^{+54}_{-74}$ &     \\
1784B & F2 & $1.26^{+0.03}_{-0.06}$ & $1365^{+146}_{-176}$ & 3.20 &  $6113^{+85}_{-105}$  & $1.04^{+0.03}_{-0.03}$  & $1292^{+51}_{-68}$ & 5.47\\
1845A & K2 & $0.94^{+0.02}_{-0.01}$ &  $411^{+24}_{-18}$   &      &                       &                         &     \\
1845B & M4 & $0.46^{+0.29}_{-0.19}$ &  $651^{+445}_{-285}$ & 0.84 &                       &                         &        \\
1845C & A7 & $1.07^{+0.17}_{-0.10}$ &$4407^{+2110}_{-1089}$& 3.67 &                       &                         &        \\
1880A & K8 & $0.78^{+0.01}_{-0.06}$ &  $206^{+3}_{-17}$    &      &                       &                         &     \\
1880B & G7 & $0.93^{+0.07}_{-0.06}$ & $2094^{+492}_{-290}$ & 6.51 &  $4717^{+418}_{-334}$ & $0.76^{+0.16}_{-0.08}$  & $2046^{+365}_{-279}$ &        \\
1884A & F9 & $1.06^{+0.07}_{-0.04}$ & $1094^{+147}_{-123}$ &      &  $5890^{+159}_{-252}$ & $1.01^{+0.04}_{-0.05}$  & $1366^{+143}_{-193}$ &     \\
1884B & K8 & $0.63^{+0.17}_{-0.22}$ & $1381^{+449}_{-514}$ & 0.54 &                       &                         &        \\
1884C & M1 & $0.51^{+0.21}_{-0.15}$ & $1330^{+587}_{-424}$ & 0.53 &                       &                         &        \\
1884D & M2 & $0.41^{+0.17}_{-0.10}$ & $1335^{+642}_{-392}$ & 0.58 &                       &                         &        \\
1891A & K0 & $0.95^{+0.01}_{-0.02}$ &  $687^{+27}_{-46}$   &      &  $5417^{+248}_{-158}$ & $0.97^{+0.08}_{-0.06}$  & $872^{+118}_{-93}$ &     \\
1891B &    &                        &                      &      &  $5802^{+284}_{-402}$ & $1.00^{+0.06}_{-0.09}$  & $8742^{+1733}_{-1568}$ & 5.00\\
1916A & F2 & $1.31^{+0.01}_{-0.01}$ & $1046^{+43}_{-36}$   &      &                       &                         &     \\
1916B & K5 & $0.89^{+0.03}_{-0.04}$ &  $672^{+85}_{-59}$   & 4.05 &  $4556^{+266}_{-276}$ & $0.73^{+0.08}_{-0.03}$  & $685^{+65}_{-84}$ &        \\
1979A & F5 & $1.23^{+0.02}_{-0.07}$ &  $568^{+27}_{-59}$   &      &  $6012^{+110}_{-199}$ & $1.02^{+0.04}_{-0.03}$  & $547^{+41}_{-59}$ &     \\
1979B & K9 & $0.69^{+0.13}_{-0.23}$ &  $454^{+98}_{-154}$  & 1.00 &  $4398^{+340}_{-569}$ & $0.70^{+0.04}_{-0.06}$  & $565^{+119}_{-182}$ & 0.10\\
1989A & F9 & $1.14^{+0.04}_{-0.03}$ &  $525^{+49}_{-23}$   &      &  $5948^{+99}_{-124}$  & $1.02^{+0.03}_{-0.03}$  & $600^{+48}_{-40}$ &     \\
1989B & K2 & $0.90^{+0.03}_{-0.04}$ & $1299^{+195}_{-156}$ & 4.73 &  $5179^{+376}_{-447}$ & $0.86^{+0.14}_{-0.10}$  & $1536^{+353}_{-243}$ & 3.78\\
2001A & K1 & $0.97^{+0.01}_{-0.01}$ &  $266^{+8}_{-6}$     &      &  $5386^{+264}_{-271}$ & $0.93^{+0.09}_{-0.07}$  & $318^{+52}_{-40}$ &     \\
2001B & G5 & $0.93^{+0.05}_{-0.03}$ & $1761^{+278}_{-200}$ & 7.47 &  $5123^{+396}_{-351}$ & $0.90^{+0.13}_{-0.09}$  & $2092^{+498}_{-270}$ & 6.45\\
2009A & F7 & $1.18^{+0.03}_{-0.03}$ &  $696^{+59}_{-47}$   &      &  $6066^{+101}_{-155}$ & $1.05^{+0.03}_{-0.03}$  & $797^{+41}_{-82}$ &     \\
2009B & K4 & $0.82^{+0.07}_{-0.17}$ & $1204^{+196}_{-264}$ & 1.88 &  $4878^{+692}_{-381}$ & $0.82^{+0.33}_{-0.09}$  & $1715^{+767}_{-298}$ & 3.05\\
2059A & K2 & $0.93^{+0.02}_{-0.01}$ &  $246^{+18}_{-14}$   &      &  $5104^{+315}_{-202}$ & $0.91^{+0.11}_{-0.07}$  & $288^{+54}_{-25}$ &     \\
2059B & K5 & $0.85^{+0.02}_{-0.03}$ &  $268^{+22}_{-23}$   & 0.75 &  $4568^{+233}_{-288}$ & $0.73^{+0.06}_{-0.04}$  & $295^{+26}_{-37}$ & 0.11\\
2069A & F7 & $1.21^{+0.02}_{-0.03}$ &  $723^{+52}_{-54}$   &      &  $5947^{+103}_{-147}$ & $1.02^{+0.03}_{-0.03}$  & $713^{+62}_{-53}$ &     \\
2069B & K8 & $0.71^{+0.13}_{-0.26}$ & $1266^{+407}_{-397}$ & 1.36 &  $4912^{+404}_{-419}$ & $0.80^{+0.15}_{-0.10}$  & $1874^{+383}_{-275}$ & 4.12\\
2083A & G1 & $1.16^{+0.10}_{-0.04}$ &  $658^{+199}_{-57}$  &      &  $5889^{+152}_{-211}$ & $1.01^{+0.03}_{-0.03}$  & $685^{+66}_{-88}$ &     \\
2083B & F7 & $1.09^{+0.18}_{-0.38}$ & $1467^{+601}_{-923}$ & 0.86 &  $5139^{+1022}_{-481}$& $0.85^{+0.26}_{-0.10}$  & $1178^{+697}_{-319}$ & 1.51\\
2117A & K5 & $0.86^{+0.01}_{-0.01}$ &  $609^{+21}_{-26}$   &      &                       &                         &     \\
2117B & K3 & $0.88^{+0.01}_{-0.01}$ &  $818^{+31}_{-33}$   & 5.34 &  $4640^{+768}_{-333}$ & $0.75^{+0.48}_{-0.08}$  & $988^{+436}_{-178}$ &        \\
2143A & G2 & $1.07^{+0.03}_{-0.03}$ &  $643^{+39}_{-36}$   &      &  $5757^{+133}_{-206}$ & $1.00^{+0.02}_{-0.03}$  & $717^{+72}_{-74}$ &     \\
2143B & A0 & $1.19^{+0.21}_{-0.14}$ &$4408^{+4189}_{-1290}$& 2.92 &  $5778^{+257}_{-364}$ & $0.99^{+0.06}_{-0.08}$  & $3590^{+667}_{-559}$ & 5.10\\
2159A & F5 & $1.26^{+0.01}_{-0.02}$ &  $745^{+30}_{-45}$   &      &  $6130^{+87}_{-112}$  & $1.06^{+0.04}_{-0.03}$  & $739^{+30}_{-44}$ &     \\
2159B & M0 & $0.65^{+0.24}_{-0.27}$ &  $742^{+373}_{-326}$ & 0.01 &  $4398^{+340}_{-569}$ & $0.70^{+0.04}_{-0.06}$  & $961^{+193}_{-306}$ & 0.72\\
2247A & K3 & $0.90^{+0.01}_{-0.01}$ &  $337^{+11}_{-20}$   &      &  $4954^{+1280}_{-222}$& $0.79^{+0.48}_{-0.07}$  & $528^{+383}_{-143}$ &     \\
2247B & M0 & $0.53^{+0.11}_{-0.10}$ & $1116^{+234}_{-234}$ & 3.33 &                       &                         &        \\
2289A & F3 & $1.30^{+0.01}_{-0.01}$ &  $842^{+23}_{-27}$   &      &  $6201^{+64}_{-137}$  & $1.06^{+0.02}_{-0.02}$  & $753^{+32}_{-42}$ &     \\
2289B & K6 & $0.80^{+0.05}_{-0.08}$ & $1094^{+130}_{-140}$ & 1.78 &  $4620^{+807}_{-349}$ & $0.72^{+0.35}_{-0.08}$  & $1540^{+715}_{-296}$ & 2.64\\
2317A & F7 & $1.17^{+0.03}_{-0.05}$ &  $817^{+58}_{-64}$   &      &  $5969^{+106}_{-128}$ & $1.01^{+0.03}_{-0.03}$  & $904^{+64}_{-67}$ &     \\
2317B & G7 & $0.88^{+0.02}_{-0.03}$ & $2661^{+219}_{-195}$ & 9.06 &  $4650^{+459}_{-296}$ & $0.75^{+0.19}_{-0.06}$  & $3029^{+603}_{-392}$ & 5.35\\
2363A & K0 & $0.95^{+0.01}_{-0.01}$ &  $448^{+17}_{-19}$   &      &  $5386^{+360}_{-179}$ & $0.95^{+0.14}_{-0.07}$  & $578^{+129}_{-69}$ &     \\
2363B & K3 & $0.65^{+0.40}_{-0.30}$ &$3298^{+5208}_{-1596}$& 1.79 &  $4923^{+515}_{-384}$ & $0.83^{+0.21}_{-0.08}$  & $4826^{+1448}_{-708}$ & 5.90\\
2377A & F9 & $1.08^{+0.10}_{-0.09}$ &  $890^{+194}_{-181}$ &      &                       &                         &     \\
2377B & K1 & $0.92^{+0.05}_{-0.06}$ &  $831^{+141}_{-104}$ & 0.26 &                       &                         &        \\
2377C & K2 & $0.64^{+0.38}_{-0.34}$ &$2449^{+3109}_{-1566}$& 0.99 &                       &                         &        \\
2377D & K7 & $0.50^{+0.34}_{-0.22}$ & $1767^{+1465}_{-883}$& 0.97 &                       &                         &        \\
2413A & G7 & $0.98^{+0.04}_{-0.03}$ &  $756^{+78}_{-53}$   &      &  $5500^{+269}_{-258}$ & $0.99^{+0.08}_{-0.07}$  & $935^{+155}_{-127}$ &     \\
2413B & M2 & $0.48^{+0.21}_{-0.14}$ &  $324^{+136}_{-113}$ & 2.96 &                       &                         &        \\
2443A & F5 & $1.21^{+0.03}_{-0.04}$ &  $838^{+77}_{-101}$  &      &  $6141^{+87}_{-88}$   & $1.05^{+0.04}_{-0.03}$  & $934^{+33}_{-50}$ &     \\
2443B & K6 & $0.70^{+0.12}_{-0.22}$ & $1659^{+361}_{-504}$ & 1.61 &  $4742^{+878}_{-576}$ & $0.81^{+0.33}_{-0.12}$  & $2790^{+1614}_{-744}$ & 2.49\\
2542A & M0 & $0.65^{+0.04}_{-0.08}$ &  $278^{+18}_{-28}$   &      &                       &                         &     \\
2542B & M3 & $0.38^{+0.07}_{-0.06}$ &  $218^{+49}_{-45}$   & 1.06 &                       &                         &        \\
2601A & F3 & $1.28^{+0.03}_{-0.05}$ & $1133^{+103}_{-141}$ &      &  $6214^{+75}_{-124}$  & $1.06^{+0.03}_{-0.03}$  & $1086^{+46}_{-58}$ &     \\
2601B & G2 & $1.07^{+0.19}_{-0.11}$ & $1193^{+560}_{-299}$ & 0.19 &  $5708^{+286}_{-393}$ & $0.98^{+0.06}_{-0.09}$  & $1283^{+249}_{-234}$ & 0.83\\
2601C & G7 & $0.93^{+0.07}_{-0.04}$ & $2076^{+411}_{-212}$ & 4.00 &  $4846^{+406}_{-299}$ & $0.78^{+0.14}_{-0.09}$  & $2236^{+443}_{-250}$ & 4.52\\
2601D &    &                        &                      &      &  $5214^{+374}_{-568}$ & $0.90^{+0.08}_{-0.13}$  & $6068^{+1393}_{-1159}$ & 4.30\\
2657A & G0 & $1.09^{+0.07}_{-0.05}$ &  $504^{+61}_{-48}$   &      &  $5859^{+127}_{-213}$ & $1.00^{+0.02}_{-0.03}$  & $585^{+56}_{-67}$ &     \\
2657B & G6 & $1.01^{+0.05}_{-0.02}$ &  $454^{+54}_{-32}$   & 0.69 &  $5688^{+249}_{-269}$ & $1.00^{+0.06}_{-0.07}$  & $566^{+92}_{-78}$ & 0.17\\
2664A & K0 & $0.94^{+0.01}_{-0.01}$ &  $868^{+50}_{-43}$   &      &  $5134^{+359}_{-167}$ & $0.87^{+0.15}_{-0.06}$  & $1025^{+208}_{-101}$ &     \\
2664B & F6 & $1.01^{+0.03}_{-0.03}$ & $1708^{+198}_{-136}$ & 5.80 &  $5620^{+252}_{-255}$ & $0.98^{+0.06}_{-0.07}$  & $2112^{+340}_{-286}$ & 3.07\\
2681A & F7 & $1.04^{+0.04}_{-0.03}$ & $1469^{+160}_{-133}$ &      &  $5763^{+202}_{-287}$ & $0.98^{+0.06}_{-0.06}$  & $1774^{+251}_{-254}$ &     \\
2681B & K3 & $0.88^{+0.01}_{-0.02}$ & $1191^{+68}_{-62}$   & 1.86 &  $4650^{+523}_{-340}$ & $0.75^{+0.26}_{-0.06}$  & $1366^{+332}_{-210}$ & 0.98\\
2705A & M3 & $0.39^{+0.13}_{-0.07}$ &   $77^{+28}_{-16}$   &      &                       &                         &     \\
2705B & M5 & $0.30^{+0.08}_{-0.01}$ &  $186^{+36}_{-11}$   & 3.62 &                       &                         &        \\
2711A & F2 & $1.28^{+0.02}_{-0.03}$ & $1193^{+77}_{-93}$   &      &  $6059^{+61}_{-129}$  & $1.03^{+0.03}_{-0.03}$  & $1072^{+40}_{-94}$ &     \\
2711B & F2 & $1.27^{+0.02}_{-0.03}$ & $1241^{+94}_{-115}$  & 0.35 &  $6059^{+57}_{-132}$  & $1.03^{+0.03}_{-0.04}$  & $1131^{+42}_{-99}$ & 0.55\\
2722A & F3 & $1.29^{+0.01}_{-0.01}$ &  $808^{+21}_{-20}$   &      &  $6188^{+86}_{-158}$  & $1.05^{+0.02}_{-0.02}$  & $732^{+41}_{-47}$ &     \\
2722B & K8 & $0.69^{+0.06}_{-0.09}$ & $1341^{+132}_{-167}$ & 3.17 &                       &                         &        \\
2779A & F3 & $1.27^{+0.03}_{-0.04}$ & $1573^{+163}_{-161}$ &      &  $6140^{+74}_{-106}$  & $1.05^{+0.03}_{-0.04}$  & $1502^{+52}_{-86}$ &     \\
2779B & G8 & $0.95^{+0.12}_{-0.06}$ & $1793^{+608}_{-260}$ & 0.72 &  $5111^{+384}_{-373}$ & $0.87^{+0.12}_{-0.09}$  & $1991^{+472}_{-263}$ & 1.82\\
2813A & K3 & $0.93^{+0.01}_{-0.01}$ &  $310^{+14}_{-12}$   &      &  $5128^{+443}_{-224}$ & $0.88^{+0.20}_{-0.07}$  & $383^{+106}_{-47}$ &     \\
2813B & F8 & $1.25^{+0.05}_{-0.09}$ & $1466^{+215}_{-265}$ & 4.36 &  $6049^{+122}_{-159}$ & $1.03^{+0.04}_{-0.03}$  & $1363^{+81}_{-140}$ & 5.58\\
2813C & M1 & $0.57^{+0.23}_{-0.21}$ &$3541^{+1995}_{-1484}$& 2.18 &  $4887^{+1083}_{-835}$& $0.89^{+0.32}_{-0.22}$  & $7753^{+5561}_{-2760}$ & 2.67\\
2837A & F0 & $1.40^{+0.05}_{-0.04}$ & $1430^{+125}_{-105}$ &      &                       &                         &     \\
2837B & F0 & $1.38^{+0.07}_{-0.04}$ & $1527^{+192}_{-122}$ & 0.56 &                       &                         &        \\
2859A & G8 & $0.98^{+0.02}_{-0.01}$ &  $432^{+18}_{-9}$    &      &  $5582^{+196}_{-256}$ & $0.95^{+0.06}_{-0.07}$  & $546^{+81}_{-63}$ &     \\
2859B & A7 & $1.05^{+0.16}_{-0.22}$ & $2142^{+822}_{-896}$ & 1.91 &  $6169^{+140}_{-137}$ & $1.05^{+0.05}_{-0.03}$  & $2914^{+224}_{-180}$ & 12.00\\
2869A & F4 & $1.28^{+0.01}_{-0.01}$ &  $919^{+29}_{-36}$   &      &  $6154^{+58}_{-145}$  & $1.05^{+0.03}_{-0.03}$  & $840^{+31}_{-60}$ &     \\
2869B & K4 & $0.63^{+0.09}_{-0.10}$ & $3401^{+481}_{-559}$ & 4.43 &                       &                         &        \\
2904A & F5 & $1.27^{+0.01}_{-0.02}$ &  $587^{+27}_{-28}$   &      &  $6116^{+63}_{-212}$  & $1.05^{+0.03}_{-0.04}$  & $537^{+26}_{-68}$ &     \\
2904B & A1 & $1.28^{+0.11}_{-0.26}$ & $1976^{+470}_{-552}$ & 2.51 &                       &                         &        \\
2971A & F4 & $1.29^{+0.01}_{-0.01}$ &  $622^{+18}_{-22}$   &      &  $6188^{+86}_{-158}$  & $1.05^{+0.02}_{-0.02}$  & $569^{+32}_{-37}$ &     \\
2971B & G2 & $0.96^{+0.04}_{-0.06}$ & $1712^{+236}_{-331}$ & 3.29 &  $5720^{+216}_{-289}$ & $0.99^{+0.05}_{-0.07}$  & $2281^{+363}_{-310}$ & 5.49\\
2971C & K7 & $0.47^{+0.33}_{-0.34}$ &$2446^{+2884}_{-1060}$& 1.72 &  $5605^{+373}_{-590}$ & $0.96^{+0.08}_{-0.14}$  & $6324^{+1639}_{-1411}$ & 4.08\\
3020A & F3 & $1.31^{+0.01}_{-0.01}$ &  $956^{+31}_{-28}$   &      &                       &                         &     \\
3020B & K9 & $0.77^{+0.03}_{-0.05}$ &  $526^{+34}_{-37}$   & 9.76 &                       &                         &        \\
3020C & K3 & $0.77^{+0.07}_{-0.14}$ & $3032^{+479}_{-508}$ & 4.08 &                       &                         &        \\
3069A & F4 & $1.20^{+0.04}_{-0.05}$ & $1227^{+162}_{-188}$ &      &  $6110^{+89}_{-83}$   & $1.04^{+0.04}_{-0.03}$  & $1402^{+51}_{-62}$ &     \\
3069B & K2 & $0.90^{+0.01}_{-0.02}$ & $1164^{+58}_{-60}$   & 0.32 &  $4876^{+173}_{-134}$ & $0.81^{+0.04}_{-0.04}$  & $1332^{+87}_{-55}$ & 0.66\\
3106A & G5 & $1.04^{+0.05}_{-0.03}$ & $1189^{+142}_{-110}$ &      &  $5855^{+202}_{-287}$ & $1.00^{+0.05}_{-0.06}$  & $1526^{+207}_{-228}$ &     \\
3106B & A3 & $1.29^{+0.05}_{-0.06}$ & $3533^{+582}_{-576}$ & 3.95 &  $6188^{+88}_{-109}$  & $1.05^{+0.04}_{-0.03}$  & $3253^{+155}_{-151}$ & 6.74\\
3377A & G9 & $0.94^{+0.01}_{-0.01}$ &  $675^{+24}_{-38}$   &      &  $5110^{+350}_{-160}$ & $0.94^{+0.16}_{-0.05}$  & $784^{+158}_{-73}$ &     \\
3377B & M7 & $0.29^{+0.01}_{-0.01}$ &  $271^{+15}_{-14}$   & 9.89 &                       &                         &        \\
3377C & M2 & $0.36^{+0.09}_{-0.07}$ & $1215^{+405}_{-296}$ & 1.82 &                       &                         &        \\
3401A & G8 & $1.01^{+0.03}_{-0.02}$ &  $665^{+60}_{-41}$   &      &  $5526^{+252}_{-277}$ & $0.93^{+0.07}_{-0.07}$  & $769^{+122}_{-110}$ &     \\
3401B & B8 & $\geq1.40            $ & $\geq3510$           & 6.69 &                       &                         &        \\
4004A & F8 & $1.14^{+0.04}_{-0.04}$ &  $400^{+32}_{-24}$   &      &  $5972^{+102}_{-135}$ & $1.02^{+0.02}_{-0.02}$  & $456^{+32}_{-35}$ &     \\
4004B & K8 & $0.75^{+0.05}_{-0.05}$ &  $524^{+48}_{-43}$   & 2.31 &                       &                         &        \\
4209A & F3 & $1.09^{+0.11}_{-0.07}$ & $1647^{+385}_{-264}$ &      &  $6021^{+119}_{-135}$ & $1.02^{+0.03}_{-0.03}$  & $2124^{+125}_{-191}$ &     \\
4209B & K7 & $0.83^{+0.22}_{-0.43}$ & $1859^{+6446}_{-929}$& 0.21 &  $4887^{+1083}_{-835}$& $0.89^{+0.32}_{-0.22}$  & $1814^{+1305}_{-654}$ & 0.24\\
4292A & G4 & $1.06^{+0.03}_{-0.02}$ &  $360^{+17}_{-16}$   &      &  $5857^{+140}_{-213}$ & $1.00^{+0.03}_{-0.03}$  & $436^{+43}_{-52}$ &     \\
4292B & M6 & $0.29^{+0.05}_{-0.01}$ &  $610^{+48}_{-38}$   & 6.01 &                       &                         &        \\
4331A & F2 & $1.34^{+0.20}_{-0.13}$ & $1177^{+499}_{-283}$ &      &  $6140^{+140}_{-119}$ & $1.04^{+0.04}_{-0.03}$  & $964^{+76}_{-51}$ &     \\
4331B & F3 & $1.28^{+0.15}_{-0.15}$ & $1109^{+404}_{-299}$ & 0.14 &  $6033^{+136}_{-134}$ & $1.03^{+0.03}_{-0.03}$  & $962^{+58}_{-90}$ & 0.03\\
4407A & F8 & $1.17^{+0.05}_{-0.06}$ &  $242^{+29}_{-24}$   &      &                       &                         &     \\
4407A & G2 & $1.14^{+0.05}_{-0.05}$ &  $230^{+29}_{-18}$   & 0.32 &                       &                         &        \\
4407B & K5 & $0.82^{+0.06}_{-0.28}$ &  $278^{+36}_{-87}$   & 0.39 &                       &                         &        \\
4407C &    &                        &                      &      &                       &                         &        \\
4463A & K8 & $0.79^{+0.01}_{-0.02}$ &  $427^{+14}_{-14}$   &      &                       &                         &     \\
4463B & K5 & $0.85^{+0.02}_{-0.02}$ &  $543^{+25}_{-28}$   & 3.71 &                       &                         &        \\
4634A & A8 & $1.33^{+0.05}_{-0.02}$ & $1223^{+136}_{-74}$  &      &                       &                         &     \\
4634B & K2 & $0.92^{+0.09}_{-0.06}$ &  $669^{+202}_{-84}$  & 2.58 &  $4670^{+403}_{-349}$ & $0.73^{+0.13}_{-0.09}$  & $668^{+116}_{-90}$ &        \\
4768A & G4 & $1.01^{+0.02}_{-0.02}$ & $1037^{+91}_{-77}$   &      &  $5696^{+250}_{-282}$ & $0.99^{+0.07}_{-0.08}$  & $1330^{+237}_{-182}$ &     \\
4768B & K5 & $0.73^{+0.09}_{-0.15}$ & $2014^{+578}_{-507}$ & 1.90 &                       &                         &        \\
4822A & F6 & $1.27^{+0.05}_{-0.03}$ &  $769^{+109}_{-67}$  &      &  $6127^{+83}_{-84}$   & $1.03^{+0.03}_{-0.04}$  & $733^{+25}_{-35}$ &     \\
4822B & K9 & $0.60^{+0.18}_{-0.17}$ & $1665^{+537}_{-564}$ & 1.56 &                       &                         &        \\
4871A & F4 & $1.28^{+0.01}_{-0.01}$ &  $724^{+18}_{-25}$   &      &                       &                         &     \\
4871B & A5 & $1.22^{+0.08}_{-0.11}$ & $2540^{+545}_{-546}$ & 3.32 &  $6030^{+133}_{-134}$ & $1.03^{+0.03}_{-0.03}$  & $2556^{+155}_{-236}$ &        \\
5578A & G5 & $1.11^{+0.05}_{-0.04}$ &  $190^{+17}_{-10}$   &      &  $6005^{+114}_{-157}$ & $1.00^{+0.02}_{-0.03}$  & $233^{+16}_{-21}$ &     \\
5578B & G5 & $1.07^{+0.12}_{-0.07}$ &  $383^{+99}_{-61}$   & 3.05 &  $5778^{+201}_{-296}$ & $0.99^{+0.05}_{-0.06}$  & $438^{+70}_{-58}$ & 3.41\\
5762A & G6 & $0.99^{+0.03}_{-0.03}$ & $1130^{+146}_{-86}$  &      &  $5551^{+255}_{-293}$ & $0.95^{+0.07}_{-0.08}$  & $1385^{+230}_{-199}$ &     \\
5762B & F4 & $1.08^{+0.12}_{-0.07}$ & $2042^{+556}_{-342}$ & 2.45 &  $5886^{+195}_{-303}$ & $1.02^{+0.05}_{-0.06}$  & $2431^{+314}_{-400}$ & 2.27\\

\end{longtable}

\subsection{On Potential Giants}
Dwarf-giant eclipsing binaries were originally expected to be approximately 200 times more abundant than detected planet-star transits in the \textit{Kepler} field \citep{brown2003}. The assembly of the \textit{Kepler} Input Catalog made use of Bayesian techniques to exclude most of these giants \citep{brown2011}. Remaining dwarf-giant eclipsing binaries are identified and screened by the \textit{Kepler} analysis pipeline, particularly by the presence of secondary eclipses in light curves \citep{batalha2010, tenenbaum2013}. Although a dwarf-giant eclipsing binary with (or as) a contaminating companion should be a relatively rare arrangement, it is reasonable to expect that some might still fall out of a dataset as large as the KIC. It has been demonstrated spectroscopically that a number of late-type \textit{Kepler} Input Catalog stars photometrically characterized as dwarfs are in fact giants, but also that an improved photometric cut exists as $K_p - J > 2$ and $K_p < 14$ that contains $96\% \pm 1\%$ giants (with the corresponding $K_p - J > 2$ and $K_p > 14$ containing only $7\% \pm 3\%$), allowing us to investigate our updated photometry for the possibility of giants hiding in blended KOIs \citep{mann2012}. None of our observed stars meet this set of criteria, and thus all objects are likely dwarfs.

\subsection{Probability of Physical Association}\label{kepmultiplicity}

Table~\ref{stellarparams} lists the confidence for each host/companion pair to be physically unassociated, derived from the respective distances and uncertainties of both fitting methods. The noted uncertainties in photometrically fitting temperature/radius ($\pm200$K and 0.2dex, respectively) to individual stars reported in B11 are ignored for the distance estimates and physical association confidences as they are functions of stellar age and metallicity, which should be consistent across all members of a physically associated system. We treat all pairs with $\geq5\sigma$ level of confidence to be inconsistent with a physically associated/gravitationally bound scenario. Note that pairs with $\leq5\sigma$ are not necessarily associated/bound but are not inconsistent with such an interpretation. The KH07 method then identifies 13 physically unassociated companions, while the B11 method finds 10.

The two methods agree to $5\sigma$ on the unboundedness of only one companion, 2001B. Of the other 12 unbound candidates via KH07, 9 do not have B11 fits, and one (2317B) has a marginal B11 $\sigma_{unassoc} = 4.90$. Only two, 1989B and 2664B, are unbound by KH07 and disputed by B11, for which the methods agree on distances but have respective B11 $\sigma_{unassoc}$ of 3.63 and 2.91 due to larger uncertainties on the B11 results.

B11 also identifies 10 additional physically unassociated companions that do not qualify by the KH07 measurement, though two (628C and 2813B) are marginal. We note that uncertainties for the B11 method are systematically underestimated by the granular dataset, and many could not be fit due to that catalog's relative sparsity, particularly for late-type stars.

The B11 results are presented to check the reproducibility of the KH07 fitting method, but given the noted issues with the former the KH07 results are preferred, and are the focus of this work.

\begin{longtable}{ r | c c | c c | c c }
\caption{Adjusted Transit Depth and Candidate Sizes}\label{planets}\\
\caption*{The transit depth and radius relative to potential host for all transit candidates, evaluated for association with all possible host stars. Evaluated only for KOIs with Kepler-band contrast observations. Candidates with radii lower limits indicate the depth of the eclipse is equal to or greater than the star's full light.}\\
\setlength{\extrarowheight}{3pt}
 & \multicolumn{2}{c|}{A} & \multicolumn{2}{c|}{B} & \multicolumn{2}{c}{C/D} \\
	object & depth (mmag) & $R_{\earth}$ & depth(mmag) & $R_{\earth}$ &	depth(mmag) & $R_{\earth}$ \\
	\hline
	\endfirsthead
 & \multicolumn{2}{c|}{A} & \multicolumn{2}{c|}{B} & \multicolumn{2}{c}{C/D} \\
	object & depth (mmag) & $R_{\earth}$ & depth(mmag) & $R_{\earth}$ &	depth(mmag) & $R_{\earth}$ \\
	\hline
	\endhead
	\endfoot
	\endlastfoot
  0190.01  &  16.61  $\pm$  0.026  & $15.99^{+0.54}_{-0.94}$ &  57.24  $\pm$  0.092  & $22.98^{+0.99}_{-0.49}$ \\
  0191.01  &  18.58  $\pm$  0.027  & $15.2^{+0.71}_{-0.43}$ &  345.0  $\pm$  0.502  & $60.31^{+7.41}_{-4.56}$ \\
  0191.02  &  0.840  $\pm$  0.013  & $3.24^{+0.15}_{-0.09}$ &  13.55  $\pm$  0.223  & $12.87^{+1.6}_{-0.99}$ \\
  0191.03  &  0.232  $\pm$  0.009  & $1.71^{+0.08}_{-0.05}$ &  3.739  $\pm$  0.147  & $6.78^{+0.86}_{-0.53}$ \\
  0191.04  &  0.725  $\pm$  0.030  & $3.01^{+0.15}_{-0.09}$ &  11.69  $\pm$  0.494  & $11.96^{+1.53}_{-0.94}$ \\
  0268.01  &  0.546  $\pm$  0.003  & $3.13^{+0.1}_{-0.12}$ &  127.9  $\pm$  0.918  & $30.54^{+1.46}_{-2.56}$ &  N/A  & $\geq 100.3$ \\
  0401.01  &  2.449  $\pm$  0.016  & $5.44^{+0.16}_{-0.1}$ &  35.30  $\pm$  0.238  & $15.21^{+0.98}_{-1.77}$ \\
  0401.02  &  1.868  $\pm$  0.043  & $4.75^{+0.14}_{-0.09}$ &  26.83  $\pm$  0.618  & $13.29^{+0.87}_{-1.57}$ \\
  0401.03  &  0.405  $\pm$  0.019  & $2.21^{+0.07}_{-0.04}$ &  5.775  $\pm$  0.282  & $6.19^{+0.42}_{-0.75}$ \\
  0425.01  &  23.09  $\pm$  0.065  & $19.46^{+0.79}_{-0.63}$ &  51.76  $\pm$  0.147  & $28.23^{+1.42}_{-1.65}$ \\
  0511.01  &  0.757  $\pm$  0.009  & $3.6^{+0.09}_{-0.15}$ &  13.15  $\pm$  0.173  & $10.05^{+0.48}_{-0.97}$ &  278.9  $\pm$  3.670  & $30.62^{+7.36}_{-7.88}$ \\
  0511.02  &  0.210  $\pm$  0.008  & $1.9^{+0.05}_{-0.08}$ &  3.645  $\pm$  0.146  & $5.3^{+0.26}_{-0.52}$ &  70.73  $\pm$  2.846  & $16.16^{+3.98}_{-4.26}$ \\
  0688.01  &  0.354  $\pm$  0.006  & $2.56^{+0.02}_{-0.04}$ &  2.608  $\pm$  0.048  & $5.39^{+0.43}_{-0.27}$ \\
  0984.01  &  2.187  $\pm$  0.013  & $5.04^{+0.15}_{-0.1}$ &  2.376  $\pm$  0.014  & $5.25^{+0.15}_{-0.1}$ \\
  0987.01  &  0.232  $\pm$  0.005  & $1.61^{+0.02}_{-0.02}$ &  6.369  $\pm$  0.139  & $6.42^{+0.26}_{-0.43}$ \\
  1066.01  &  15.39  $\pm$  0.032  & $14.62^{+0.65}_{-0.65}$ &  1261.  $\pm$  2.662  & $76.84^{+2.72}_{-3.62}$ \\
  1067.01  &  50.95  $\pm$  0.069  & $30.12^{+0.23}_{-0.23}$ &          N/A          & $\geq93.78$ \\
  1112.01  &  0.689  $\pm$  0.022  & $3.38^{+0.11}_{-0.17}$ &  50.69  $\pm$  1.648  & $17.0^{+1.2}_{-0.96}$ \\
  1214.01  &  0.294  $\pm$  0.018  & $1.79^{+0.04}_{-0.04}$ &  0.890  $\pm$  0.056  & $5.0^{+0.3}_{-0.33}$ \\
  1447.01  &  228.6  $\pm$  0.074  & $68.91^{+1.9}_{-1.43}$ &          N/A          & $\geq97.05$ \\
  1447.02  &  17.96  $\pm$  0.038  & $20.25^{+0.56}_{-0.42}$ &  125.5  $\pm$  0.267  & $32.79^{+1.08}_{-0.72}$ \\
  1700.01  &  0.442  $\pm$  0.016  & $2.13^{+0.05}_{-0.05}$ &  1.174  $\pm$  0.042  & $3.19^{+0.11}_{-0.11}$ \\
  1784.01  &  7.479  $\pm$  0.092  & $10.48^{+0.46}_{-0.55}$ &  12.74  $\pm$  0.157  & $14.96^{+0.36}_{-0.6}$ \\
  1880.01  &  0.680  $\pm$  0.009  & $2.13^{+0.03}_{-0.14}$ &  18.33  $\pm$  0.248  & $13.83^{+1.29}_{-0.71}$ \\
  1884.01  &  3.201  $\pm$  0.049  & $6.27^{+0.42}_{-0.24}$ &  123.3  $\pm$  1.919  & $26.09^{+5.08}_{-9.79}$ &  228.7  $\pm$  3.558  & $27.57^{+11.09}_{-9.64}$ \\ 
	         &                       &                        &                       &                        &  356.4  $\pm$  5.543  & $23.07^{+10.38}_{-5.77}$ \\
  1884.02  &  0.618  $\pm$  0.039  & $2.76^{+0.19}_{-0.11}$ &  22.80  $\pm$  1.459  & $11.47^{+2.33}_{-4.5}$ &  40.66  $\pm$  2.603  & $12.13^{+5.1}_{-4.44}$ \\  
	         &                       &                        &                       &                        &  60.42  $\pm$  3.868  & $10.15^{+4.84}_{-2.69}$ \\
  1916.01  &  0.395  $\pm$  0.009  & $2.73^{+0.02}_{-0.02}$ &  4.866  $\pm$  0.115  & $7.0^{+0.45}_{-0.37}$ \\
  1916.02  &  0.305  $\pm$  0.006  & $2.39^{+0.02}_{-0.02}$ &  3.754  $\pm$  0.079  & $6.15^{+0.39}_{-0.33}$ \\
  1916.03  &  0.079  $\pm$  0.004  & $1.22^{+0.01}_{-0.01}$ &  0.980  $\pm$  0.051  & $3.14^{+0.21}_{-0.17}$ \\
  1989.01  &  0.534  $\pm$  0.020  & $2.76^{+0.1}_{-0.08}$ &  13.31  $\pm$  0.505  & $11.08^{+0.37}_{-0.37}$ \\
  2001.01  &  0.205  $\pm$  0.007  & $1.45^{+0.02}_{-0.02}$ &  13.88  $\pm$  0.502  & $11.43^{+0.64}_{-0.38}$ \\
  2009.01  &  0.626  $\pm$  0.022  & $3.06^{+0.08}_{-0.11}$ &  25.22  $\pm$  0.901  & $13.72^{+1.2}_{-3.59}$ \\
  2059.01  &  0.186  $\pm$  0.007  & $1.33^{+0.03}_{-0.01}$ &  0.509  $\pm$  0.020  & $2.01^{+0.05}_{-0.07}$ \\
  2059.02  &  0.057  $\pm$  0.005  & $0.74^{+0.02}_{-0.01}$ &  0.156  $\pm$  0.015  & $1.11^{+0.03}_{-0.04}$ \\
  2069.01  &  0.678  $\pm$  0.013  & $3.3^{+0.06}_{-0.08}$ &  29.60  $\pm$  0.605  & $12.7^{+2.19}_{-4.92}$ \\
  2083.01  &  0.399  $\pm$  0.015  & $2.49^{+0.17}_{-0.13}$ &  1.027  $\pm$  0.039  & $4.16^{+0.24}_{-1.01}$ \\
  2117.01  &  1.519  $\pm$  0.074  & $3.51^{+0.04}_{-0.04}$ &  2.060  $\pm$  0.101  & $4.18^{+0.05}_{-0.05}$ \\
  2247.01  &  0.205  $\pm$  0.011  & $1.35^{+0.02}_{-0.02}$ &  20.34  $\pm$  1.121  & $8.02^{+1.72}_{-1.72}$ \\
  2289.01  &  0.369  $\pm$  0.018  & $2.61^{+0.02}_{-0.02}$ &  28.81  $\pm$  1.470  & $14.12^{+0.92}_{-1.48}$ \\
  2289.02  &  0.175  $\pm$  0.009  & $1.8^{+0.01}_{-0.01}$ &  13.57  $\pm$  0.761  & $9.72^{+0.64}_{-1.02}$ \\
  2317.01  &  0.149  $\pm$  0.009  & $1.5^{+0.04}_{-0.07}$ &  16.21  $\pm$  1.076  & $11.68^{+0.28}_{-0.42}$ \\
  2363.01  &  0.204  $\pm$  0.012  & $1.42^{+0.02}_{-0.02}$ &  45.44  $\pm$  2.821  & $14.13^{+8.65}_{-6.78}$ \\
  2413.01  &  0.531  $\pm$  0.029  & $2.39^{+0.15}_{-0.08}$ &  3.329  $\pm$  0.184  & $2.53^{+1.27}_{-0.89}$ \\
  2413.02  &  0.457  $\pm$  0.038  & $2.22^{+0.14}_{-0.07}$ &  2.868  $\pm$  0.238  & $2.35^{+1.21}_{-0.85}$ \\
  2443.01  &  0.110  $\pm$  0.007  & $1.33^{+0.04}_{-0.05}$ &  13.41  $\pm$  0.877  & $8.7^{+1.41}_{-2.82}$ \\
  2443.02  &  0.105  $\pm$  0.008  & $1.3^{+0.03}_{-0.05}$ &  12.79  $\pm$  1.084  & $8.5^{+1.4}_{-2.81}$ \\
  2542.01  &  0.576  $\pm$  0.033  & $1.63^{+0.13}_{-0.21}$ &  1.710  $\pm$  0.100  & $1.64^{+0.37}_{-0.27}$ \\
  2657.01  &  0.091  $\pm$  0.007  & $1.09^{+0.08}_{-0.05}$ &  0.117  $\pm$  0.010  & $1.16^{+0.07}_{-0.04}$ \\
  2664.01  &  1.377  $\pm$  0.105  & $3.65^{+0.04}_{-0.04}$ &  2.954  $\pm$  0.226  & $5.74^{+0.18}_{-0.12}$ \\
  2681.01  &  8.139  $\pm$  0.129  & $9.8^{+0.38}_{-0.29}$ &  26.00  $\pm$  0.413  & $14.76^{+0.17}_{-0.34}$ \\
  2681.02  &  1.006  $\pm$  0.111  & $3.45^{+0.15}_{-0.11}$ &  3.193  $\pm$  0.353  & $5.2^{+0.07}_{-0.13}$ \\
  2705.01  &  0.745  $\pm$  0.027  & $1.31^{+0.53}_{-0.27}$ &  11.40  $\pm$  0.416  & $3.34^{+0.69}_{-0.12}$ \\
  2711.01  &  0.442  $\pm$  0.013  & $2.82^{+0.05}_{-0.07}$ &  0.493  $\pm$  0.015  & $2.95^{+0.05}_{-0.1}$ \\
  2711.02  &  0.351  $\pm$  0.016  & $2.51^{+0.04}_{-0.06}$ &  0.392  $\pm$  0.018  & $2.63^{+0.04}_{-0.09}$ \\
  2722.01  &  0.161  $\pm$  0.004  & $1.72^{+0.01}_{-0.01}$ &  60.00  $\pm$  1.601  & $17.45^{+1.55}_{-2.33}$ \\
  2722.02  &  0.154  $\pm$  0.005  & $1.68^{+0.01}_{-0.01}$ &  57.25  $\pm$  1.971  & $17.05^{+1.53}_{-2.3}$ \\
  2722.03  &  0.109  $\pm$  0.003  & $1.41^{+0.01}_{-0.01}$ &  40.34  $\pm$  1.439  & $14.37^{+1.29}_{-1.94}$ \\
  2722.04  &  0.115  $\pm$  0.005  & $1.45^{+0.01}_{-0.01}$ &  42.41  $\pm$  1.847  & $14.73^{+1.33}_{-2.0}$ \\
  2722.05  &  0.105  $\pm$  0.006  & $1.39^{+0.01}_{-0.01}$ &  38.80  $\pm$  2.471  & $14.1^{+1.3}_{-1.95}$ \\
  2779.01  &  0.592  $\pm$  0.028  & $3.23^{+0.08}_{-0.08}$ &  6.113  $\pm$  0.292  & $7.76^{+1.02}_{-0.51}$ \\
  2813.01  &  0.205  $\pm$  0.017  & $1.4^{+0.02}_{-0.02}$ &  0.506  $\pm$  0.042  & $2.97^{+0.13}_{-0.23}$ &  38.89  $\pm$  3.261  & $11.66^{+4.86}_{-4.64}$ \\
  2837.01  &  0.259  $\pm$  0.010  & $2.36^{+0.09}_{-0.07}$ &  0.320  $\pm$  0.013  & $2.59^{+0.14}_{-0.08}$ \\
  2849.01  &  0.205  $\pm$  0.014  & $1.38^{+0.02}_{-0.03}$ &  0.435  $\pm$  0.029  & $4.54^{+0.65}_{-0.81}$ \\
  2859.01  &  0.100  $\pm$  0.007  & $1.04^{+0.02}_{-0.01}$ &  0.707  $\pm$  0.054  & $2.53^{+0.42}_{-0.54}$ \\
  2859.02  &  0.068  $\pm$  0.006  & $0.86^{+0.02}_{-0.01}$ &  0.482  $\pm$  0.044  & $2.09^{+0.35}_{-0.45}$ \\
  2859.03  &  0.078  $\pm$  0.007  & $0.92^{+0.02}_{-0.01}$ &  0.556  $\pm$  0.053  & $2.25^{+0.38}_{-0.48}$ \\
  2859.04  &  0.077  $\pm$  0.006  & $0.91^{+0.02}_{-0.01}$ &  0.544  $\pm$  0.044  & $2.22^{+0.37}_{-0.47}$ \\
  2859.05  &  0.107  $\pm$  0.008  & $1.07^{+0.02}_{-0.01}$ &  0.756  $\pm$  0.062  & $2.62^{+0.43}_{-0.56}$ \\
  2869.01  &  0.139  $\pm$  0.009  & $1.58^{+0.01}_{-0.01}$ &          N/A          & $\geq58.89$ \\
  2904.01  &  0.144  $\pm$  0.005  & $1.6^{+0.01}_{-0.03}$ &  0.895  $\pm$  0.033  & $3.6^{+0.45}_{-0.71}$ \\
  2971.01  &  0.071  $\pm$  0.004  & $1.14^{+0.01}_{-0.01}$ &  2.705  $\pm$  0.155  & $5.28^{+0.23}_{-0.34}$ &  21.15  $\pm$  1.214  & $4.85^{+12.94}_{-0.8}$ \\
  2971.02  &  0.103  $\pm$  0.006  & $1.37^{+0.01}_{-0.01}$ &  3.938  $\pm$  0.260  & $6.36^{+0.28}_{-0.42}$ &  30.91  $\pm$  2.042  & $5.85^{+15.73}_{-0.97}$ \\
  3020.01  &  0.106  $\pm$  0.006  & $1.42^{+0.01}_{-0.01}$ &  2.122  $\pm$  0.134  & $3.71^{+0.15}_{-0.26}$ &  137.1  $\pm$  8.661  & $28.92^{+2.63}_{-4.88}$ \\
  3069.01  &  0.387  $\pm$  0.031  & $2.47^{+0.09}_{-0.13}$ &  2.996  $\pm$  0.244  & $5.15^{+0.06}_{-0.12}$ \\
  3377.01  &  0.476  $\pm$  0.044  & $2.15^{+0.02}_{-0.02}$ &  15.94  $\pm$  1.488  & $3.82^{+0.01}_{-0.01}$ &  100.9  $\pm$  9.425  & $11.37^{+2.92}_{-2.27}$ \\
  3401.01  &  0.200  $\pm$  0.022  & $1.5^{+0.05}_{-0.03}$ &  0.452  $\pm$  0.049  & $3.32^{+0.25}_{-0.22}$ \\
  3401.02  &  0.603  $\pm$  0.061  & $2.6^{+0.08}_{-0.06}$ &  1.362  $\pm$  0.139  & $5.75^{+0.42}_{-0.38}$ \\
  4004.01  &  0.151  $\pm$  0.010  & $1.47^{+0.05}_{-0.05}$ &  6.263  $\pm$  0.440  & $6.2^{+0.44}_{-0.44}$ \\
  4209.01  &  1.439  $\pm$  0.350  & $4.25^{+0.45}_{-0.4}$ &  13.51  $\pm$  3.292  & $13.95^{+2.43}_{-2.28}$ \\
  4292.01  &  0.045  $\pm$  0.004  & $0.75^{+0.03}_{-0.02}$ &  50.35  $\pm$  4.668  & $6.73^{+1.52}_{-0.0}$ \\
  4331.01  &  0.125  $\pm$  0.011  & $1.59^{+0.37}_{-0.19}$ &  0.157  $\pm$  0.013  & $1.72^{+0.38}_{-0.21}$ \\
  4463.01  &  0.372  $\pm$  0.024  & $1.6^{+0.02}_{-0.06}$ &  0.376  $\pm$  0.024  & $1.7^{+0.04}_{-0.04}$ \\
  4634.01  &  0.118  $\pm$  0.012  & $1.51^{+0.06}_{-0.03}$ &  0.618  $\pm$  0.066  & $2.39^{+0.29}_{-0.2}$ \\
  4768.01  &  0.598  $\pm$  0.062  & $2.58^{+0.06}_{-0.06}$ &  24.00  $\pm$  2.489  & $11.77^{+1.77}_{-2.66}$ \\
  4822.01  &  0.035  $\pm$  0.003  & $0.79^{+0.03}_{-0.02}$ &  19.46  $\pm$  2.083  & $8.57^{+2.89}_{-2.73}$ \\
  4871.01  &  0.029  $\pm$  0.004  & $0.73^{+0.01}_{-0.01}$ &  0.524  $\pm$  0.070  & $2.87^{+0.22}_{-0.3}$ \\
  4871.02  &  0.038  $\pm$  0.004  & $0.83^{+0.01}_{-0.01}$ &  0.671  $\pm$  0.081  & $3.25^{+0.24}_{-0.33}$ \\
  5578.01  &  0.193  $\pm$  0.025  & $1.62^{+0.08}_{-0.07}$ &  0.997  $\pm$  0.133  & $3.54^{+0.49}_{-0.26}$ \\
  5762.01  &  0.484  $\pm$  0.065  & $2.28^{+0.08}_{-0.08}$ &  0.876  $\pm$  0.118  & $3.35^{+0.42}_{-0.25}$ \\
	
\end{longtable}

\begin{table}
\caption{Probability of $R > 15R_{\earth}$ for Each Planet Candidate}\label{FPprobs}
\centering
Estimated probabilities that each KOI planet candidate has a radius $R > 15R_\Earth$, for each potential host and summed across all. Only candidates for with $P(R > 15R_\Earth) \geq 0.01$ are listed. For full transit depth and planet size estimates, see Table~\ref{planets}.

$ ^{a}$ Disposition is FALSE POSITIVE in the Exoplanet Archive as of 18 Sep 2015.

$ ^{b}$ Disposition is CONFIRMED in the Exoplanet Archive as of 18 Sep 2015.

$ ^{c}$ Other literature indicates candidate is false positive.
\begin{tabular}{ c c c c c c }
KOI & $P_{A}$ & $P_{B}$& $P_{C}$& $P_{D}$& $P_{total}$\\
\hline
0190.01$^{a}$ & 0.85  & 1.00 &       &       & 0.93 \\
0191.01       & 0.68  & 1.00 &       &       & 0.84 \\
0191.02       &   0   & 0.09 &       &       & 0.05 \\
0191.04       &   0   & 0.02 &       &       & 0.01 \\
0268.01       &   0   & 1.00 & 1.00  &       & 0.67 \\
0401.01$^{b}$ &   0   & 0.55 &       &       & 0.27 \\
0401.02$^{b}$ &   0   & 0.03 &       &       & 0.01 \\
0425.01       & 1.00  & 1.00 &       &       & 1.00 \\
0511.01$^{b}$ &   0   &   0  & 0.98  &       & 0.33 \\
0511.02$^{b}$ &   0   &   0  & 0.61  &       & 0.20 \\
1066.01       & 0.30  & 1.00 &       &       & 0.65 \\
1067.01       & 1.00  & 1.00 &       &       & 1.00 \\
1112.01$^{c}$ &   0   & 0.95 &       &       & 0.48 \\
1447.01$^{c}$ & 1.00  & 1.00 &       &       & 1.00 \\
1447.02       & 1.00  & 1.00 &       &       & 1.00 \\
1784.01       &   0   & 0.46 &       &       & 0.23 \\
1880.01       &   0   & 0.18 &       &       & 0.09 \\
1884.01       &   0   & 0.87 & 0.90  & 0.92  & 0.65 \\
1884.02       &   0   & 0.07 & 0.29  & 0.14  & 0.21 \\
2009.01       &   0   & 0.14 &       &       & 0.07 \\
2069.01       &   0   & 0.15 &       &       & 0.07 \\
2289.01$^{b}$ &   0   & 0.17 &       &       & 0.09 \\
2363.01       &   0   & 0.46 &       &       & 0.23 \\
2681.01				&   0   & 0.08 &       &       & 0.04 \\
2722.01$^{b}$ &   0   & 0.85 &       &       & 0.43 \\
2722.02$^{b}$ &   0   & 0.81 &       &       & 0.41 \\
2722.03$^{b}$ &   0   & 0.31 &       &       & 0.16 \\
2722.04$^{b}$ &   0   & 0.42 &       &       & 0.21 \\
2722.05       &   0   & 0.24 &       &       & 0.12 \\
2813.01       &   0   &   0  & 0.25  &       & 0.08 \\
2869.01       &   0   & 1.00 &       &       & 0.50 \\
2971.01       &   0   &   0  & 0.22  &       & 0.07 \\
2971.02       &   0   &   0  & 0.28  &       & 0.09 \\
3020.01       &   0   &   0  & 1.00  &       & 0.33 \\
3377.01       &   0   &   0  & 0.11  &       & 0.04 \\
4209.01       &   0   & 0.33 &       &       & 0.16 \\
4768.01       &   0   & 0.03 &       &       & 0.02 \\
4822.01       &   0   & 0.01 &       &       & 0.01 \\
\end{tabular}

\end{table}

\subsection{Updated Transiting Object Parameters}
For KOI systems observed in the \textit{Kepler} band, we reinterpret the relative depth and size of all transit candidates in Table~\ref{planets}, relying on the KH07 results as new stellar characteristics. As we lack the ability to determine whether the primary or a companion is the host of the transiting object, all possible scenarios are presented. Note that these derivations require knowledge of each KOI component's luminosity in the transit band, and thus only candidates with resolved LP600 photometry are shown. As mentioned in section~\ref{kepobservations}, the LP600 combined with the EMCCD's sensitivity curve approximates the \textit{Kepler} passband, suppressing blue wavelengths that experience less benefit from adaptive optics correction.

With the new planet candidate sizes we estimate the probability each has a radius $R > 15 R_\Earth$, the rough position of the boundary between gas giants and late-type stars. We then assume that every star in a blended KOI is an equally likely host for the transit, and measure an overall P$(R > 15 R_\Earth)$ as the mean average of probabilities for all possible hosts. This identifies potential false positives but does not constitute a full false positive calculation. Table~\ref{FPprobs} shows the results for all candidates for with P$(R > 15 R_\Earth) \geq 0.01$. We did not take into account the relative prevalence of planetary bodies and brown dwarfs, which might indicate that planets are generally more likely system members and invalidate the assumption that all possible configurations are equally likely.

Three candidates have been identified elsewhere as false positives. Likely the largest transiting object, KOI1447.01 appears in the \textit{Kepler} Eclipsing Binary Catalog \citep{slawson2011}. KOI0190.01 has a disposition of FALSE POSITIVE in the Exoplanet Archive from radial velocity measurements. KOI1112.01 has been identified as a false positive via ephemeris matching with the nearby KOI4720 by \citet{coughlin2014}, which notes that two stars are separated by only 4\farcs8, and states that the transit host is believed to be a third then-unobserved object. This implies the host is KOI1112B, which elevates our estimate to P$(R > 15 R_\Earth) = 0.95$.

\section{Discussion}\label{kepdiscussion}
Probabilities and uncertainties in this section are computed binomially by the method described in \citet{burgasser2003}.

Of the 93 companions with sufficient photometry, 13 (or $14.0\%^{+4.4\%}_{-2.9\%}$) are inconsistent with physical association with their primaries via KH07, while the B11 method gives 10 unassociated companions out of 53 examined (or $18.9\%^{+6.5\%}_{-4.2\%}$). All others are $<5\sigma$ consistent with a bound interpretation. Simulations have previously demonstrated that the vast majority (96\%) of narrowly-separated companions ($<1.0\arcsec$) are physically associated \citep{horch2014}. As 6 out of 40 (or $15\%^{+7.3\%}_{-4.0\%}$) narrowly-separated primary/companion pairs with fit results are inconsistent with a bound interpretation to $5\sigma$, our results are inconsistent with the Horch prediction to $\sim2.3\sigma$ and we see no evidence that narrowly separated ($<1.0\arcsec$) companions are more likely to be physically associated than KOI companions in general or than widely-separated companions, for which we determine 7 of 53 or $13.2\%^{+6.0\%}_{-3.3\%}$ are unbound.

Via transit reinterpretation we have 38 potential non-planetary objects out of 88 reinterpreted transiting objects. By summing the computed P$(R > 15 R_\Earth)$ values this sample has a mean of $12.8^{+3.5}_{-3.1}$, or $14.5\%^{+4.0\%}_{-3.5\%}$ of candidates with $(R > 15 R_\Earth)$. Considered with the previously reported $17.6\% \pm 1.5\%$ nearby-star (companion) probability of \citet{baranec2016}, we estimate a $(R > 15 R_\Earth)$ rate due to unresolved companions to be $2.6\% \pm 0.4\%$. This is a rough measurement of false positives in the KOI catalog and is easily consistent with the broad $<10\%$ false positive rate predicted by modeling \citep{morton2011}.

On the whole, the derived planet candidate sizes are only slightly larger than estimates from \citet{law2014}, henceforth L14. The exceptions are primarily those candidates listed in Table~\ref{FPprobs}. These are much larger than the L14 predictions as our analysis includes new sizes for the host stars and accounts for the change in distance estimates, whereas L14 used the original \textit{Kepler} predictions derived from blended light.

\subsection{KOI0191: Possible Coincident Multiple}
L14 noted that this system is \textit{a priori} unusual as the only multi-candidate KOI to have a large Jupiter-class candidate ($>10R_\Earth$) in a very close orbit ($P < 20$d). Assuming binarity (of KOI0191A/B), L14 calculated a planetary candidate size of 11.3$R_\Earth$ for the A scenario and 29.3$R_\Earth$ for B, making the potential KOI0191B/KOI0191.01 system a close eclipsing binary in a hierarchical triple. With the inclusion of \textit{JHK} photometry, we revise these estimates upward to 13.9$R_\Earth$ and 55.9$R_\Earth$, respectively. 

Although hot Jupiters were previously thought inconsistent with other short-period planets, the discovery of multiple planets in the WASP-47 system proves the arrangement does exist in nature \citep{becker2015}. Thus, we can not rule out that all four candidates are hosted by KOI0191A and are then planets.

\subsection{KOI0268: No Longer Habitable}
Both companions of KOI0268 have also been reported in \citet{adams2012}. KOI0268.01 was originally identified as a potentially habitable super-Earth with a radius of 1.7 $R_\Earth$ and equilibrium temperature of 295 K. L14 reports both companions, and notes that if the planet orbits either of them rather than the target A, the equilibrium temperature of the planet will probably not be in the habitable range. The candidate's equilibrium temperature in literature has since been revised upward from 295K to 470K as reported in the Exoplanet Archive. Our fitted stellar types (from KH07) yield an uninhabitable surface temperature of 650K if hosted by A. For B and C both we estimate a surface temperature of $\sim$350K, marginally allowing for the presence of liquid water, but the reinterpreted sizes imply a gas giant hosted by B or an eclipsing binary at C. Exomoons notwithstanding, this rules out habitability for KOI0268.01.

\subsection{KOI1447: Double Eclipsing Binary}
Both KOI1447.01 and KOI1447.02 are likely too large to be planets for either potential host, and .01 was included in the second release of the Kepler Eclipsing Binary Stars catalog \citep{slawson2011}. Given the size of both candidates and their short orbital periods (40.2d and 2.3d, respectively), it seems likely they would conflict with each other if in the same system. KOI1447 is then a unique double false positive, consisting of two merely visually associated eclipsing binaries with coincidentally low inclination.

\section{Conclusions}\label{kepconclusions}
We have obtained visible and near-infrared multi-wavelength photometry of 104 blended companions to 84 KOIs, validating the original detections by Robo-AO. We report additional companions not originally detected by Robo-AO's original investigation. We find that $14.5\%^{+3.8\%}_{-3.4\%}$ of the investigated companions are physically unassociated with their KOI primaries at the $5\sigma$ level. Additional follow-up is recommended to confirm this result, with spectroscopy of both targets the best means of measuring log $g$ to confirm actual sizes and distances, and to provide improved constraints on transit candidate size. We also find no evidence that narrowly separated KOI companions are more likely to be physically associated than widely separated companions, contrary to prior modeling work.

We have also reinterpreted 88 transit candidates, refining estimates of size given possible hosts and identifying 43 candidates potentially too large for planetary interpretation. With some assumptions, this produces an overall P$(R > 15 R_\earth)$ for transits with detected contaminating companions of $17.5\%^{+4.1\%}_{-3.7\%}$, or an overall P$(R > 15 R_\earth)$ for all KOIs (as a result of undetected companions) of $2.5\% \pm 0.4\%$. A more complete set of \textit{JHK} follow-up on KOI companions would refine this result.

Given the termination of \textit{Kepler}'s primary mission, solving host ambiguity for individual transit candidates is difficult. A close review of extant \textit{Kepler} data for astrometric motion or light curve re-analysis may detect centroid motion correlated with transit that would identify the host star. Independent investigations like radial velocity and ground-based AO transit imaging are possible but difficult and limited to bright targets and deep transits, respectively.

\begin{longtable}{ c c c c c c }
\setlength{\extrarowheight}{3pt}\\
\caption{Measured \textit{JHK} Contrasts}\label{contrasts}\\
\caption*{Relative locations and NIR contrast measurements of observed \textit{Kepler} Objects of Interest. Contrast uncertainties are systematically measured by varying the photometric aperture size. Use of co-added images reduces separation/angle measurement uncertainties to the single-pixel level.}\\
	object	 & sep($\arcsec$) & ang($^{o}$) &$\Delta m_J$ (mag)&$\Delta m_H$ (mag)&$\Delta m_{K}$ (mag)\\
	\hline
	\endfirsthead
	object	 & sep($\arcsec$) & ang($^{o}$) &$\Delta m_J$ (mag)&$\Delta m_H$ (mag)&$\Delta m_{K}$ (mag)\\
	\hline
	\endhead
	\endfoot
	\endlastfoot
	0190B & 0.180 $\pm$ 0.010 & 109.4 $\pm$ 3.2 &                  &                  & 0.642 $\pm$0.137 \\
  0191B & 1.660 $\pm$ 0.002 &  96.6 $\pm$ 0.1 & 2.588 $\pm$0.057 & 2.615 $\pm$0.054 & 2.626 $\pm$0.055 \\
	0268B & 1.753 $\pm$ 0.003 & 267.6 $\pm$ 0.1 & 3.056 $\pm$0.059 & 2.654 $\pm$0.057 & 2.553 $\pm$0.056 \\
	0268C & 2.528 $\pm$ 0.007 & 310.2 $\pm$ 0.1 & 3.810 $\pm$0.118 & 3.353 $\pm$0.127 & 3.984 $\pm$0.145 \\
  0401B & 1.986 $\pm$ 0.002 & 270.0 $\pm$ 0.1 & 2.066 $\pm$0.059 &                  & 1.635 $\pm$0.055 \\
  0425B & 0.491 $\pm$ 0.001 & 343.4 $\pm$ 0.1 &                  &                  & 0.831 $\pm$0.054 \\
  0511B & 1.300 $\pm$ 0.002 & 123.4 $\pm$ 0.1 & 2.221 $\pm$0.058 & 1.817 $\pm$0.007 & 1.707 $\pm$0.008 \\
  0511C & 3.865 $\pm$ 0.005 & 348.6 $\pm$ 0.1 & 5.055 $\pm$0.122 & 4.493 $\pm$0.077 & 4.308 $\pm$0.069 \\
  0628B & 2.748 $\pm$ 0.002 & 238.9 $\pm$ 0.1 &                  &                  & 3.000 $\pm$0.058 \\
  0628C & 1.828 $\pm$ 0.003 & 311.5 $\pm$ 0.2 &                  &                  & 3.871 $\pm$0.057 \\
  0687B & 0.680 $\pm$ 0.003 &  13.4 $\pm$ 0.4 &                  &                  & 1.251 $\pm$0.054 \\
  0688B & 1.734 $\pm$ 0.001 & 141.8 $\pm$ 0.1 & 1.552 $\pm$0.060 &                  & 1.373 $\pm$0.056 \\
  0712B & 0.470 $\pm$ 0.002 & 174.2 $\pm$ 0.3 & 0.435 $\pm$0.055 &                  & 0.351 $\pm$0.056 \\
  0931B & 1.263 $\pm$ 0.002 & 177.7 $\pm$ 0.1 &                  &                  & 3.227 $\pm$0.063 \\
  0984B & 1.764 $\pm$ 0.005 & 221.3 $\pm$ 1.4 & 0.064 $\pm$0.058 & 0.050 $\pm$0.054 & 0.059 $\pm$0.056 \\
  0987B & 1.974 $\pm$ 0.002 & 225.7 $\pm$ 0.3 & 2.612 $\pm$0.077 & 2.381 $\pm$0.058 & 2.239 $\pm$0.055 \\
	1066B & 1.690 $\pm$ 0.002 & 231.3 $\pm$ 0.1 &                  &                  & 2.949 $\pm$0.070 \\
	1067B & 2.932 $\pm$ 0.005 & 142.6 $\pm$ 0.1 &                  &                  & 2.785 $\pm$0.106 \\
	1112B & 3.068 $\pm$ 0.005 & 172.2 $\pm$ 0.1 & 3.607 $\pm$0.138 & 2.956 $\pm$0.081 & 2.758 $\pm$0.070 \\
	1151B & 0.758 $\pm$ 0.002 & 307.5 $\pm$ 0.7 &                  & 2.554 $\pm$0.055 & 2.407 $\pm$0.055 \\
	1214B & 0.371 $\pm$ 0.029 & 136.3 $\pm$ 0.3 &                  & 2.584 $\pm$0.055 & 2.455 $\pm$0.055 \\
  1274B & 1.085 $\pm$ 0.001 & 242.0 $\pm$ 0.1 & 2.801 $\pm$0.056 &                  & 2.506 $\pm$0.055 \\
	1359B & 1.387 $\pm$ 0.003 & 331.6 $\pm$ 0.4 &                  &                  & 2.168 $\pm$0.057 \\
	1375B & 0.784 $\pm$ 0.001 & 270.0 $\pm$ 0.1 &                  & 3.303 $\pm$0.069 & 3.393 $\pm$0.065 \\
  1442B & 2.114 $\pm$ 0.006 &  70.8 $\pm$ 0.1 & 4.155 $\pm$0.065 & 3.802 $\pm$0.056 & 3.631 $\pm$0.055 \\
	1447B & 0.282 $\pm$ 0.001 & 212.0 $\pm$ 0.1 &                  &                  & 0.625 $\pm$0.061 \\
  1536B & 0.580 $\pm$ 0.001 &  97.9 $\pm$ 0.1 &                  & 4.262 $\pm$0.136 & 4.177 $\pm$0.112 \\
  1546B & 0.603 $\pm$ 0.002 &  89.5 $\pm$ 0.1 & 0.940 $\pm$0.055 & 0.784 $\pm$0.055 & 0.726 $\pm$0.054 \\
  1546C & 2.915 $\pm$ 0.001 &   4.0 $\pm$ 0.1 & 3.224 $\pm$0.059 & 3.021 $\pm$0.071 & 2.945 $\pm$0.081 \\
	1546D & 4.119 $\pm$ 0.011 & 164.7 $\pm$ 0.1 & 3.338 $\pm$0.073 & 3.253 $\pm$0.077 & 3.479 $\pm$0.058 \\
	1613B & 0.214 $\pm$ 0.004 & 185.5 $\pm$ 1.3 & 1.136 $\pm$0.055 & 0.996 $\pm$0.055 & 0.997 $\pm$0.055 \\
	1700B & 0.274 $\pm$ 0.048 & 288.1 $\pm$10.7 &                  &                  & 0.551 $\pm$0.055 \\
	1784B & 0.278 $\pm$ 0.001 & 291.1 $\pm$ 0.1 &                  &                  & 0.781 $\pm$0.058 \\
  1845B & 1.999 $\pm$ 0.011 &  78.9 $\pm$ 0.5 & 3.238 $\pm$0.055 &                  & 2.886 $\pm$0.055 \\
  1845C & 2.958 $\pm$ 0.025 & 348.0 $\pm$ 0.2 & 4.264 $\pm$0.069 &                  & 4.400 $\pm$0.092 \\
  1880B & 1.713 $\pm$ 0.001 & 100.9 $\pm$ 0.1 & 3.936 $\pm$0.058 & 4.149 $\pm$0.057 & 4.282 $\pm$0.058 \\
  1884B & 0.934 $\pm$ 0.001 &  95.6 $\pm$ 0.1 & 2.642 $\pm$0.056 & 2.410 $\pm$0.056 & 2.305 $\pm$0.055 \\
  1884C & 1.838 $\pm$ 0.001 &  81.9 $\pm$ 0.1 & 3.075 $\pm$0.056 & 2.867 $\pm$0.057 & 2.731 $\pm$0.055 \\
  1884D & 2.567 $\pm$ 0.002 & 327.5 $\pm$ 0.1 & 3.590 $\pm$0.164 & 3.536 $\pm$0.141 & 3.204 $\pm$0.141 \\
  1891B & 2.066 $\pm$ 0.003 & 211.4 $\pm$ 0.1 & 4.340 $\pm$0.077 & 4.561 $\pm$0.060 & 4.596 $\pm$0.066 \\
  1916B & 0.252 $\pm$ 0.001 & 146.3 $\pm$ 0.1 & 1.201 $\pm$0.056 &                  & 1.054 $\pm$0.055 \\
  1979B & 0.842 $\pm$ 0.002 & 193.4 $\pm$ 0.1 & 2.291 $\pm$0.055 &                  & 1.822 $\pm$0.055 \\
	1989B & 0.816 $\pm$ 0.001 &  39.5 $\pm$ 0.1 &                  &                  & 2.921 $\pm$0.055 \\
  2001B & 1.167 $\pm$ 0.001 & 342.0 $\pm$ 0.1 &                  &                  & 4.320 $\pm$0.060 \\
  2009B & 1.513 $\pm$ 0.004 & 178.0 $\pm$ 0.1 & 3.042 $\pm$0.092 & 2.950 $\pm$0.061 & 2.750 $\pm$0.055 \\
  2059B & 0.394 $\pm$ 0.001 & 290.0 $\pm$ 0.1 &                  &                  & 0.539 $\pm$0.151 \\
	2069B & 1.128 $\pm$ 0.001 & 107.0 $\pm$ 0.1 &                  &                  & 3.195 $\pm$0.059 \\
  2083B & 0.255 $\pm$ 0.002 & 166.1 $\pm$ 0.3 &                  & 1.687 $\pm$0.056 & 1.600 $\pm$0.054 \\
	2117B & 0.334 $\pm$ 0.001 & 111.5 $\pm$ 0.1 &                  &                  & 0.531 $\pm$0.055 \\
  2143B & 2.184 $\pm$ 0.005 & 317.4 $\pm$ 0.1 & 3.200 $\pm$0.120 &                  & 3.457 $\pm$0.087 \\
  2159B & 2.009 $\pm$ 0.001 & 323.8 $\pm$ 0.1 &                  & 2.638 $\pm$0.063 & 2.476 $\pm$0.060 \\
	2247B & 1.917 $\pm$ 0.002 & 350.3 $\pm$ 0.1 &                  &                  & 3.867 $\pm$0.068 \\
  2289B & 0.948 $\pm$ 0.001 & 221.7 $\pm$ 0.1 &                  &                  & 2.938 $\pm$0.055 \\
	2317B & 1.512 $\pm$ 0.002 & 112.2 $\pm$ 0.1 &                  &                  & 3.923 $\pm$0.058 \\
  2363B & 1.952 $\pm$ 0.001 & 357.3 $\pm$ 0.1 &                  &                  & 5.041 $\pm$0.086 \\
  2377B & 2.185 $\pm$ 0.002 & 335.2 $\pm$ 0.1 & 0.828 $\pm$0.080 & 0.671 $\pm$0.073 & 0.629 $\pm$0.068 \\
  2377C & 3.903 $\pm$ 0.008 & 315.9 $\pm$ 0.1 & 3.925 $\pm$0.193 & 3.816 $\pm$0.146 & 3.551 $\pm$0.170 \\
  2377D & 2.540 $\pm$ 0.002 &  41.5 $\pm$ 0.1 & 4.234 $\pm$0.096 & 4.029 $\pm$0.116 & 3.752 $\pm$0.117 \\
	2413B & 0.308 $\pm$ 0.036 & 250.1 $\pm$ 8.7 &                  & 0.470 $\pm$0.109 & 0.170 $\pm$0.059 \\
  2443B & 1.384 $\pm$ 0.002 & 164.0 $\pm$ 0.1 & 4.133 $\pm$0.066 &                  & 3.632 $\pm$0.060 \\
  2542B & 0.769 $\pm$ 0.002 &  29.1 $\pm$ 0.2 & 0.896 $\pm$0.055 &                  & 0.602 $\pm$0.054 \\
	2554B & 0.372 $\pm$ 0.010 & 149.3 $\pm$ 1.6 &                  &                  & 0.267 $\pm$0.054 \\
	2554C & 3.547 $\pm$ 0.005 & 203.6 $\pm$ 0.1 &                  &                  & 2.960 $\pm$0.098 \\
	2601B & 1.598 $\pm$ 0.002 &  14.1 $\pm$ 0.1 &                  &                  & 0.966 $\pm$0.057 \\
	2601C & 1.480 $\pm$ 0.002 & 295.2 $\pm$ 0.1 &                  &                  & 2.979 $\pm$0.057 \\
	2601D & 3.059 $\pm$ 0.003 &  30.1 $\pm$ 0.2 &                  &                  & 4.899 $\pm$0.135 \\
	2657B & 0.744 $\pm$ 0.365 & 131.7 $\pm$ 1.8 & 0.145 $\pm$0.055 & 0.126 $\pm$0.055 & 0.106 $\pm$0.054 \\
	2664B & 1.190 $\pm$ 0.005 &  90.5 $\pm$ 0.2 &                  &                  & 1.103 $\pm$0.055 \\
	2681B & 1.132 $\pm$ 0.005 & 148.0 $\pm$ 0.3 &                  &                  & 0.431 $\pm$0.056 \\
  2705B & 1.900 $\pm$ 0.003 & 304.3 $\pm$ 0.2 & 2.565 $\pm$0.097 & 2.672 $\pm$0.099 & 2.584 $\pm$0.067 \\
  2711B & 0.472 $\pm$ 0.006 & 148.9 $\pm$ 0.2 & 0.149 $\pm$0.055 & 0.122 $\pm$0.055 & 0.118 $\pm$0.055 \\
  2722B & 3.224 $\pm$ 0.001 & 283.3 $\pm$ 0.2 &	                 & 3.937 $\pm$0.084 & 3.770 $\pm$0.064 \\
	2779B & 0.965 $\pm$ 0.010 &  66.5 $\pm$ 0.6 &                  &                  & 1.752 $\pm$0.055 \\
  2813B & 1.062 $\pm$ 0.001 & 261.1 $\pm$ 0.1 &                  &                  & 1.842 $\pm$0.055 \\
	2813C & 1.842 $\pm$ 0.005 & 187.8 $\pm$ 0.2 &                  &                  & 6.547 $\pm$ 0.237\\
	2837B & 0.355 $\pm$ 0.018 & 140.5 $\pm$ 2.8 & 0.218 $\pm$0.056 & 0.199 $\pm$0.055 & 0.200 $\pm$0.055 \\
	2859B & 0.454 $\pm$ 0.001 & 290.9 $\pm$ 0.1 & 3.262 $\pm$0.067 & 3.138 $\pm$0.066 & 2.890 $\pm$0.059 \\
  2869B & 1.625 $\pm$ 0.001 & 205.2 $\pm$ 0.1 &                  &                  & 5.670 $\pm$0.074 \\
  2904B & 0.699 $\pm$ 0.001 & 225.6 $\pm$ 0.1 & 2.705 $\pm$0.055 & 2.501 $\pm$0.055 & 2.446 $\pm$0.054 \\
	2971B & 0.300 $\pm$ 0.001 & 273.9 $\pm$ 0.1 & 4.503 $\pm$0.130 &                  & 3.568 $\pm$0.057 \\
	2971C & 3.561 $\pm$ 0.004 &  37.7 $\pm$ 0.1 & 7.656 $\pm$0.219 &                  & 5.931 $\pm$0.170 \\
  3020B & 0.379 $\pm$ 0.001 & 271.6 $\pm$ 0.1 &                  &                  & 1.266 $\pm$0.057 \\
  3020C & 3.862 $\pm$ 0.001 & 231.3 $\pm$ 0.1 &                  &                  & 5.008 $\pm$0.069 \\
  3029B & 0.251 $\pm$ 0.010 & 264.3 $\pm$ 2.3 &                  &                  & 0.135 $\pm$0.060 \\
  3029C & 2.543 $\pm$ 0.005 &   4.1 $\pm$ 0.1 &                  &                  & 3.440 $\pm$0.068 \\
  3029D & 1.734 $\pm$ 0.005 & 356.2 $\pm$ 0.2 &                  &                  & 4.489 $\pm$0.071 \\
  3069B & 1.790 $\pm$ 0.002 & 108.3 $\pm$ 0.1 & 1.579 $\pm$0.056 & 1.310 $\pm$0.055 & 1.265 $\pm$0.056 \\
  3106B & 0.272 $\pm$ 0.010 & 186.3 $\pm$ 2.2 &                  &                  & 1.221 $\pm$0.131 \\
  3377B & 0.265 $\pm$ 0.001 & 334.7 $\pm$ 0.1 &                  &                  & 0.485 $\pm$0.058 \\
  3377C & 1.406 $\pm$ 0.005 &  50.2 $\pm$ 0.2 &                  &                  & 3.741 $\pm$0.063 \\
	3401B & 0.648 $\pm$ 0.010 &  98.9 $\pm$ 0.9 &                  &                  & 1.877 $\pm$0.057 \\
  4004B & 1.954 $\pm$ 0.003 & 218.1 $\pm$ 0.1 &                  &                  & 2.373 $\pm$0.076 \\
  4209B & 0.976 $\pm$ 0.001 & 205.1 $\pm$ 0.1 & 0.322 $\pm$0.059 & 0.539 $\pm$0.056 & 0.570 $\pm$0.055 \\
  4292B & 1.950 $\pm$ 0.002 &  29.9 $\pm$ 0.1 &                  & 4.813 $\pm$0.075 & 4.542 $\pm$0.079 \\
  4331B & 0.335 $\pm$ 0.005 & 100.9 $\pm$ 1.0 &                  & 0.118 $\pm$0.055 & 0.125 $\pm$0.054 \\
  4407B & 2.453 $\pm$ 0.003 & 299.9 $\pm$ 0.1 & 2.286 $\pm$0.056 & 1.956 $\pm$0.056 & 1.893 $\pm$0.058 \\
  4407C & 2.660 $\pm$ 0.003 & 311.0 $\pm$ 0.1 & 4.479 $\pm$0.490 & 4.140 $\pm$0.712 & 4.654 $\pm$0.344 \\
  4463B & 2.457 $\pm$ 0.003 & 323.9 $\pm$ 0.1 & 0.160 $\pm$0.102 & 0.242 $\pm$0.082 & 0.259 $\pm$0.068 \\
  4634B & 0.281 $\pm$ 0.001 & 276.1 $\pm$ 0.1 &                  &                  & 0.653 $\pm$0.055 \\
  4768B & 1.339 $\pm$ 0.005 & 159.0 $\pm$ 0.2 &                  &                  & 2.608 $\pm$0.071 \\
  4822B & 0.559 $\pm$ 0.010 &  63.2 $\pm$ 1.0 &                  &                  & 4.503 $\pm$0.147 \\
  4871B & 0.922 $\pm$ 0.001 & 333.6 $\pm$ 0.1 & 3.126 $\pm$0.058 & 3.026 $\pm$0.057 & 3.038 $\pm$0.055 \\
	5578B & 0.322 $\pm$ 0.001 &  97.2 $\pm$ 0.1 &                  &                  & 1.681 $\pm$0.055 \\
  5762B & 0.221 $\pm$ 0.010 & 100.3 $\pm$ 2.5 &                  &                  & 0.833 $\pm$0.076 \\

\end{longtable}

\begin{longtable}{ c c c c c c }
\setlength{\extrarowheight}{3pt}\\
\caption{Apparent Magnitudes of Resolved KOI Components}\label{magnitudes}\\
\caption*{Apparent magnitudes of individual stars from contrast measurements and literature values of blended system. The great majority of JHK values are from the 2MASS catalog \citep{liebert1995}, while sources for $i$ and $Kepler$ are more varied. All values are as reported in the Exoplanet Archive, except $JHK$ for KOI0268, which is linked to a spurious 2MASS entry.
* indicates that not all companions were detected by Robo-AO. Without contrast measurements for all objects we can not accurately determine the apparent magnitudes.
$\dagger$ indicates that although a contrast measurement has been made, there is no blended measurement, and we can not determine the apparent magnitude.}\\
	KOI   &        $m_J$      &        $m_H$      &         $m_K$      &          $m_i$     &        $m_{Kep}$ \\
	\hline
	\endfirsthead
	KOI   &        $m_J$      &        $m_H$      &         $m_K$      &          $m_i$     &        $m_{Kep}$ \\
	\hline
	\endhead
	\endfoot
	\endlastfoot
0190A &         &         & 12.876 $\pm$0.053 &         & 14.419 $\pm$0.050\\ 
0190B &         &         & 13.517 $\pm$0.089 &         & 15.748 $\pm$0.137\\ 
0191A & 13.827 $\pm$0.023 & 13.419 $\pm$0.026 & 13.340 $\pm$0.037 &         & 15.057 $\pm$0.042\\ 
0191B & 16.414 $\pm$0.057 & 16.032 $\pm$0.056 & 15.961 $\pm$0.062 &         & 18.156 $\pm$0.453\\ 
0268A & 9.773  $\pm$0.021 & 9.609  $\pm$0.023 & 9.518  $\pm$0.019 &         & 10.599 $\pm$0.301\\ 
0268B & 12.826 $\pm$0.059 & 12.259 $\pm$0.056 & 12.074 $\pm$0.056 &         & 14.603 $\pm$0.317\\ 
0268C & 13.573 $\pm$0.117 & 12.963 $\pm$0.123 & 13.499 $\pm$0.144 &         & 16.199 $\pm$0.314\\ 
0401A & 12.845 $\pm$0.023 &         & 12.402 $\pm$0.027 &         & 14.076 $\pm$0.036\\ 
0401B & 14.909 $\pm$0.056 &         & 14.037 $\pm$0.051 &         & 16.975 $\pm$0.261\\ 
0425A &         &         & 13.571 $\pm$0.043 &         & 15.101 $\pm$0.043\\ 
0425B &         &         & 14.401 $\pm$0.056 &         & 15.957 $\pm$0.076\\ 
0511A & 13.222 $\pm$0.023 & 12.957 $\pm$0.035 & 12.883 $\pm$0.034 &         & 14.276 $\pm$0.036\\ 
0511B & 15.444 $\pm$0.059 & 14.816 $\pm$0.056 & 14.647 $\pm$0.055 &         & 17.387 $\pm$0.355\\ 
0511C & 18.280 $\pm$0.127 & 17.449 $\pm$0.084 & 17.194 $\pm$0.077 &         & 20.649 $\pm$0.383\\ 
0628A &         &         & 12.484 $\pm$0.025 & 13.773 $\pm$0.020 &         \\ 
0628B &         &         & 15.486 $\pm$0.061 & 18.287 $\pm$0.138 &         \\ 
0628C &         &         & 16.356 $\pm$0.059 & 18.591 $\pm$0.217 &         \\ 
0687A &         &         & 12.415 $\pm$0.026 & 13.771 $\pm$0.042 &         \\ 
0687B &         &         & 13.667 $\pm$0.046 & 15.808 $\pm$0.238 &         \\ 
0688A & 13.158 $\pm$0.028 &         & 12.858 $\pm$0.023 &         & 14.129 $\pm$0.044\\ 
0688B & 14.708 $\pm$0.054 &         & 14.231 $\pm$0.049 &         & 16.316 $\pm$0.244\\ 
0712A & 13.176 $\pm$0.032 &         & 12.771 $\pm$0.030 & 13.831 $\pm$0.072 &         \\ 
0712B & 13.615 $\pm$0.040 &         & 13.123 $\pm$0.037 & 15.010 $\pm$0.207 &         \\ 
0931A &         &         & 13.828 $\pm$0.049 &         & 15.318 $\pm$0.030\\ 
0931B &         &         & 17.055 $\pm$0.079 &         & 18.719 $\pm$0.136\\ 
0984A & 11.178 $\pm$0.037 & 10.823 $\pm$0.036 & 10.741 $\pm$0.034 & 12.100 $\pm$0.072 & 12.340 $\pm$0.049\\ 
0984B & 11.242 $\pm$0.036 & 10.873 $\pm$0.036 & 10.797 $\pm$0.036 & 12.113 $\pm$0.074 & 12.430 $\pm$0.049\\ 
0987A & 11.328 $\pm$0.021 & 10.951 $\pm$0.020 & 10.906 $\pm$0.013 & 12.356 $\pm$0.027 & 12.590 $\pm$0.030\\ 
0987B & 13.940 $\pm$0.074 & 13.331 $\pm$0.055 & 13.146 $\pm$0.050 & 16.423 $\pm$0.629 & 16.158 $\pm$0.103\\ 
1066A &         &         & 13.922 $\pm$0.040 &         & 15.647 $\pm$0.030\\ 
1066B &         &         & 16.950 $\pm$0.077 &         & 19.837 $\pm$0.184\\ 
1067A &         &         & 13.313 $\pm$0.035 &         & 14.710 $\pm$0.029\\ 
1067B &         &         & 16.095 $\pm$0.105 &         & 18.762 $\pm$0.157\\ 
1112A & 13.509 $\pm$0.050 & 13.251 $\pm$0.039 & 13.155 $\pm$0.049 &         & 14.650 $\pm$0.029\\ 
1112B & 17.122 $\pm$0.146 & 16.204 $\pm$0.085 & 15.913 $\pm$0.081 &         & 19.222 $\pm$0.057\\ 
1151A &         & 11.917 $\pm$0.020 & 11.857 $\pm$0.021 & 13.249 $\pm$0.035 &         \\ 
1151B &         & 14.474 $\pm$0.053 & 14.263 $\pm$0.054 & 16.716 $\pm$0.562 &         \\ 
1214A &         & 13.092 $\pm$0.027 & 12.978 $\pm$0.028 &         & 14.933 $\pm$0.051\\ 
1214B &         & 15.678 $\pm$0.054 & 15.431 $\pm$0.054 &         & 16.139 $\pm$0.134\\ 
1274A & 12.088 $\pm$0.020 &         & 11.638 $\pm$0.017 & 13.142 $\pm$0.025 &         \\ 
1274B & 14.892 $\pm$0.055 &         & 14.143 $\pm$0.052 & 16.911 $\pm$0.429 &         \\ 
1375A &         & 12.310 $\pm$0.020 & 12.279 $\pm$0.018 & 13.554 $\pm$0.022 &         \\ 
1375B &         & 15.613 $\pm$0.070 & 15.668 $\pm$0.064 & 17.936 $\pm$0.506 &         \\ 
1442A & 11.354 $\pm$0.024 & 11.015 $\pm$0.028 & 10.947 $\pm$0.022 & 12.298 $\pm$0.020 &         \\ 
1442B & 15.507 $\pm$0.067 & 14.819 $\pm$0.060 & 14.581 $\pm$0.058 & 18.971 $\pm$0.563 &         \\ 
1447A &         &         & 12.292 $\pm$0.030 &         & 13.248 $\pm$0.039\\ 
1447B &         &         & 12.917 $\pm$0.043 &         & 15.288 $\pm$0.160\\ 
1536A &         & 11.405 $\pm$0.017 & 11.349 $\pm$0.018 & 12.550 $\pm$0.020 &         \\ 
1536B &         & 15.669 $\pm$0.135 & 15.532 $\pm$0.114 & 17.816 $\pm$0.570 &         \\ 
1546A & 13.783 $\pm$0.030 & 13.459 $\pm$0.029 & 13.373 $\pm$0.030 &         & 14.760 $\pm$0.095\\ 
1546B & 14.725 $\pm$0.047 & 14.245 $\pm$0.045 & 14.096 $\pm$0.045 &         & 16.318 $\pm$0.377\\ 
1546C & 17.121 $\pm$0.077 & 16.711 $\pm$0.079 & 16.851 $\pm$0.062 &         & 18.587 $\pm$0.530\\ 
1546D & 17.010 $\pm$0.064 & 16.479 $\pm$0.072 & 16.319 $\pm$0.085 &         & 18.615 $\pm$0.560\\ 
1613A & 10.915 $\pm$0.025 & 10.680 $\pm$0.027 & 10.647 $\pm$0.024 &    $\dagger$     &         \\ 
1613B & 12.049 $\pm$0.046 & 11.677 $\pm$0.048 & 11.643 $\pm$0.043 &    $\dagger$     &         \\ 
1700A &         &         & 12.890 $\pm$0.030 &         & 14.822 $\pm$0.079\\ 
1700B &         &         & 13.446 $\pm$0.040 &         & 15.893 $\pm$0.192\\ 
1784A &         &         & 12.533 $\pm$0.027 &         & 14.093 $\pm$0.055\\ 
1784B &         &         & 13.310 $\pm$0.043 &         & 14.674 $\pm$0.087\\ 
1845A & 12.881 $\pm$0.023 &         & 12.281 $\pm$0.021 &    *    &    *    \\ 
1845B & 16.119 $\pm$0.057 &         & 15.166 $\pm$0.053 &    *    &    *    \\ 
1845C & 17.146 $\pm$0.071 &         & 16.680 $\pm$0.092 &    *    &    *    \\ 
1880A & 12.293 $\pm$0.022 & 11.634 $\pm$0.018 & 11.474 $\pm$0.012 &         & 14.480 $\pm$0.033\\ 
1880B & 16.231 $\pm$0.059 & 15.785 $\pm$0.059 & 15.754 $\pm$0.058 &         & 18.146 $\pm$0.441\\ 
1884A & 14.196 $\pm$0.045 & 13.789 $\pm$0.057 & 13.738 $\pm$0.060 &         & 15.509 $\pm$0.035\\ 
1884B & 16.836 $\pm$0.067 & 16.200 $\pm$0.079 & 16.041 $\pm$0.078 &         & 19.617 $\pm$0.588\\ 
1884C & 17.268 $\pm$0.068 & 16.652 $\pm$0.079 & 16.468 $\pm$0.079 &         & 20.383 $\pm$0.436\\ 
1884D & 17.794 $\pm$0.165 & 17.321 $\pm$0.144 & 16.943 $\pm$0.146 &         & 21.075 $\pm$0.349\\ 
1891A & 13.879 $\pm$0.021 & 13.331 $\pm$0.028 & 13.274 $\pm$0.029 &         & 15.284 $\pm$0.031\\ 
1891B & 18.220 $\pm$0.077 & 17.895 $\pm$0.062 & 17.872 $\pm$0.068 &         & 19.545 $\pm$0.444\\ 
1916A & 12.797 $\pm$0.023 &         & 12.493 $\pm$0.026 &         & 13.684 $\pm$0.035\\ 
1916B & 14.002 $\pm$0.047 &         & 13.547 $\pm$0.046 &         & 16.421 $\pm$0.229\\ 
1979A & 11.941 $\pm$0.025 &         & 11.601 $\pm$0.014 & 12.845 $\pm$0.029 &         \\ 
1979B & 14.235 $\pm$0.057 &         & 13.423 $\pm$0.045 & 16.047 $\pm$0.366 &         \\ 
1989A &         &         & 11.841 $\pm$0.018 & 13.144 $\pm$0.024 & 13.372 $\pm$0.030\\ 
1989B &         &         & 14.766 $\pm$0.054 & 17.363 $\pm$0.535 & 16.866 $\pm$0.151\\ 
2001A &         &         & 11.144 $\pm$0.019 & 12.835 $\pm$0.020 & 13.135 $\pm$0.030\\ 
2001B &         &         & 15.463 $\pm$0.060 & 17.411 $\pm$0.258 & 17.617 $\pm$0.219\\ 
2009A & 12.714 $\pm$0.021 & 12.387 $\pm$0.021 & 12.333 $\pm$0.022 &         & 13.848 $\pm$0.033\\ 
2009B & 15.751 $\pm$0.089 & 15.335 $\pm$0.061 & 15.085 $\pm$0.055 &         & 17.929 $\pm$0.491\\ 
2059A &         &         & 11.182 $\pm$0.063 &         & 13.246 $\pm$0.048\\ 
2059B &         &         & 11.724 $\pm$0.098 &         & 14.339 $\pm$0.104\\ 
2069A &         &         & 12.231 $\pm$0.020 & 13.582 $\pm$0.020 & 13.777 $\pm$0.031\\ 
2069B &         &         & 15.422 $\pm$0.060 & 18.874 $\pm$0.508 & 18.026 $\pm$0.504\\ 
2083A &         & 12.263 $\pm$0.020 & 12.230 $\pm$0.024 & 13.446 $\pm$0.037 & 13.871 $\pm$0.056\\ 
2083B &         & 13.948 $\pm$0.050 & 13.827 $\pm$0.050 & 16.164 $\pm$0.359 & 14.907 $\pm$0.134\\ 
2117A &         &         & 13.516 $\pm$0.037 &         & 16.236 $\pm$0.032\\ 
2117B &         &         & 14.047 $\pm$0.046 &         & 16.567 $\pm$0.034\\ 
2143A & 12.924 $\pm$0.025 &         & 12.505 $\pm$0.026 &         & 14.145 $\pm$0.031\\ 
2143B & 16.122 $\pm$0.118 &         & 15.965 $\pm$0.088 &         & 17.654 $\pm$0.283\\ 
2159A &         & 12.098 $\pm$0.019 & 12.087 $\pm$0.021 & 13.322 $\pm$0.025 &         \\ 
2159B &         & 14.731 $\pm$0.058 & 14.563 $\pm$0.059 & 17.340 $\pm$0.512 &         \\ 
2247A &         &         & 12.046 $\pm$0.022 &         & 14.384 $\pm$0.029\\ 
2247B &         &         & 15.916 $\pm$0.069 &         & 19.508 $\pm$0.213\\ 
2289A &         &         & 12.075 $\pm$0.018 & 13.214 $\pm$0.021 & 13.374 $\pm$0.030\\ 
2289B &         &         & 15.012 $\pm$0.055 & 17.540 $\pm$0.285 & 18.008 $\pm$0.297\\ 
2317A &         &         & 12.704 $\pm$0.026 &         & 14.298 $\pm$0.031\\ 
2317B &         &         & 16.628 $\pm$0.063 &         & 19.227 $\pm$0.194\\ 
2363A &         &         & 12.360 $\pm$0.018 &         & 14.369 $\pm$0.031\\ 
2363B &         &         & 17.407 $\pm$0.085 &         & 20.945 $\pm$1.338\\ 
2377A & 13.673 $\pm$0.038 & 13.286 $\pm$0.045 & 13.245 $\pm$0.040 &    *    &    *    \\ 
2377B & 14.501 $\pm$0.063 & 13.958 $\pm$0.062 & 13.876 $\pm$0.057 &    *    &    *    \\ 
2377C & 17.598 $\pm$0.196 & 17.112 $\pm$0.148 & 16.785 $\pm$0.173 &    *    &    *    \\ 
2377D & 17.909 $\pm$0.105 & 17.318 $\pm$0.126 & 17.004 $\pm$0.120 &    *    &    *    \\ 
2413A &         & 13.352 $\pm$0.046 & 13.345 $\pm$0.041 &         & 15.236 $\pm$0.101\\ 
2413B &         & 13.820 $\pm$0.068 & 13.515 $\pm$0.044 &         & 17.342 $\pm$0.554\\ 
2443A & 12.933 $\pm$0.024 &         & 12.574 $\pm$0.022 &         & 13.995 $\pm$0.030\\ 
2443B & 17.070 $\pm$0.070 &         & 16.209 $\pm$0.063 &         & 19.395 $\pm$0.540\\ 
2542A & 13.253 $\pm$0.027 &         & 12.525 $\pm$0.034 &         & 15.841 $\pm$0.056\\ 
2542B & 14.147 $\pm$0.043 &         & 13.125 $\pm$0.043 &         & 17.037 $\pm$0.142\\ 
2554A &         &         & 13.708 $\pm$0.042 &    *    &    *    \\ 
2554B &         &         & 13.977 $\pm$0.049 &    *    &    *    \\ 
2554C &         &         & 16.667 $\pm$0.101 &    *    &    *    \\ 
2601A &         &         & 12.850 $\pm$0.032 &         & 14.222 $\pm$0.111\\ 
2601B &         &         & 13.816 $\pm$0.051 &         & 15.646 $\pm$0.384\\ 
2601C &         &         & 15.828 $\pm$0.065 &         & 18.129 $\pm$0.245\\ 
2601D &         &         & 17.752 $\pm$0.137 &         & 19.872 $\pm$1.153\\ 
2657A & 12.333 $\pm$0.032 & 11.978 $\pm$0.031 & 11.936 $\pm$0.031 &         & 13.497 $\pm$0.084\\ 
2657B & 12.477 $\pm$0.035 & 12.104 $\pm$0.033 & 12.041 $\pm$0.033 &         & 13.770 $\pm$0.103\\ 
2664A &         &         & 13.877 $\pm$0.041 &         & 16.065 $\pm$0.040\\ 
2664B &         &         & 14.979 $\pm$0.056 &         & 16.897 $\pm$0.065\\ 
2681A &         &         & 14.460 $\pm$0.057 &         & 16.295 $\pm$0.040\\ 
2681B &         &         & 14.890 $\pm$0.063 &         & 17.547 $\pm$0.091\\ 
2705A & 11.667 $\pm$0.025 & 11.016 $\pm$0.028 & 10.822 $\pm$0.024 &         & 14.765 $\pm$0.297\\ 
2705B & 14.232 $\pm$0.091 & 13.690 $\pm$0.094 & 13.404 $\pm$0.064 &         & 17.956 $\pm$0.328\\ 
2711A & 13.248 $\pm$0.033 & 12.982 $\pm$0.033 & 12.992 $\pm$0.033 &         & 14.337 $\pm$0.046\\ 
2711B & 13.397 $\pm$0.038 & 13.103 $\pm$0.035 & 13.111 $\pm$0.038 &         & 14.457 $\pm$0.052\\ 
2722A &         & 12.110 $\pm$0.023 & 12.026 $\pm$0.018 &         & 13.274 $\pm$0.029\\ 
2722B &         & 16.049 $\pm$0.082 & 15.795 $\pm$0.067 &         & 19.149 $\pm$0.138\\ 
2779A &         &         & 13.623 $\pm$0.039 &         & 15.040 $\pm$0.048\\ 
2779B &         &         & 15.376 $\pm$0.060 &         & 17.586 $\pm$0.332\\ 
2813A &         &         & 11.697 $\pm$0.020 &         & 13.951 $\pm$0.068\\ 
2813B &         &         & 13.540 $\pm$0.050 &         & 14.958 $\pm$0.155\\ 
2813C &         &         & 18.263 $\pm$0.244 &         & 22.013 $\pm$0.680\\ 
2837A & 12.997 $\pm$0.033 & 12.822 $\pm$0.031 & 12.800 $\pm$0.032 &         & 13.873 $\pm$0.035\\ 
2837B & 13.212 $\pm$0.036 & 13.022 $\pm$0.035 & 13.000 $\pm$0.036 &         & 14.102 $\pm$0.037\\ 
2859A & 12.589 $\pm$0.021 & 12.181 $\pm$0.016 & 12.125 $\pm$0.016 &         & 13.997 $\pm$0.044\\ 
2859B & 15.849 $\pm$0.067 & 15.323 $\pm$0.066 & 15.013 $\pm$0.056 &         & 16.118 $\pm$0.215\\ 
2869A &         &         & 12.367 $\pm$0.018 &         & 13.750 $\pm$0.029\\ 
2869B &         &         & 18.039 $\pm$0.074 &         & 21.629 $\pm$0.163\\ 
2904A & 11.766 $\pm$0.021 & 11.518 $\pm$0.022 & 11.467 $\pm$0.018 &         & 12.848 $\pm$0.045\\ 
2904B & 14.474 $\pm$0.053 & 14.017 $\pm$0.054 & 13.911 $\pm$0.051 &         & 14.833 $\pm$0.206\\ 
2971A & 11.792 $\pm$0.021 &         & 11.482 $\pm$0.016 &         & 12.769 $\pm$0.031\\ 
2971B & 16.289 $\pm$0.126 &         & 15.053 $\pm$0.056 &         & 16.840 $\pm$0.094\\ 
2971C & 19.444 $\pm$0.225 &         & 17.411 $\pm$0.167 &         & 20.654 $\pm$1.884\\ 
3020A &         &         & 12.323 $\pm$0.023 &         & 13.591 $\pm$0.031\\ 
3020B &         &         & 13.586 $\pm$0.046 &         & 16.818 $\pm$0.104\\ 
3020C &         &         & 17.328 $\pm$0.072 &         & 20.511 $\pm$0.335\\ 
3029A &         &         & 13.869 $\pm$0.040 &    *    &    *    \\ 
3029B &         &         & 14.004 $\pm$0.043 &    *    &    *    \\ 
3029C &         &         & 17.309 $\pm$0.076 &    *    &    *    \\ 
3029D &         &         & 18.358 $\pm$0.083 &    *    &    *    \\ 
3069A & 13.916 $\pm$0.033 & 13.561 $\pm$0.032 & 13.482 $\pm$0.037 &         & 15.049 $\pm$0.032\\ 
3069B & 15.496 $\pm$0.054 & 14.872 $\pm$0.052 & 14.747 $\pm$0.054 &         & 17.249 $\pm$0.065\\ 
3106A &         &         & 14.018 $\pm$0.052 &         & 15.858 $\pm$0.060\\ 
3106B &         &         & 15.233 $\pm$0.108 &         & 16.605 $\pm$0.111\\ 
3377A &         &         & 13.321 $\pm$0.034 &         & 15.355 $\pm$0.029\\ 
3377B &         &         & 13.805 $\pm$0.043 &         & 19.175 $\pm$0.190\\ 
3377C &         &         & 17.062 $\pm$0.070 &         & 21.181 $\pm$0.284\\ 
3401A &         &         & 12.909 $\pm$0.026 &         & 14.788 $\pm$0.069\\ 
3401B &         &         & 14.782 $\pm$0.054 &         & 15.679 $\pm$0.144\\ 
4004A &         &         & 11.221 $\pm$0.021 &         & 12.722 $\pm$0.031\\ 
4004B &         &         & 13.596 $\pm$0.069 &         & 16.732 $\pm$0.111\\ 
4209A & 15.098 $\pm$0.042 & 14.635 $\pm$0.060 & 14.496 $\pm$0.056 &         & 16.073 $\pm$0.095\\ 
4209B & 15.421 $\pm$0.047 & 15.172 $\pm$0.066 & 15.066 $\pm$0.063 &         & 18.759 $\pm$0.758\\ 
4292A &         & 11.340 $\pm$0.017 & 11.294 $\pm$0.011 &         & 12.897 $\pm$0.030\\ 
4292B &         & 16.152 $\pm$0.076 & 15.837 $\pm$0.075 &         & 20.932 $\pm$0.343\\ 
4331A &         & 12.697 $\pm$0.033 & 12.629 $\pm$0.032 &         & 13.869 $\pm$0.296\\ 
4331B &         & 12.816 $\pm$0.037 & 12.755 $\pm$0.035 &         & 14.118 $\pm$0.296\\ 
4407A & 10.292 $\pm$0.026 & 10.071 $\pm$0.029 & 9.9666 $\pm$0.018 &    *    &    *    \\ 
4407B & 12.579 $\pm$0.055 & 12.029 $\pm$0.057 & 11.859 $\pm$0.053 &    *    &    *    \\ 
4407C & 14.782 $\pm$0.475 & 14.180 $\pm$0.695 & 14.610 $\pm$0.341 &    *    &    *    \\ 
4463A & 13.575 $\pm$0.054 & 13.189 $\pm$0.046 & 13.175 $\pm$0.043 &         & 16.347 $\pm$0.033\\ 
4463B & 13.728 $\pm$0.059 & 13.428 $\pm$0.052 & 13.435 $\pm$0.048 &         & 16.356 $\pm$0.033\\ 
4634A &         &         & 12.725 $\pm$0.031 &         & 13.876 $\pm$0.080\\ 
4634B &         &         & 13.380 $\pm$0.042 &         & 15.753 $\pm$0.386\\ 
4768A &         &         & 13.886 $\pm$0.050 &         & 15.738 $\pm$0.029\\ 
4768B &         &         & 16.495 $\pm$0.265 &         & 19.726 $\pm$0.260\\ 
4822A &         &         & 12.062 $\pm$0.016 &         & 13.475 $\pm$0.030\\ 
4822B &         &         & 16.573 $\pm$0.144 &         & 20.183 $\pm$0.443\\ 
4871A & 12.125 $\pm$0.023 & 11.884 $\pm$0.022 & 11.843 $\pm$0.013 &         & 13.107 $\pm$0.032\\ 
4871B & 15.254 $\pm$0.060 & 14.908 $\pm$0.058 & 14.880 $\pm$0.054 &         & 16.225 $\pm$0.189\\ 
5578A &         &         & 9.7288 $\pm$0.018 &         & 11.281 $\pm$0.048\\ 
5578B &         &         & 11.410 $\pm$0.047 &         & 13.057 $\pm$0.193\\ 
5762A &         &         & 14.158 $\pm$0.056 &         & 16.115 $\pm$0.097\\ 
5762B &         &         & 14.993 $\pm$0.073 &         & 16.764 $\pm$0.169\\ 

\end{longtable}

\chapter{NASA Site Experiences}\label{ch:siteexperiences} 
As part of my NASA Space Technology Research Fellowship, I spent 1-2 months a year visiting a site engaged in research for NASA. The fellowship refers to this as a NASA Site Experience. This was working visits over 2 months to Goddard Space Flight Center (GSFC) in 2014, over  a month at the University of California at Santa Barbara (UCSB) in 2015, and a month back at GSFC in 2016.

The first NASA Site Experience was from July to September 2014 at GSFC with Bernie Rauscher. The goal was to modify an existing test setup to explore the fundamental noise limitations of NIR detector arrays. Today's NIR arrays are hybrids, as the detector layer is fabricated in a material that is adapted to the wavelength of interest. For many NIR arrays that material is HgCdTe. The detector layer is then hybridized by bonding it to a ROIC. Noise can arise in the detector layer, in the ROIC, and at later stages of the detection process in the readout electronics. My focus was on noise originating in the detector layer and ROIC.

Noise in the detector layer comes from several sources. Some, such as kTC noise, are easily calibrated out using multiple NDROs. However, in modern low dark current astronomical arrays, the resistive interconnect usually dominates. The interconnect interjects Johnson noise prior to the first amplifier in the ROIC. Johnson noise is recognizable because its power scales proportionally with temperature and bandwidth. One of the goals for the test setup that I worked on was to be able to make these measurements.

After the HgCdTe detector layer, the ROIC can also inject noise. One source is expected to be $1/f$ noise from the pixel readout metal-oxide-semiconductor field effect transistor (MOSFET) and other MOSFETs in the ROIC.

An H2RG was taken from the JWST team, found to be inoperable, and swapped with an H1RG from the Detector Characterization Laboratory. A custom circuit board was designed and ordered for connection allowing operation of a detector in the RIMAS cryostat. This required learning to design multi-layer circuit boards and ordering from an online manufacturer. At length I developed code to operate the detector on a new system with an unfamiliar controller. It was eventually proved to work, though parts of cabling were found to be disconnected and the detector could not be operated at that time. Additional parts of existing coding were found not to work well enough to allow for experiment to be concluded with damaged cabling.

The work at UCSB August to September 2015 was focused on the quantum efficiency (QE) measurement bench in a lab focused on the MKIDs run by Ben Mazin. An MKID is an arrayed collection of resonators with independent frequencies. Incident photons activate the resonator and put a signal on the output, which also determines the wavelength of the photon. Operating MKIDs require demanding temperatures around 100 mK. The group under Mazin has previously deployed MKIDs to the Palomar 5.1-m telescope.

\begin{figure}
   \begin{center}
   \includegraphics[height = 9.0cm]{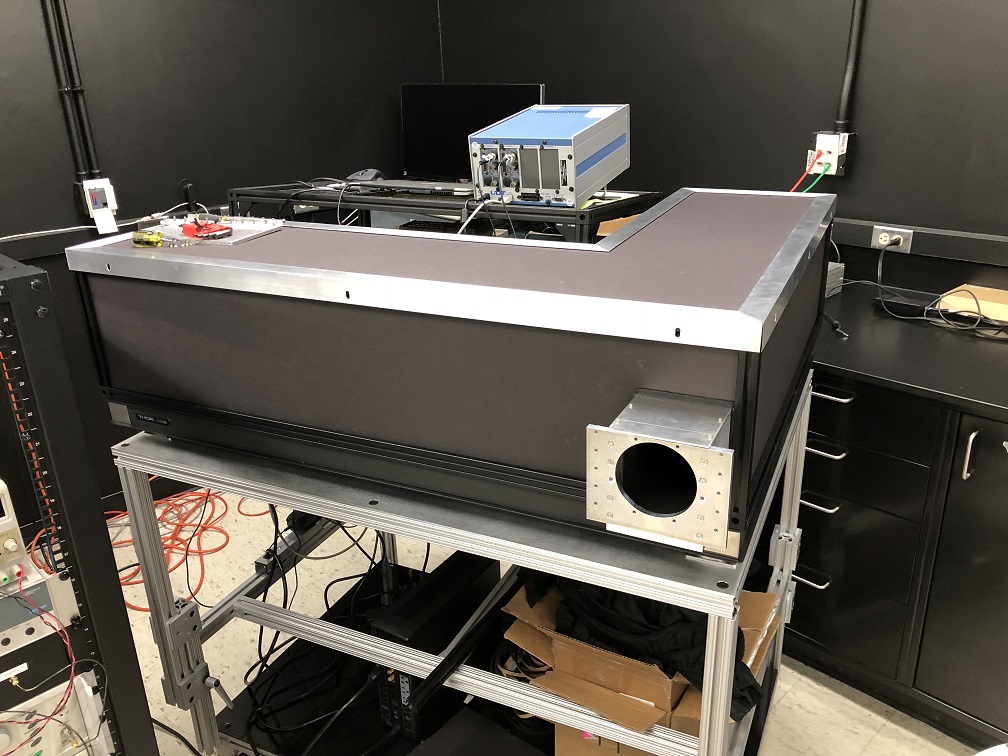}
   \end{center}
   \caption[Quantum Efficiency Bench at UCSB]{ \label{fig:QEbench} 
Equipment for measuring quantum efficiency at the Mazin Lab in UCSB, which I put together in 2015. The bench is now being used for more advanced MKIDs than were available while I worked there. (\textit{Photograph courtesy Ben Mazin.})}
   \end{figure}

The QE bench includes a calibration source with a variable wavelength and slit size. Calibration of the source was required and a new configuration to produce collimated light was to be assembled. A new physical apparatus was then required to connect the QE system to the DARKNESS cryostat. This required a full design of a new framing system and ordering of $80/20$ parts thereof. Acquisition was complicated by multiple shipments with the wrong parts. Once working, this system was able to characterize sensitivity in the MKIDs. The QE bench as assembled is still in use at the Mazin Lab of UCSB, and is currently testing new MKIDs (see Figure~\ref{fig:QEbench}). These sport an anti-reflection coating and now have QE up to $\sim70\%$, an improvement from the previous $35\%$. This work also explored an issue in MKIDs called 'collisions'. Given that each pixel has a unique frequency it outputs on, pixels for which manufacturing tolerances cause their output frequencies to overlap are colliding, and one of the pixels must be deactivated.

The last visit to GSFC occurred in August 2016 and focused on programmatics. This included participation in the LUVIOR Design Meeting. During the LUVOIR work I was able to learn how NASA science groups define very high level requirements. I worked with Bernie Rauscher to better understand future aims of detector development and how development could meet future LUVOIR requirements. During the stay I also worked on data analysis of NIR detectors, including some poorly understood outcomes of averaging data. 

\section{Summary}
This work covered 5 months over the last 4 years of my graduate school. At GSFC I worked to perform new measurements on the ubiquitous HxRG devices and participated in a design meeting for the upcoming major NASA telescope LUVOIR with NIR detector expert Bernie Rauscher. At UCSB I made a QE testbench for use with the MKIDs created by Ben Mazin that is still used. Not only was this work productive but it also helps with my post-graduate school employment at GSFC with Bernie Rauscher, where I will be performing further work on NIR detectors for future space telescopes.

\chapter{Dissertation Conclusions}\label{ch:conclusions} 
Over the last 5 years I've worked to operate the SAPHIRA in the lab, on telescopes, and extensively characterize its performance in both. My work at UH-IfA has covered the development of the SAPHIRA from the initial Mark 2 all the way to the current Mark 19. The SAPHIRA has come down from a dark current of $10\esp$ to $\sim0.025\esp$, comparable to other NIR arrays. It has proved capable of photon-counting. I've worked in deployments of the SAPHIRA to 3 different telescopes, integrating it to improve 2 existing AO instruments. I produced new characterizations of \textit{Kepler} Objects of Interest to refine our understanding of over 100 exoplanets. I've worked with teams on both ends of the mainland US on other NIR detectors to improve my understanding of the field. I worked the SAPHIRA during the first ever on-sky use of the array in the world. I believe over the last several years I've contributed a great deal to the development of necessary NIR APD arrays for astronomy, to our understanding of exoplanets, published 3 first-author papers, contributed to several others, written several conference publications, contributed to several co-author papers, and helped develop other NIR devices. I'm thankful to the Institute for Astronomy at the University of Hawai'i at M\={a}noa for letting me contribute to this extremely valuable work, and I'm excited to start my postdoctoral position at NASA Goddard Space Flight Center and continue contributing to astronomy.

\bibliographystyle{astron}
\bibliography{biblio}

\end{document}